\documentclass[10pt]{article}

\usepackage{amsmath}
\usepackage{amssymb}
\usepackage{graphicx}
\usepackage{cite}
\usepackage{color}
\usepackage[labelfont=bf]{caption}
\usepackage{setspace}
\usepackage{tabularx}
\usepackage{reaction}
\usepackage{longtable}

\makeatletter
\renewcommand{\@biblabel}[1]{\quad#1.}
\makeatother

\newcommand{\ie}{{\emph i.e.}}

\begin{document}
\begin{flushleft}
{\Large
\textbf{The origin of large molecules in primordial autocatalytic reaction networks}
}
\\
Varun Giri$^{1}$,
Sanjay Jain$^{1,2,3\ast}$
\\
\bf{1} Department of Physics and Astrophysics, University of Delhi, Delhi 110007, India
\\
\bf{2} Jawaharlal Nehru Centre for Advanced Scientific Research, Bangalore 560064, India
\\
\bf{3} Santa Fe Institute, 1399 Hyde Park Road, Santa Fe, NM 87501, USA
\\
$\ast$ E-mail: jain@physics.du.ac.in
\end{flushleft}

\section*{Abstract}
Large molecules such as proteins and nucleic acids are crucial for life, yet their primordial origin remains a major puzzle. The production of large molecules, as we know it today, requires good catalysts, and the only good catalysts we know that can accomplish this task consist of large molecules. Thus the origin of large molecules is a chicken and egg problem in chemistry. Here we present a mechanism, based on autocatalytic sets (ACSs), that is a possible solution to this problem. We discuss a mathematical model describing the population dynamics of molecules in a stylized but prebiotically plausible chemistry. Large molecules can be produced in this chemistry by the coalescing of smaller ones, with the smallest molecules, the `food set', being buffered. Some of the reactions can be catalyzed by molecules within the chemistry with varying catalytic strengths. Normally the concentrations of large molecules in such a scenario are very small, diminishing exponentially with their size. ACSs, if present in the catalytic network, can focus the resources of the system into a sparse set of molecules. ACSs can produce a bistability in the population dynamics and, in particular, steady states wherein the ACS molecules dominate the population. However to reach these steady states from initial conditions that contain only the food set typically requires very large catalytic strengths, growing exponentially with the size of the catalyst molecule. We present a solution to this problem by studying `nested ACSs', a structure in which a small ACS is connected to a larger one and reinforces it. We show that when the network contains a cascade of nested ACSs with the catalytic strengths of molecules increasing gradually with their size (e.g., as a power law), a sparse subset of molecules including some very large molecules can come to dominate the system.

\section*{Introduction}
One of the puzzles in the origin of life is the question: How did large molecules, which are essential for all cells to function, first arise? Macromolecules such as RNA and protein molecules, which contain from about a hundred to several thousand monomers, are produced in cells with the help of two crucial catalysts (a) the RNA polymerase which reads the genes on DNA molecules and produces the corresponding messenger RNA molecules and (b) the ribosome which reads the messenger RNA molecules and produces the corresponding protein molecules. These two powerful catalysts, RNA polymerase and ribosome, are themselves made up of proteins and RNA molecules, each of which is produced by the process mentioned above. When cells produce daughter cells, the latter are already endowed with these catalysts at birth, from which they synthesize other molecules. Nowhere in the living world is there a natural process we know of that produces macromolecules and that does {\it not} itself use macromolecules. Hence the puzzle. We expect that the answer to the question lies in the processes that occurred before life originated.

The Miller experiment \cite{Miller1953} and subsequent work \cite{Miller1959, Ricardo2004, Powner2009, Parker2011} were successful in synthesizing monomer building blocks of large molecules in simulated prebiotic environments. Those experiments suggested that amino acids and nucleotides, monomer building blocks of macromolecules, could be produced on the prebiotic earth. Subsequently there has been much experimental work to explore mechanisms that could enhance the concentrations of monomers and synthesize long polymers \cite{Rode1999, Ferris2004, Orgel2004, Budin2010}. While there is interesting progress, as yet there is no compelling scenario for the primordial origin of large molecules.

Meanwhile what has been observed is that catalysis is a fairly ubiquitous property that arises in different kinds of molecules and even at small sizes. Organocatalysts \cite{Barbas2008, MacMillan2008, Bertelsen2009}, peptides \cite{Severin1997, Brack2007}, and RNA molecules \cite{Cech1986a, Symons1992, Chen2007} are known to have catalytic properties. Cofactors play an important role in catalyzing metabolic reactions and they (or their evolutionary predecessors) may have had a role in prebiotic catalysis \cite{Srinivasan2009}.

The ubiquity of catalysis motivates the main idea behind the present paper. Here we attempt to investigate theoretically, using a mathematical model, whether one can construct a chemical organization that produces large molecules from small ones, using the property of catalysis. Apart from the specific question of the origin of large molecules the present work is also motivated by a larger question of how complex structures and organizations are built incrementally from simpler ones. In systems where catalysis is possible an important self-organizing structure that can appear is an autocatalytic set (ACS). ACSs were proposed by Eigen \cite{Eigen1971}, Kauffman \cite{Kauffman1971} and Rossler \cite{Rossler1971} and have been used by many authors to study various aspects of self-organization, evolution and the origin of metabolism \cite{Farmer1986, Bagley1991, Bagley1991a, Stadler1993, Kauffman1993, Fontana1994, Jain1998, Jain2001, Hanel2005, Piedrafita2010}, the origin of replication \cite{Eigen1977, Szathmary2006, Ohtsuki2009}, and the origin and dynamics of  protocells \cite{Segre2000, Furusawa2006, Carletti2008, Kamimura2010}. In order to separate the issues, the present model only has catalysis and no replication or spatial enclosures; we wish to see what can be achieved by catalysis alone.

Farmer et al \cite{Farmer1986}, Bagley et al \cite{Bagley1989}, and Bagley and Farmer \cite{Bagley1991} proposed and analyzed a model of an artificial chemistry in which polymers could form by ligation of shorter polymers through spontaneous reactions as well as reactions catalyzed by other polymers in the chemistry. Bagley and Farmer \cite{Bagley1991} analyzed the population dynamics of the molecular species and established some important properties of autocatalytic self-organization. When the food set (monomers) were supplied at a fixed input rate and the chemistry contained an ACS they showed that in a suitable range of parameters the concentrations of the ACS molecules dominated over the rest of the molecules (the background), thereby focusing the chemical resources of the system into a small subset of molecules comprising the ACS. However the largest polymers in the ACSs they considered had about 15-20 monomers; they did not systematically investigate the problems that arise in generating much larger molecules in their chemistry.

These problems were sharply articulated in the work of Ohtsuki and Nowak \cite{Ohtsuki2009}, in which they considered a much simpler model that could be analytically solved. In this model, which they refer to as `symmetric prelife' with a catalyst, they showed that in order for the catalyst to acquire a significant concentration in a prebiotic scenario its catalytic strength should be very large, growing exponentially with its length. The inference from the model, therefore, was that it is difficult for a large catalyst molecule to arise in a prebiotic scenario.

In this work we consider a model of artificial chemistry similar in structure to that of Bagley and Farmer. This model is intermediate in complexity and realism between the model of Bagley and Farmer (which is slightly more complex) and model of Ohtsuki and Nowak (which is much simpler). We study the dynamics of this model in the presence of ACSs and in particular a structure that we refer to as a `nested ACS' in which a small ACS helps trigger a larger one. We show that this mechanism when iterated across a cascade of nested ACSs avoids the problem of exponentially growing catalyst strengths. This mechanism, therefore, provides a possible route to the construction of large molecules in a pre-biotic scenario. Apart from these results our work provides an insight, based on the analytic treatment of the system under certain approximations as well as numerical work, of certain ACS properties and questions such as why ACSs dominate, why nested ACSs work, etc.

\section*{The Model}
The model is specified by describing the set of molecular species, their reactions, and the dynamical rate equations for their population dynamics. A special set of molecules, the `food set', denoted ${\mathcal F}$, consists of small molecular species, $f$ in number, that are presumed to be abundantly present in a prebiotic niche. The simplest version of the model ($f=1$) contains only a single monomer species A (or A(1)) whose concentration $x_1$ in a well stirred prebiotic region will be assumed to be buffered (constant). The other molecules, A(2), A(3),... (dimers, trimers, etc.), whose concentrations are denoted $x_2,x_3,\ldots$, are all made through ligation and cleavage reactions of the type $\mathrm{A}(i) + \mathrm{A}(j) \rightleftharpoons \mathrm{A}(i+j)$ with forward (ligation) rate constant denoted $k^F_{ij}$ and reverse (cleavage) rate constant $k^R_{ij}$. The net forward flux of this reaction pair is given by $v_{ij} = (k^F_{ij}x_ix_j - k^R_{ij}x_{i+j})$. The rate equations for the system are given by $\dot{x}_1 = 0$, and, for $n=2,3,\ldots$,
\begin{eqnarray}\label{rateequation1}
\dot{x}_n &=& \sum_{i\leq j,i+j=n}v_{ij} -
 \sum_{i=1, i \neq n}^\infty v_{in} - 2v_{nn} - \phi_n x_n \\
&=& \sum_{i\leq j, i+j=n}(k^F_{ij}x_ix_j - k^R_{ij}x_n) + \sum_{i=1,
i \neq n}^\infty(k^R_{in}x_{i+n} - k^F_{in}x_ix_n) +
2(k^R_{nn}x_{2n} - k^F_{nn}x_n^2) - \phi_n x_n,
\end{eqnarray}
where $\phi_n$ represents a loss rate of species $n$ from the region in question. The two terms in the first sum represent the formation (respectively, cleavage) of $\mathrm{A}(n)$ from (into) smaller molecules. The two terms in the second sum and the following bracket represent the cleavage (respectively, formation) of larger molecules via reactions that produce (consume) $\mathrm{A}(n)$. The stoichiometric factor of 2 before the bracket arises because two molecules of $\mathrm{A}(n)$ are involved in the corresponding reaction pair. The set of parameters $k^F_{ij}$, $k^R_{ij}$ that are non-zero define the set of possible reactions; collectively they define the `spontaneous chemistry'(`spontaneous' in the sense that the reactions are possible even in the absence of catalysts). A pair of ligation and cleavage reactions can be excluded from the chemistry by setting both $k^F_{ij}$ and $k^R_{ij}$ to zero. The scheme permits chemistries in which some reactions proceed in only one direction (ligation or cleavage) by setting only one of $k^F_{ij}$ and $k^R_{ij}$ to zero. However, we will primarily be interested in a chemistry in which each reaction is reversible. The existence of the cleavage reactions makes it more difficult for the long molecules to survive; thus it is more significant to demonstrate the appearance of long molecules in a model in which cleavage reactions are permitted than in one where only the forward (ligation) reactions are.

We consider a simple scheme for catalyzed reactions, assuming that a molecule enhances the rate of a reaction that it catalyzes in proportion to its own concentration. Thus, if $\mathrm{A}(m)$ is a catalyst of the reaction pair $\mathrm{A}(i) + \mathrm{A}(j) \rightleftharpoons \mathrm{A}(i+j)$, then the rate constants of this reaction pair, $k^F_{ij}$ and $k^R_{ij}$, are replaced $k^F_{ij} \rightarrow k^F_{ij}(1 + \kappa^{ij}_m x_m)$ and $k^R_{ij} \rightarrow k^R_{ij}(1 + \kappa^{ij}_m x_m)$, where $\kappa^{ij}_m$ is the `catalytic strength' of the catalyst for this reaction pair. The first term in the bracket, unity, represents the spontaneous reaction rate (which is present irrespective of whether the reaction is catalyzed or not), and the second term $\kappa^{ij}_m x_m$ represents the enhancement of the reaction rate due to the catalyst. Note that in this scheme a catalyst enhances both the forward and reverse reaction rates by the same factor. If a reaction has multiple catalysts, $\kappa^{ij}_m x_m$ is replaced by $\sum_m \kappa^{ij}_m x_m$, where the sum runs over all catalysts $\mathrm{A}(m)$ of the reaction in question. Typically, only a small subset of the spontaneous reactions will be catalyzed. The set of catalyzed reactions together with the catalysts and their catalytic strengths will be referred to as the `catalyzed chemistry'.

When there are $f$ food set (or `monomer') species a general molecule A is represented as an $f$-tuple of non-negative integers: $\mathrm{A} = (i_1, i_2, \ldots, i_f)$, where $i_l$ is number of monomers of type $l$ contained in A. The identity of a molecule in the model is completely determined by the number of monomers of each type contained in the molecule; the order in which they appear is irrelevant. Thus the combinatorial diversity of distinct compounds containing a total of $n$ monomers (of all types) grows only as a power of $n$ ($\sim n^{f-1}$) instead of exponentially ($\sim f^n$ for strings) if the order had mattered. This simplification helps in picturizing the chemistry and significantly reducing the computational power needed to explore large values of $n$. The reaction scheme and rate equations are similar to the `1-dimensional' version above. Details of the general model and explicit examples of rate equations for $f=1$ and 2 are discussed in the Supporting Appendix S1.

The main differences between the present model and that of Bagley and Farmer are (a) a simpler representation of molecules (we do not consider molecules as strings), (b) a simpler treatment of catalysis (we do not consider intermediate complexes), and (c) we ignore the effects coming from small populations containing a discrete number of molecules. We reproduce the main phenomenon of ACS dominance that Bagley and Farmer observed, but the relative simplicity of the present model allows us to explore other phenomena that they do not report about (this includes a multistability in the dynamics and the possibility of building large molecules through nested ACSs).

The main differences with the model of Ohtsuki and Nowak are (a) a much richer spontaneous chemistry of ligation reactions and the inclusion of reverse reactions (which makes an analytical treatment more difficult), and (b) a much more general class of catalyzed chemistries, instead of a single catalyst (which allows us to talk of nested ACSs, in particular). With a specific choice of parameters our $f=1$ model reduces exactly to their `symmetric prelife' model with a catalyst. In spite of greater complexity we are able to numerically reproduce their main results in a much more general setting, and also provide approximate analytical understanding of the results.

\subsection*{Autocatalytic Sets (ACSs)}

The dynamics of the above system is particularly interesting when autocatalytic sets (ACSs) are present in the catalyzed chemistry. Consider a set $S$ of catalyzed one-way reactions. `One-way' means that each reaction in $S$ is either a ligation or cleavage reaction. Thus the set of reactants and the set of products are unambiguously defined for each reaction and the two sets are distinguished. The presence of a given ligation or cleavage reaction in $S$ does not mean that its reverse is also necessarily a member of $S$. Let $P(S)$ be the union of sets of products of all reactions in $S$, and $R(S)$ the union of sets of reactants of all reactions in $S$. We exclude the food set molecules from both $P(S)$ and $R(S)$. We will refer to the set $S$ of catalyzed reactions as an ACS if (a) $P(S)$ includes a catalyst for every reaction in $S$, and (b) $R(S) \subset P(S)$. The latter condition implies that all members of $R(S)$ can be produced from the food set by (recursively) applying reactions from within $S$. An ACS thus ensures the existence of a catalyzed pathway, starting from the food set, for the production of each of its products \cite{Eigen1971,Kauffman1971,Rossler1971}. Alternative valid definitions of an ACS can be given (see \cite{Wachtershauser1990, Morowitz2000} for one such); the above definition suffices for our present purposes and we hope to return to consequences of other kinds of ACSs in the future. Note that if $S$ is an ACS, then its extension, $S'$, that additionally includes the reverse of some reactions in $S$, is also trivially an ACS, as in our scheme a catalyst works for both forward and reverse reactions if both exist in the chemistry.

\section*{Results}

\subsection*{Spontaneous (uncatalyzed) chemistries}
\subsubsection*{Nomenclature}
We first consider the case when none of the reactions is catalyzed, $\kappa^{ij}_m =0$ for all $ij$ pairs. For concreteness we first consider spontaneous chemistries that are `reversible', `homogeneous' and `fully connected'. A `reversible' chemistry is one for which each allowed reaction is reversible, i.e., $k^F_{ij} \neq 0 \iff k^R_{ij} \neq 0$. A `homogeneous' chemistry is one in which all the nonzero rate constants are independent of the species labels: $\phi_n = \phi$ for all $n$, $k^F_{ij} \neq 0 \implies k^F_{ij} = k_f$ independent of $i$ and $j$, and $k^R_{ij} \neq 0 \implies k^R_{ij} = k_r$ independent of $i$ and $j$. A `fully connected' chemistry is one in which all possible ligation and cleavage reactions are allowed: $k^F_{ij} \neq 0$ and $k^R_{ij} \neq 0$ for all $i,j$. A chemistry is `connected' if every molecule can be produced from the food set in some pathway consisting of a sequence of allowed reactions. In this paper we discuss spontaneous chemistries that are reversible and homogeneous. We have checked that introducing irreversible reactions and bringing in a small amount of heterogeneity does not change the conclusions. Some results for sparse chemistries are discussed later. For homogeneous and fully connected chemistries the model has 4 parameters, $k_f$, $k_r$, $\phi$, and the concentration of the monomer, $x_1 \equiv A$.

We explore the model numerically and, to a limited extent, analytically. While the chemistry under consideration is infinite, numerical simulations were done by choosing a finite number $N$ for the size of the largest molecule in the simulation. In simulating Eq. (\ref{rateequation1}) all terms corresponding to reactions in which any molecule larger than $\mathrm{A}(N)$ is produced or consumed were omitted. In principle this introduces another parameter, $N$, an artifact of the simulation. However, one expects that most properties of physical interest should become independent of $N$ when $N$ is sufficiently large. Evidence for this is presented in Supporting Appendix S2. Our numerical work was mostly done using the CVODE solver library of the SUNDIALS (Suite of Nonlinear and Differential/Algebraic Equation Solvers) package \cite{SUNDIALS}, and, for smaller $N$ values, using XPPAUT \cite{XPPAUT}.

\subsubsection*{Steady state properties of the spontaneous chemistry: Populations decline exponentially with the size of molecules}
Starting from the initial condition in which all concentrations other than the food set are zero (we refer to this as the standard initial condition), the concentrations were found to increase monotonically and reach a steady state (Fig. \ref{noacs}A). Numerically the graph of steady state $x_n$ versus $n$ on a semi-log plot was found to be approximately a straight line for large $n$, consistent with the expression
\begin{equation}\label{ss1}
x_n = c\Lambda^{n} = ce^{-\gamma n},
\end{equation}
where, $c$ and $\Lambda=e^{-\gamma}$ are constants. $\Lambda$, determined by numerically fitting the slope, decreases monotonically as $\phi$ increases (Fig. \ref{noacs}B).

For $\phi = 0$, the following exact analytical solution for the steady state concentrations exists for homogeneous and connected uncatalyzed chemistries:
\begin{equation}\label{sszerophi}
x_n = A \left(\frac{k_f A}{k_r}\right)^{n-1}.
\end{equation}
To see that this is a fixed point, note that when Eq. (\ref{sszerophi}) holds, then $v_{ij} = k_fx_ix_j - k_r x_{i+j} = 0$ for all $i,j = 1,2,\ldots$; hence the r.h.s. of Eq. (\ref{rateequation1}) vanishes (at $\phi = 0$). Thus $\Lambda(\phi =0) = k_f A/k_r$. Hence, whenever $k_f A < k_r$, the steady state concentrations of large molecules are exponentially damped, $\Lambda < 1$.

When $\phi > 0$ we do not have an analytic solution. Numerically, we find that $\Lambda (\phi>0)$ drops to below 1 even when $\frac{k_fA}{k_r} > 1$. $\Lambda$ is found to be a monotonically increasing function of $k_f$ and $A$, and a monotonically decreasing function of $k_r$ and $\phi$. This corresponds to the intuition that an increased ligation rate favours large molecules and an increased cleavage or dissipation rate disfavours them. By casting the rate equation in terms of dimensionless variables one can easily see that there are only two independent parameters, which may be taken to be $k' = k_f A/k_r$ and $\phi' = \phi/k_r$ whenever $k_r \neq 0$ (for details see Supporting Appendix S3). Alternatively when $\phi\neq0$, we can take the two dimensionless parameters to be $\frac{k_fA}{\phi}$ and $\frac{k_r}{\phi}$. The dependence of $\Lambda$ on these two sets of parameters is also shown in Supporting Appendix S3. The uncatalyzed chemistry seems to have a global fixed point attractor (all initial conditions tested lead to the same steady state).

Similar results hold when two food sources are present in the system ($f=2$) with buffered concentrations of the monomers (1,0) and (0,1). Simulations are done with all possible reaction and cleavage reactions allowed between molecules containing a maximum of $N$ monomers, all with the same forward rate constant $k_f$ and reverse rate constant $k_r$ and a common dissipation rate $\phi$ for the molecules. A steady state concentration profile is shown in Fig. \ref{noacs-2d}. `Diagonal entries' ($n_1 = n_2$) have higher concentrations in homogeneous chemistries because there are more reaction pathways to build molecules with equal numbers of both monomers than unequal. Since the number of species goes as $N^2/2$ and the number of reactions as $\sim N^4$, computational limitations require us to work with a smaller $N$ than for $f=1$. Qualitative conclusions nevertheless appear to be $N$ independent.

\subsection*{Chemistries with autocatalytic sets}
\subsubsection*{ACS molecules dominate the population in certain parameter regions}
We now consider chemistries which contain some catalyzed reactions in addition to the spontaneous reactions described above. As a specific example to display certain generic properties, we consider the following catalyzed chemistry:
\begin{subequations}
 \label{acs65-definition}
 \begin{eqnarray}
    \nonumber   & \mathrm{A}(9) & \\
    \label {acs65-rct1} \mathrm{A}(1) + \mathrm{A}(1) & \rightleftharpoons & \mathrm{A}(2) \\
    \nonumber   & \mathrm{A}(5) & \\
    \mathrm{A}(2) + \mathrm{A}(2) & \rightleftharpoons & \mathrm{A}(4) \\
    \nonumber   & \mathrm{A}(28) & \\
    \mathrm{A}(1) + \mathrm{A}(4) & \rightleftharpoons & \mathrm{A}(5) \\
    \nonumber   & \mathrm{A}(14) & \\
    \mathrm{A}(4) + \mathrm{A}(5) & \rightleftharpoons & \mathrm{A}(9) \\
    \nonumber   & \mathrm{A}(37) & \\
    \mathrm{A}(5) + \mathrm{A}(9) & \rightleftharpoons & \mathrm{A}(14) \\
    \nonumber   & \mathrm{A}(37) & \\
    \mathrm{A}(14) + \mathrm{A}(14) & \rightleftharpoons & \mathrm{A}(28) \\
    \nonumber   & \mathrm{A}(65) & \\
    \mathrm{A}(9) + \mathrm{A}(28) & \rightleftharpoons & \mathrm{A}(37) \\
    \nonumber   & \mathrm{A}(14) & \\
`    \mathrm{A}(28) + \mathrm{A}(37) & \rightleftharpoons & \mathrm{A}(65).
\end{eqnarray}
\end{subequations}

Note that this set of reactions constitutes an ACS (which we will refer to as ACS65). If any one reaction pair is deleted from the set, it is no longer an ACS. For the moment, for simplicity, we consider the case where the catalytic strengths of all the catalyzed reactions are equal (`homogeneous' catalytic strengths): $\kappa^{1,1}_9 = \kappa^{2,2}_{5} =
\kappa^{1,4}_{28} =\kappa^{4,5}_{14} =\kappa^{5,9}_{37} =\kappa^{14,14}_{37} =\kappa^{9,28}_{65} =\kappa^{28,37}_{14} =\kappa$, and all other $\kappa^{ij}_m = 0$. (For clarity, in view of double digit indices, we have introduced a comma between the pair
of indices in the superscript.) Fig. \ref{acs-eg}A describes the steady state concentrations, starting from the standard initial condition, for the chemistry that contains these eight catalyzed reactions in addition to all the reactions of the fully connected spontaneous chemistry. At $\kappa=2.5 \times 10^6$ the ACS product molecules dominate over the background (the `background' being defined as the set of all molecules except the ACS product molecules and the food set), in the sense that the ACS molecules have significantly larger populations than the background molecules of similar size \cite{Bagley1991}. There is a fairly sharp threshold value of $\kappa$ above which ACS domination appears, as evident from the comparison with the lower curve in Fig. \ref{acs-eg}A drawn for $\kappa=2.0 \times 10^6$. Fig. \ref{acs-eg}B shows that the steady state background concentrations decline as $\phi$ increases, while the ACS concentrations are relatively unaffected in this regime (thus ACS domination increases).  If catalyzed production pathways from the food set to other molecules are broken somewhere, the concentration of the latter molecules declines significantly. This is evident from Fig. \ref{acs-eg}C for which only one reaction pair (\ref{acs65-rct1}) is deleted from the catalyzed chemistry (which now contains no ACS) while others are catalyzed at the same strength as before.

ACS domination at a sufficiently high catalytic strength also occurs when there is more than one monomer. An example with $f=2$ is shown in Fig. \ref{acs-eg-2d}.

\subsubsection*{Understanding why ACS concentrations are large (the $\kappa \rightarrow \infty$ limit)}
The above features are generic for a large class of ACSs. It is instructive to consider the $\kappa \rightarrow \infty$ limit which we discuss analytically. When $\kappa$ is nonzero, the terms in Eq. (\ref{rateequation1}) corresponding to catalyzed reactions get modified. The net flux $v$ of such reaction pairs on the r.h.s. (for brevity we are omitting the subscript $ij$ in $v_{ij}$) is replaced by $(1 + \kappa \sum_m x_m)v$, where the sum over $m$ is a sum over all catalysts of the reaction pair. Now let the set $S$ of catalyzed reactions be an ACS. Then, if $\mathrm{A}(n) \in P(S)$ the r.h.s. of
$\dot{x}_n$ contains at least one such catalyzed term, while if $\mathrm{A}(n) \notin P(S)$ $\dot{x}_n$ contains no such term. For example, for the above mentioned ACS, we have
\begin{subequations}
  \label{kappainftyrateeqns}
  \begin{eqnarray}
  \label{kappainftyrateeqns-x2dot}
  \dot{x}_2 & \simeq & \kappa (x_9 v_{1,1} - 2 x_5 v_{2,2}) + (\mathrm{terms\ independent\ of \ }\kappa)\\
  \label{kappainftyrateeqns-x4dot}
  \dot{x}_4 & \simeq & \kappa (x_5 v_{2,2} - x_{28} v_{1,4} - x_{14} v_{4,5}) + (\kappa^0 \mathrm {\ terms})\\
  \label{kappainftyrateeqns-x5dot}
  \dot{x}_5 & \simeq & \kappa (x_{28} v_{1,4} - x_{14} v_{4,5} - x_{37} v_{5,9}) + (\kappa^0 \mathrm{\ terms})\\
  \label{kappainftyrateeqns-x9dot}
  \dot{x}_9 & \simeq & \kappa (x_{14} v_{4,5} - x_{37} v_{5,9} - x_{65} v_{9,28}) + (\kappa^0 \mathrm {\ terms})\\
  \label{kappainftyrateeqns-x14dot}
  \dot{x}_{14} & \simeq & \kappa (x_{37} v_{5,9} - 2 x_{37} v_{14,14}) + (\kappa^0 \mathrm{\ terms})\\
  \label{kappainftyrateeqns-x28dot}
  \dot{x}_{28} & \simeq & \kappa (x_{37} v_{14,14} - x_{65} v_{9,28} - x_{14} v_{28,37}) + (\kappa^0 \mathrm{\ terms})\\
  \label{kappainftyrateeqns-x37dot}
  \dot{x}_{37} & \simeq & \kappa (x_{65} v_{9,28} - x_{14} v_{28,37}) + (\kappa^0 \mathrm{\ terms})\\
  \label{kappainftyrateeqns-x65dot}
  \dot{x}_{65} & \simeq & \kappa (x_{14} v_{28,37}) + (\kappa^0 \mathrm{\ terms}),
 \end{eqnarray}
\end{subequations}
while the rate equations for all other (non ACS) molecules ($\dot{x}_3$, $\dot{x}_6$, etc.) have no terms proportional to $\kappa$. In a steady state solution the r.h.s. of Eqs. (\ref{kappainftyrateeqns}) is zero, and to leading order in the $\kappa \rightarrow \infty$ limit we must set the coefficients of $\kappa$ to zero. The coefficients involve only the ACS fluxes $v_{ij}$ and catalyst concentrations. Each coefficient is a sum of terms, and each term is proportional to an ACS flux $v_{ij}$. Thus $v_{ij} = 0$ for the ACS fluxes provides a steady state solution in the $\kappa \rightarrow \infty$ limit. Numerically we find that when $\kappa$ is sufficiently high the rate equations converge to this solution starting from the standard initial condition. Now $v_{ij} = k_f x_i x_j - k_r x_{i+j}$, therefore $v_{ij} = 0$ implies $x_{i+j} = k_f x_i x_j / k_r$ for the members of $P(S)$. Since by definition there is a catalyzed pathway from the food set to every ACS product, we can recursively express the steady state concentration of every ACS molecule in terms of $x_1 = A$: $x_n = A (k_f A/k_r)^{n-1}$.

It is evident that this argument applies whenever the set $S$ of catalyzed reactions is an ACS; thus for every member of $P(S)$, $x_n \simeq A (k_f A/k_r)^{n-1}$ is a steady state solution of the rate equations in the limit $\kappa \rightarrow \infty$. This is corroborated numerically: in Fig. \ref{acs-eg}B since $A = k_f = k_r = 1$, all the eight ACS products should have $x_n = 1$ in this limit; the numerical result at $\kappa = 3 \times 10^6$ is not too far from this limiting analytical value.

\subsubsection*{A strong ACS counteracts dissipation}
Recall from Eq. (\ref{sszerophi}) and the discussion following it that every molecule
in a homogeneous connected uncatalyzed chemistry has the steady state concentration $x_n = A (k_f A/k_r)^{n-1}$ when there is no dilution flux or dissipation ($\phi = 0$), and a smaller concentration when there is dissipation ($\phi > 0$). We have observed above that an ACS with a sufficiently large $\kappa$ can boost the steady state concentrations of its members, even when $\phi > 0$, to the same level. The expression $x_n = A (k_f A/k_r)^{n-1}$ seems to represent an upper limit on the steady state concentration of $\mathrm{A}(n)$, which can be approached either when dissipation goes to zero, or, when there is dissipation, by membership of an ACS whose catalytic strength becomes very large.

When the reaction pair $\mathrm{A}(1) + \mathrm{A}(1) \rightleftharpoons \mathrm{A}(2)$ is
not catalyzed the production of A(2) takes place at a much smaller rate, the spontaneous rate. Therefore its concentration is much smaller, and hence so are the concentrations of the larger molecules.

When $\mathrm{A}(n)$ belongs to the background the r.h.s. of $\dot{x}_n$ contains no term proportional to $\kappa$, and all the $\kappa$-independent terms have to be kept, including the $\phi x_n$ term. Thus its steady state concentration depends upon $\phi$, and as in the case of the uncatalyzed chemistry, declines more rapidly with $n$ when $\phi$ increases.

\subsubsection*{Multistability in the ACS dynamics and ACS domination}
The reason for the sudden change in the qualitative character of the steady state profile as $\kappa$ is increased is a bistability in the chemical dynamics due to the
presence of the ACS. Fig. \ref{acs-eg-bis} shows three regions in the phase diagram of the system, separated by values $\kappa^{I}$ and $\kappa^{II}$ of $\kappa$. For $0 \leq \kappa < \kappa^{I}$ (region I), the dynamics starting from both the initial conditions mentioned in the figure caption converged to the same attractor configuration, which is a fixed point in which the large ACS molecules have a very small concentration (the concentration declines exponentially with $n$). For $\kappa^{II} < \kappa$ (region III), again they converge to a single attractor, a fixed point in which the ACS molecules have a significant concentration which approaches $x_n = A (k_f A/k_r)^{n-1}$ as $\kappa \rightarrow \infty$. In the range $\kappa^{I} \leq \kappa \leq \kappa^{II}$ (region II), they converge to two different stable attractors, both fixed points for the ACS under discussion. (We remark that using other initial conditions we have found at least one more stable fixed point in a part of region II which has intermediate values of $x_{65}$, indicating that this system has multistability.)

This phase structure implies that if we start from the standard initial condition and consider the steady state profile to which the system converges for different values of $\kappa$, we will see a sharp change in the steady state profile as $\kappa$ is increased from a value
slightly below $\kappa^{II}$ to a value slightly above $\kappa^{II}$. Below $\kappa^{II}$ the large ACS molecules will be essentially absent in the steady state, and above $\kappa^{II}$ they will be present in large numbers and will dominate over the background.

Therefore, following the nomenclature of Ohtsuki and Nowak \cite{Ohtsuki2009}, who observed a similar bistability in their model with a single catalyst, we refer to $\kappa^{II}$ as the `initiation threshold' of the ACS. Similarly $\kappa^{I}$ will be referred to as the `maintenance threshold' of the ACS, because once the ACS has been initiated, $\kappa$ can come down to as low a value as $\kappa^{I}$, and the ACS will continue to dominate.

\subsubsection*{Bistability in simple ACSs}
In general $\kappa^{I}$ and $\kappa^{II}$ depend upon the other parameters, as well as the
topology of the catalyzed and spontaneous chemistries. The phase structure is exhibited in more detail for a simpler example in Fig. \ref{acs4}, where the catalyzed chemistry consists of only two reaction pairs:
\begin{subequations}
 \label{acs4-definition}
 \begin{eqnarray}
    \nonumber   & \mathrm{A}(4) & \\
    \label {acs4-rct1} \mathrm{A}(1) + \mathrm{A}(1) & \rightleftharpoons & \mathrm{A}(2) \\
    \nonumber   & \mathrm{A}(4) & \\
    \label {acs4-rct2} \mathrm{A}(2) + \mathrm{A}(2) & \rightleftharpoons & \mathrm{A}(4),
\end{eqnarray}
\end{subequations}
which constitute an ACS (called ACS4). This system, investigated numerically using XPPAUT, shows bistability. For a fixed $\phi$ the bistability diagram is shown in Fig. \ref{acs4}A. The dependence of $\kappa^{I},\kappa^{II}$ on $\phi$ is exhibited in Fig. \ref{acs4}B, and
on both $\phi$ and $k_f$ in Fig. \ref{acs4}C. For a given $k_f$, there is critical value of $\phi(=\bar\phi)$ at which the $\kappa^{I}$ and $\kappa^{I}$ curves meet, below which there is no bistability. The locations of the phase boundaries, the $\kappa^{I}$ and $\kappa^{II}$ curves, depend upon the specific underlying chemistries (catalyzed and spontaneous) as well as the ACS topology. The steady state profiles are shown at sample points in the phase space in Fig. \ref{acs4}D. For $\phi>\bar\phi$ it can be seen, that as in the case of the larger ACS discussed earlier, if we start from the standard initial condition, the largest molecule of the catalyzed chemistry, here A(4), dominates over the background in the steady state only for $\kappa > \kappa^{II}$ (e.g., the panel marked 3 in Fig. \ref{acs4}D). In the range $\kappa^{I} \leq \kappa \leq \kappa^{I}$, it dominates only if we start from initial conditions where it has a large enough value to begin with (panel 2b), but not if we start from the standard initial condition (panel 2a). It does not dominate for any initial condition if $\kappa < \kappa^{I}$ (panel 1). If $\phi$ is below $\bar\phi$, there is a single attractor with no significant ACS dominance if $\kappa$ is small (panel 4), or if $\kappa$ is large (panel 5), ACS dominance exists but is not very pronounced as the background concentrations are also substantial.

We remark that while bistability seems to be quite generic in homogeneous chemistries containing ACSs, the existence of an ACS does not guarantee that bistability exists somewhere in phase space. For example consider the simplest possible chemistry ($N=2$) containing only the monomer (which is buffered) and the dimer. If we assume that the sole reaction pair $\mathrm{A}(1) + \mathrm{A}(1) \rightleftharpoons \mathrm{A}(2)$ is catalyzed by $\mathrm{A}(2)$, the catalyzed chemistry is trivially an ACS and the only rate equation is $\dot{x}_2 = k_f A^2 (1 + \kappa x_2) - k_r x_2 (1 + \kappa x_2) - \phi x_2$. The system can be solved exactly and always goes to a global fixed point attractor starting from any initial condition $x_2(0) \geq 0$. However, the $N=3$ chemistry defined by the two catalyzed reactions
\begin{subequations}
 \label{acs3-definition-eg}
 \begin{eqnarray}
    \nonumber   & \mathrm{A}(3) & \\
    \label {acs3-eg-rct1} \mathrm{A}(1) + \mathrm{A}(1) & \rightleftharpoons & \mathrm{A}(2) \\
    \nonumber   & \mathrm{A}(3) & \\
    \label {acs3-eg-rct2} \mathrm{A}(1) + \mathrm{A}(2) & \rightleftharpoons & \mathrm{A}(3),
\end{eqnarray}
\end{subequations}
does exhibit bistability at a sufficiently large $\phi$. Ohtsuki and Nowak \cite{Ohtsuki2009} had also found a lower limit on catalyst size for bistability to exist in their model. Similar results hold for the $f=2$ case. From our simulations a general observation seems to be that bistability is ubiquitous at sufficiently large values of $\phi$ in homogeneous chemistries whenever the smallest catalyst is large enough compared to the food set. When it does exist it seems to provide a crisp criterion for `ACS domination', including `initiation' ($\kappa > \kappa^{II}$) and `maintenance' (($\kappa \geq \kappa^{I}$).

We must mention that there exists a substantial literature including results on the conditions for bistability in chemical systems\cite{Craciun2006,Wilhelm2007,Ramakrishnan2008}. It would be interesting to apply some of those results to models of the kind being studied here, which involve a large number of molecular species.

We now make a brief remark about catalyzed chemistries that do not contain an ACS. A particularly interesting case arises when the set of catalyzed reactions forms unbroken pathways from the food set to higher molecules, but the catalysts are not drawn from the product set of these reactions (i.e., the catalyzed chemistry satisfies condition (b) for an ACS but not the condition (a)). In this case we again observe the domination of the molecules produced by the catalyzed pathways, but this occurs at even higher catalytic strengths than for ACSs. The reasons are similar to those discussed above in the context of the $\kappa \rightarrow \infty$ limit for ACSs (note that in that limit the identity of the catalyst is irrelevant for the argument as long as the catalyst has a nonzero concentration). We have also observed multistabilities in these chemistries as well as a dependence on $\phi$ similar to the ACS case. In general, catalyzed chemistries (with and without ACSs) seem to have a rich and complex phase structure but more comprehensive investigations than we have done are needed to nail down the possible range of interesting behaviours.

\subsection*{A problem for primordial ACSs to produce large molecules:
The requirement of exponentially large catalytic strength}
A natural initial condition for the origin of life scenario is one where only the food set molecules, and perhaps a few other not very large molecules (dimers, trimers, etc.) have nonzero concentrations, while the large molecules have zero concentrations. It is from such an initial condition that we would like to see the emergence of large molecules through the dynamics. We have seen that in uncatalyzed chemistries, the concentrations of the large molecules remain exponentially small ($x_n \sim e^{-\gamma n}, \gamma > 0$). In catalyzed chemistries, especially in the presence of an ACS, a few specific large molecules produced by the ACS can acquire a high population. However, this seems to require a large catalytic strength for the catalysts. For example, for ACS65 this happens at $\kappa > \kappa^{II} = 2226342$, starting from the standard initial condition. The fact that such a large catalytic strength is needed to produce appreciable concentrations of molecules of even moderate length like $n=65$ could be a problem for the ACS mechanism to produce large molecules in the kind of prebiotic scenario we are considering. In this section we characterize the problem somewhat more quantitatively by determining how $\kappa^{II}$ depends upon the size $n$ of the catalysts in the ACS.

As mentioned earlier, the values of $\kappa^{I}$, $\kappa^{II}$ depend on the topology of the ACS. The topology of the ACS includes the set of catalyzed reactions and the assignment of catalysts to each of the catalyzed reactions. Define the `length' $L$ of an ACS as the size of (i.e., the total number of monomers of all types in) the largest molecule produced in the ACS. An `extremal' ACS of length $L$ will be referred to as one in which all reactions belonging to the ACS are catalyzed by the same molecule which is the largest molecule (of size $L$) in the ACS. For concreteness, since we are interested in the dependence of $\kappa^{II}$ on the catalyst size, we consider only extremal ACSs of length $L$. We assume that the catalyst has the same catalytic strength $\kappa$ for all the reactions in the ACS. We wish to determine the bistable region for such ACSs and in particular how the values of $\kappa^{I}$ and $\kappa^{II}$ depend upon $L$. These values depend upon the precise set of catalyzed reactions constituting the ACS. For illustrative purposes we consider three different ways of generating the ACS described under Methods as Algorithm 1, 2 and 3, which generate ACSs with different characteristic structure.

We determine the $\kappa^{I}$ and $\kappa^{II}$ values for ACSs of different values of $L$ numerically. These are plotted in Fig. \ref{kappa1-2-vs-L}. It is evident that $\kappa^{I}$ increases with $L$ according to a power law $\kappa^{I} \sim L^{\alpha}$ (with
$\alpha$ ranging from 2.1 to 2.8 for the three algorithms), while $\kappa^{II}$ increases exponentially,
\begin{equation}\label{kappa2-exponential-L}
\kappa^{II} \sim e^{\rho L},
\end{equation}
with $\rho \approx 0.64$ for all the algorithms. $\alpha$ and $\rho$ depend upon the other parameters. In particular we find that $\rho$ increases with $\phi$, i.e., the catalytic strength needed for large molecules to arise increases faster with the size of catalyst at larger values of dissipation. This generalizes, to a much larger class of models, the results of Ohtsuki and Nowak \cite{Ohtsuki2009}, who found a linear dependence of $\kappa^{I}$ on $L$ and an exponential dependence of $\kappa^{II}$.

The exponential increase of the initiation threshold, $\kappa^{II}$, with $L$, quantifies the difficulty in using ACSs to generate large molecules in the primordial scenario of the type modeled above. This means that one needs large molecules with unreasonably high catalytic strengths to exist in the chemistry in order to get them to appear with appreciable concentrations starting from physically reasonable primordial initial conditions.

\subsection*{Nested ACSs: Using a small ACS to reinforce a larger one}
We now discuss a mechanism that may overcome the barrier of large catalytic strengths, and may enable large molecules to arise from primordial initial conditions without exponentially increasing catalytic strengths. This mechanism relies on the existence of multiple ACSs of different sizes in the catalyzed chemistry, in a topology such that the smaller ACSs reinforce the larger ones, thereby enabling large molecules to appear with significant concentrations without exponentially increasing their catalytic strength.

To illustrate the basic idea we consider the following simple example where the catalyzed chemistry contains only two ACSs, one of length three and the other of length eight (which we refer to as ACS3 and ACS8, respectively), each generated by the Algorithm 2 mentioned above. All reactions of the former are catalyzed by $\mathrm{A}(3)$ with a catalytic strength $\kappa_3$, and of the latter by $\mathrm{A}(8)$ with the catalytic strength $\kappa_8$. Thus the two ACSs are:
\begin{subequations}
  \label{acs3-definition}
  \begin{eqnarray}
    \nonumber   & \mathrm{A}(3) & \\
    \label {acs3-rct1} \mathrm{A}(1) + \mathrm{A}(1) & \rightleftharpoons & \mathrm{A}(2) \\
    \nonumber   & \mathrm{A}(3) & \\
    \label {acs3-rct2} \mathrm{A}(1) + \mathrm{A}(2) & \rightleftharpoons & \mathrm{A}(3),
  \end{eqnarray}
\end{subequations}
and
\begin{subequations}
  \label{acs8-definition}
  \begin{eqnarray}
    \nonumber   & \mathrm{A}(8) & \\
    \label {acs8-rct1} \mathrm{A}(1) + \mathrm{A}(1) & \rightleftharpoons & \mathrm{A}(2) \\
    \nonumber   & \mathrm{A}(8) & \\
    \label {acs8-rct2} \mathrm{A}(2) + \mathrm{A}(2) & \rightleftharpoons & \mathrm{A}(4) \\
    \nonumber   & \mathrm{A}(8) & \\
    \label {acs8-rct3} \mathrm{A}(4) + \mathrm{A}(4) & \rightleftharpoons & \mathrm{A}(8).
  \end{eqnarray}
\end{subequations}
The catalyzed chemistry consists of the above five catalyzed reaction pairs (we will refer to this catalyzed chemistry as ACS3+8). The system also exhibits bistability, and the
concentration of $\mathrm{A}(8)$ in the two fixed point attractors is exhibited in Fig. \ref{nested-3-8} as a function of $\kappa_3$ and $\kappa_8$.

When $\kappa_3$ is small the two pictures show the usual bistability of ACS8 along the $\kappa_8$ axis. The initiation and maintenance thresholds are $\kappa_8^{II} = 1.78 \times 10^7$ and $\kappa_8^{I} = 1145$ given by the location of the boundary between the low concentration region (blue, $x_8 \sim 10^{-7}$) and the high concentration region (yellow $x_8 \sim 1$) along the $\kappa_8$ axis in Figs. \ref{nested-3-8}A and \ref{nested-3-8}B respectively. As
$\kappa_3$ increases, the initiation threshold of ACS8 decreases slowly for a while, then drops sharply near $\kappa_3 = 141$. This value of $\kappa_3$ is the initiation threshold of ACS3 when $\kappa_8=0$. When $\kappa_3$ exceeds this value, the steady state value of $x_8$ is either high (yellow, $x_8 \sim 1$) or intermediate (orange, $x_8 \sim 10^{-3}$), depending upon the value of $\kappa_8$.

The key point is that the initiation threshold of the larger catalyst depends on the catalytic strength of the smaller catalyst. The former plummets sharply when the latter approaches the initiation threshold of the smaller catalyst, dropping to a much lower value than before (compare the lower limit of the yellow region in Fig. \ref{nested-3-8}A to the left and right of $\kappa_3 = \kappa_3^{II} = 141$; the value of $\kappa_8^{II}$ plunges several orders of magnitude from $1.78 \times 10^7$ at $\kappa_3 = 0$ down to $2178$ at $\kappa_3 = 141$). Starting from the standard initial condition, thus, the larger catalyst can acquire a significant concentration at a much lower value of its catalytic strength in the presence of a smaller ACS operating above its initiation threshold than in its absence.

\subsubsection*{Why a small ACS reinforces a larger one}
We now present an intuitive explanation of the above mentioned property. The argument rests on two observations.\\
(a) {\bf Why the initiation threshold is exponentially large}: The first observation attempts to explain $\kappa^{II}$ is so large in the first place. The contribution of a catalyst to the rate of the reaction it catalyzes appears through the factor $1 + \kappa x$, where $\kappa$ is the catalytic strength of the catalyst and $x$ its concentration. The term unity in the above factor is the relative contribution of the spontaneous (uncatalyzed) reaction rate. If the catalyst is to play a significant role in the reaction, the catalytic contribution to the reaction rate should be at least comparable to the spontaneous rate, i.e., $\kappa x$ should be at least comparable to unity. As we have seen earlier the concentration of large molecules is typically damped exponentially with their size. Therefore the compensating factor $\kappa$ needs to increase exponentially in order for the catalyzed reaction rate to be comparable to the spontaneous reaction rate. For concreteness consider the extremal ACSs of length $L$ and consider the steady state population $x_L$ of the catalyst $\mathrm{A}(L)$ in the low fixed point as $\kappa$ is increased. In the spirit of this rough argument one expects that at the initiation threshold the term $\kappa^{II} x_L$ should be of order unity. In Fig. \ref{kappa2-xL-relationship} we display this product for different values of $L$. Though there is a secular decreasing trend with $L$, this product remains of order unity (Fig. \ref{kappa2-xL-relationship}A) even as the individual factors change over several orders of magnitude (Fig. \ref{kappa2-xL-relationship}B). This lends numerical support to the above explanation for the exponential dependence of $\kappa^{II}$ on $L$.

\noindent (b) {\bf Role of the background and spontaneous reactions}:
The second observation is that when $\kappa$ exceeds the initiation threshold for a catalyzed chemistry containing an ACS, not only do the steady state concentrations of the ACS product molecules rise by several orders of magnitude, but also those of the background molecules rise. As an example compare the two steady state profiles of ACS65 in Fig. \ref{acs-eg}A, which correspond to values of $\kappa$ below and above the initiation threshold. As one goes from the lower to the upper curve, the concentration of the ACS members of course increases dramatically (as shown by the sharp peaks), but note that the concentrations of other molecules not produced by catalyzed reactions also goes up significantly. Thus in the chemistry containing two ACSs (ACS3+8) as one moves along the $\kappa_3$ axis in Fig. \ref{nested-3-8}A and crosses the initiation threshold of ACS3 (i.e., $\kappa_3$ exceeds $\kappa_3^{II} = 141$), the concentration of $\mathrm{A}(8)$ (a molecule belonging to the background of ACS3 as its production is not catalysed by ACS3) increases from $\sim 10^{-7}$ (blue region) to $\sim 10^{-3}$ (orange region). This increase in the concentration of $\mathrm{A}(8)$ by a factor of $\sim 10^4$ makes it easier for ACS8 to function and its initiation threshold drops by a corresponding factor of about $10^4$ (from $\sim 10^7$ to
$\sim 10^3$).

This fact highlights the role of spontaneous reactions in the overall dynamics. The background molecules are connected to the ACS through spontaneous reactions, and if it were not for the latter, an ACS would not be able to push up the concentrations of its nearby background. We shall refer to a structure such as the one described above containing ACSs of different sizes with the smaller ACS feeding into the larger one through the spontaneous reactions as a `nested ACS' structure.

\subsubsection*{The role of `overlapping' catalyzed pathways in nested ACSs}
The above example also serves to highlight some other features of catalyzed chemistries containing multiple ACSs. Note that the production pathway of $\mathrm{A}(8)$ in ACS8 (Eqs. \ref{acs8-definition}) contains one reaction pair in common with ACS3, namely the reaction
pair $\mathrm{A}(1) + \mathrm{A}(1) \rightleftharpoons \mathrm{A}(2)$. One can consider a situation wherein the overlap is greater. E.g., consider the ACS8' defined by
\begin{subequations}
  \label{acs8'-definition}
  \begin{eqnarray}
    \nonumber   & \mathrm{A}(8) & \\
    \label {acs8'-rct1} \mathrm{A}(1) + \mathrm{A}(1) & \rightleftharpoons & \mathrm{A}(2) \\
    \nonumber   & \mathrm{A}(8) & \\
    \label {acs8'-rct2} \mathrm{A}(1) + \mathrm{A}(2) & \rightleftharpoons & \mathrm{A}(3) \\
    \nonumber   & \mathrm{A}(8) & \\
    \label {acs8'-rct3} \mathrm{A}(2) + \mathrm{A}(3) & \rightleftharpoons & \mathrm{A}(5) \\
    \nonumber   & \mathrm{A}(8) & \\
    \label {acs8'-rct4} \mathrm{A}(3) + \mathrm{A}(5) & \rightleftharpoons & \mathrm{A}(8).
  \end{eqnarray}
\end{subequations}
Now the set of reactions in ACS3 is a subset of ACS8' (ignoring the catalyst, which is different in the two cases). The degree of overlap of the catalyzed reaction sets  between a pair of nested ACSs makes a difference in the dynamics. Consider, for example, the catalyzed chemistry consisting of ACS3 and ACS8', i.e., the set of catalyzed reactions given by Eqs. (\ref{acs3-definition}) and (\ref{acs8'-definition}), which we refer to as ACS3+8'. Like ACS3+8, this chemistry also shows a reduction of $\kappa_8^{II}$, when $\kappa_3$ exceeds its initiation threshold. We find that while at $\kappa_3=0$ the value of $\kappa_8^{II}$ for the two chemistries is not too different ($1.6 \times 10^7$ for ACS3+8' versus $1.8 \times 10^7$ for ACS3+8), at $\kappa_3 = 141$, $\kappa_8^{II}$ reduces to a value 920 in ACS3+8', which is less than half of the value 2178 that it reduces to in ACS3+8. Thus a larger degree of overlap between the catalyzed reaction sets of nested ACSs causes more effective reinforcement.

Another example with this behaviour for $f=2$ is described in Fig. \ref{acs-2d-nested-4-8}. In each of the three ACS pairs shown in the figure, the smaller ACS, of length 4, is the same, (it will be referred to as ACS(2,2)) and is defined by the reactions (each catalyzed by (2,2))
\begin{subequations}
  \label{ACS(2,2)-definition}
  \begin{eqnarray}
    \nonumber & (2,2) & \\
    (0,1) + (1,0) & \rightleftharpoons & (1,1) \\
    \nonumber & (2,2) & \\
    (1,1) + (1,1) & \rightleftharpoons & (2,2).
  \end{eqnarray}
\end{subequations}

\noindent The three larger ACSs, called ACS(5,3)(a), ACS(5,3)(b) and ACS(5,3)(c), respectively, can essentially be determined from the
figure. For example, ACS(5,3)(a) consists of the two reaction pairs given by Eqs. (\ref{ACS(2,2)-definition}), both catalyzed by (5,3) as well as the three reactions
\begin{subequations}
  \label{ACS(5,3)(a)-definition}
  \begin{eqnarray}
    \nonumber & (5,3) & \\
    (1,0) + (2,2) & \rightleftharpoons & (3,2) \\
    \nonumber & (5,3) & \\
    (1,1) + (3,2) & \rightleftharpoons & (4,3) \\
    \nonumber & (5,3) & \\
    (1,0) + (4,3) & \rightleftharpoons & (5,3).
  \end{eqnarray}
\end{subequations}

\noindent ACS(5,3)(b) consists of the single reaction pair given by the first of Eqs. (\ref{ACS(2,2)-definition}), catalyzed by (5,3), as well as the four reactions
\begin{subequations}
  \label{ACS(5,3)(b)-definition}
  \begin{eqnarray}
    \nonumber & (5,3) & \\
    (1,0) + (1,1) & \rightleftharpoons & (2,1) \\
    \nonumber & (5,3) & \\
    (1,0) + (2,1) & \rightleftharpoons & (3,1) \\
    \nonumber & (5,3) & \\
    (1,1) + (3,1) & \rightleftharpoons & (4,2) \\
    \nonumber & (5,3) & \\
    (1,1) + (4,2) & \rightleftharpoons & (5,3),
  \end{eqnarray}
\end{subequations}
and ACS(5,3)(c) consists of the five reaction pairs
\begin{subequations}
  \label{ACS(5,3)(c)-definition}
  \begin{eqnarray}
    \nonumber & (5,3) & \\
    (1,0) + (1,0) & \rightleftharpoons & (2,0) \\
    \nonumber & (5,3) & \\
    (0,1) + (2,0) & \rightleftharpoons & (2,1) \\
    \nonumber & (5,3) & \\
    (1,0) + (2,1) & \rightleftharpoons & (3,1) \\
    \nonumber & (5,3) & \\
    (0,1) + (3,1) & \rightleftharpoons & (3,2) \\
    \nonumber & (5,3) & \\
    (2,1) + (3,2) & \rightleftharpoons & (5,3).
\end{eqnarray}
\end{subequations}

We consider the population dynamics of chemistries in which the spontaneous part includes all possible ligation and cleavage reactions involving molecules with upto $N=15$ monomers with homogeneous rate constants $k_f = k_r = 1$, $\phi = 15$, and the catalyzed part containing one or more of the above mentioned ACSs. When ACS(2,2) is the only ACS present, the system shows bistability with the initiation threshold being $\kappa_{(2,2)}^{II} = 551$. When ACS(5,3)(a), (b) or (c) are the only ACSs present, the initiation thresholds for them are 1125197, 1031082, and 1000112, respectively. When ACS(2,2) and one of ACS(5,3) (a), (b) or (c) are both present, and the catalytic strength of (2,2) is 552, the initiation thresholds of the
three larger ACSs reduce to 941, 1256, and 2482, respectively. Again, it is seen that the larger the degree of overlap of the two nested ACSs, the more effective is the reinforcement.

\subsection*{A hierarchy of nested ACSs: A possible route for the appearance of large molecules}
The process of nesting discussed above for two ACSs can be extended to multiple levels of ACSs connected to each other. Here we discuss sequences of ACSs of increasing size, with the catalyzed reaction set of each ACS in the sequence partially or completely contained within the next one, and the catalytic strength of molecules increasing with size in a controlled manner. We construct examples of such sequences in which large catalyst molecules containing several hundred monomers can acquire significant concentrations starting from the standard initial condition, even though all catalysts have moderate catalytic strengths.

In order to construct a cascade of nested ACSs in which reaction sets of smaller ACSs are completely contained in the larger ones (maximal overlap), we used Algorithm 4 described in Methods. This algorithm produces a cascade of ACSs with $g$ steps (generations), with the $k^\mathrm{th}$ generation ACS containing $n_k$ new reactions. We studied several catalyzed chemistries containing a cascade of nested ACSs for $f=1$ and 2. One example of each type is presented below; other examples gave qualitatively similar results.

\subsubsection*{Dominance of an ACS of length 441 (ACS441)}
For $f=1$ we describe a cascade with $g=15$ and $n_1=1, n_2 = n_3 = \dots = n_{15} =2$. This catalyzed chemistry had 29 product molecules, the largest of which was $A(441)$ having 441 monomers. The list of molecules and reactions is given in Supporting Table S2. The catalytic strength $\kappa$ of each molecule was chosen by an explicit length dependent rule
\begin{equation}
  \label{cascade-kappa-L-dependence1}
  \kappa(L) = K L^{\beta},
\end{equation}
where $K$ and $\beta$ are constants. We describe a simulation with $K=500$ and $\beta = 1.5$. This particular rule was chosen to contrast with Eq. (\ref{kappa2-exponential-L}) which characterizes the initiation threshold of an extremal ACS of length $L$. For a value of $L$ such as 441, the exponential function in Eq. (\ref{kappa2-exponential-L}) would have given an astronomically large catalytic strength, whereas the much slower growing power law in Eq. (\ref{cascade-kappa-L-dependence1}) gives $\kappa(441) = 4.6 \times 10^6$ for the above mentioned values of the constants. Starting from the standard initial condition, the steady state concentration profile of this catalyzed chemistry embedded in a fully connected spontaneous chemistry with $N=800$ is shown in Fig. \ref{1d-algo4}A. This example shows that with the nested ACS structure in the catalyzed chemistry, large catalyst molecules can acquire significant concentrations starting from an initial condition containing only the monomers, even when catalytic strengths grow quite slowly with the length of the catalyst. It is worth noting that product of the catalytic strength of $A(441)$ and its steady state concentration ($x_{441} = 0.0077$) is about 36000, and this is the factor by which it speeds up the reactions it catalyzes over the spontaneous rate. In view of the fact that enzymes containing a few hundred amino acids speed up the reactions they catalyze within cells by factors of about $10^5$ and greater, the catalytic efficiency demanded of $A(441)$ does not seem unreasonably high.

Eq. (\ref{cascade-kappa-L-dependence1}), which gives a particular functional form for $\kappa(L)$, is ad-hoc, and, at this stage, merely an example given to quantify the level of catalytic strengths that is sufficient for large molecules to arise in appreciable concentrations in the prebiotic scenario under consideration if chemistry has the nested ACS structure of the kind discussed. One may ask if an even weaker requirement would suffice. We have considered smaller values of $\alpha$ (1.2 and 1.0) keeping $K$ fixed, and found that ACS molecules upto a particular size (depending on $\alpha$) do well but that the concentration of larger ACS molecules trails off and merges with the background. The size range of ACS molecules that do well can be increased by increasing the coefficient $K$. Since the results depend upon several factors, including the topology of the ACS and the uncatalyzed chemistry, a detailed investigation has not been carried out.

\subsubsection*{Effect of a `sparse' spontaneous chemistry}
Fig. \ref{1d-algo4} also shows another aspect of ACS dynamics -- the relationship between ACS domination and the sparseness of the uncatalyzed chemistry. A fully connected uncatalyzed chemistry is one in which all possible ligation and cleavage reactions are allowed. A chemistry with average degree $k$ is one in which the average number of ligation reactions in which a molecule can be produced is $k$. In a fully connected $f=1$ chemistry, a molecule of size $n$ can be produced in about $n/2$ ligation reactions ($\mathrm{A}(1) + \mathrm{A}(n-1) \rightarrow \mathrm{A}(n)$, $\mathrm{A}(2) + \mathrm{A}(n-2) \rightarrow \mathrm{A}(n)$, etc.); therefore the average degree of a chemistry containing all molecules upto size $N$ is about $N/4$ (=200 for the chemistry in Fig. \ref{1d-algo4}A). In Figs. \ref{1d-algo4}B and \ref{1d-algo4}C, we pruned the uncatalyzed reaction set to only $k$ ligation reactions per molecule ($k=20$ and 2, respectively), randomly chosen from all possible ligation reactions producing the molecule. (For molecules too small to have $k$ ligation reactions, all ligation reactions were retained. For the ACS molecules the ligation reaction producing them in the catalyzed chemistry was included as one of the $k$ uncatalyzed reactions.) Note that while we refer to only the ligation reactions and not cleavage reactions for the purpose of defining the degree of a molecule, in our simulations all reactions are treated as reverse reactions. That is, for every ligation reaction included in the chemistry the reverse (cleavage) reaction is also present. It is seen in Fig. \ref{1d-algo4} that the increase of sparseness causes the background concentrations to decline. This is because there are fewer pathways to produce the background molecules. There is also a larger variation in their concentrations because their production reactions are chosen randomly, and background molecules produced in reactions that happen to involve the ACS molecules as reactants do better than others. The ACS molecules are seen to dominate more strongly over the background in sparser chemistries; this is because there are fewer production pathways to the background that drain their concentrations.

\subsubsection*{Cascading nested ACSs with $f=2$}
An example of a nested ACS with two food sources, ACS(36,28), is presented in Fig. \ref{2d-algo4}. This is also generated by Algorithm 4 and has 7 generations with $n_k =3$ for each generation, the largest molecule being (36,28) (the full list of molecules and reactions is given is Supporting Table S3). Again starting from the standard initial condition the larger ACS molecules acquire appreciable concentrations with a moderate demand on their catalytic strengths.

As a final example we present in Fig. \ref{2d-partial-nest} a cascade of ACSs, named ACS(18,27) after its largest molecule, in which smaller ACSs have only a partial overlap with longer ones. This is generated using Algorithm 5 (see Methods) and consists of a series of 10 ACSs of increasing length. The detailed list of molecules, reactions and catalytic strengths is given in Supporting Table S4. Each successive ACS in the cascade has only a few reactions in common with other ACSs. Unlike in the examples discussed above, generated by Algorithm 4, in the present case molecules (except the small molecules) produced in the catalyzed chemistry have typically only one or two catalyzed ligation reactions producing them. The chemistry also contains a number of catalyzed `side reactions', which produce molecules that are neither catalysts nor reactants in any pathway leading to the largest molecule. In fact there is a `side pathway' consisting of several reactions that may be viewed as `draining the resources' of the main ACS. ACS dominance at moderate catalytic strengths occurs for this chemistry also. The largest $\kappa$ is 50000 for the molecule (18,27), and at a steady state population of 0.26 enhances the rate of a reaction by a factor of 13000 over the spontaneous rate. This shows that the mechanism outlined by us is not restricted to maximally overlapping nested ACSs but is more generic.

\section*{Discussion}
Our work discusses a possible mechanism by which large molecules can arise in a prebiotic scenario. In the context of the present model the appearance of large molecules is a natural dynamical consequence of chemistry possessing the structure described above -- a cascading nested ACS structure (with a not very demanding set of catalytic strengths) embedded in a spontaneous chemistry -- together with the buffered presence of the food set molecules in a well stirred region of space. The mechanism is an incremental one: at each step successive step the system is able to access new states made available by the previous step while making only an incremental demand on molecular catalytic capabilities.

The kind of mathematical model we have studied, inspired by the work of Bagley and Farmer, is quite abstract; its virtue is the economy of assumptions that go into its structure. The main ingredients are that objects can combine with each other in processes or `reactions' to form other objects and certain objects can facilitate certain processes, i.e., `catalyze' certain reactions. The population dynamics implements a simple scheme for how the abundances of the objects would change with time assuming that the probability of objects combining is proportional to their abundances. Such a generic scheme while it applies in detail to no particular situation allows us to imagine mechanisms at a conceptual level. It is significant that in this scheme an ACS can direct the flows towards itself and cause a certain sparse subset of objects, including some specific large composite ones, to capture a large fraction of the chemical resources.

At this level of abstraction the model (or a variant with qualitatively similar features) could apply to the peptide chemistry as well as an RNA chemistry and to a prebiotic metabolism, as already noted by Bagley and Farmer. Indeed it would be equally applicable if a prebiotic environment actually had a mixture of ingredients from all these classes of chemistries, a possibility that has been advocated in, e.g., \cite{Copley2007, Powner2011}. Copley, Smith and Morowitz \cite{Copley2007} have proposed a scenario which seeks to explain how the RNA world might have originated through a series of incremental steps starting from a primitive metabolism. The food sources for this supposed metabolism are $\mathrm{CO}_2$, $\mathrm{H}_2$, $\mathrm{H}_2\mathrm{S}$, $\mathrm{NH}_3$, etc., in a hydrothermal vent. Their scenario envisages multiple stages of increasing complexity which they refer to as (i) the monomer stage, in which metabolism, possibly powered by an autocatalytic set such as the reverse TCA cycle, produces nucleotides and simple amino acids, (ii) the multimer stage, which produces dimers and small cofactors, (iii) the micro-RNA stage, producing of oligonucleotides of length 3-10, (iv) the mini-RNA stage, with 11-40mers, followed by (v) the macro-RNA stage, or the RNA world. In their scenario each successive stage produced better catalysts that collectively catalyzed not only their own production from the molecules of the previous stage, but also the reactions of the lower stages. This structure is very similar to the cascade of nested ACSs that we have discussed. A suitably modified version of our model could be constructed to explore the dynamics of this scenario in more detail. At a general level, in the fact that the dynamics of our model results in the stable domination, or concentrating, of the large catalysts, our mathematical work perhaps lends support to the workability of such a scenario.

There is another level at which the present model (or its variants) might talk to prebiotic chemistry. Morowitz \cite{Morowitz1999} has suggested that the metabolic network itself has a shell like structure which can convert simple molecules like $\mathrm{CO}_2$, $\mathrm{H}_2$, $\mathrm{NH}_3$, etc., through ``a hierarchy of nested reaction networks involving increasing complexity" into purines, pyrimidines, complex cofactors, etc. Reaction sets created by our Algorithms 4 and 5 are reminiscent of this picture. Missing from Morowitz's picture is a catalyst assignment for each reaction from among the molecules in the various shells or from among other catalysts accessible prebiotically (e.g., surfaces in hydrothermal vents). It might be worthwhile to attempt to add in that information for a more complete scenario and for potential modeling.

\subsection*{Caveats and future directions}
(1) We have studied the properties of autocatalytic systems, by choosing specific examples of ACS topology and special algorithms for constructing them. This has allowed us to systematically investigate the dynamical properties that ACSs offer. We believe that similar dynamical properties would hold for more general topological structures than we have considered. Nevertheless the question arises as to whether all these structures are very special structures and whether or not they are likely to arise within real chemistry and `generic' artificial chemistries. ACSs have been shown to be quite generic in a large class of randomly constructed artificial chemistries \cite{Steel2000, Hordijk2011}, and a similar analysis could be extended to cascading nested ACSs. This would help parametrize or characterize chemistries that would contain such structures and those that would not. In this context it would useful to go beyond the simple case we have considered in which a molecule is defined by the number of monomers of each type and not their sequence. It may also be interesting to look for structures similar to nested ACSs in real metabolic networks using methods similar to those in \cite{Kun2008}.

(2) An important related question is one of side reactions (discussed in \cite{Orgel2000, Orgel2008}) which might destroy the efficacy of ACSs. In the real chemistry one expects that even if cascading nested ACSs exist, there would also exist other catalyzed reactions channelizing the ACS products into pathways leading in other directions. Whether substantial populations of large molecules in the nested ACSs can be maintained in the presence of such diversion is a question that remains to be systematically investigated. Our last example of cascading nested ACSs in fact has several side reactions and it may be noted that large ACSs still dominated in that case. We remark that while side reactions can drain resources from ACSs, they also help the system to explore new directions in chemical space in an evolutionary scenario.

(3) We have considered deterministic dynamics in this paper. Stochastic fluctuations are important when molecular populations are small. For a chemistry containing multiple ACSs, Bagley, Farmer and Fontana \cite{Bagley1991a} used stochasticity to produce examples of trajectories that differed from each other in the sequence of ACSs that came to dominate the reactor. It would be interesting to explore such effects in the context of the present model.

(4) Another simplification we have made is that of homogeneous chemistries, wherein the rate constants of all spontaneous reactions have been taken to be the same, and even catalytic strengths, where variable, have been taken to be smooth functions of the length. We have checked that introducing a small amount of heterogeneity or randomness in the rate constants does not change the qualitative behaviour significantly. However, the effect of cranking up the heterogeneity has not been studied. From studies of disordered systems in statistical mechanics and condensed matter physics it has become clear that such heterogeneity can lead to rugged landscapes, multiple attractors and timescales, and paths that are difficult to locate \cite{Wales2004}. The dynamics of such systems when they are driven by a non-equilibrium flux or buffering of food set molecules is an open question. It is possible that the constraints placed by the ruggedness of the landscape will reduce the number of accessible paths. The populating of molecules at different levels in a nested hierarchy of ACSs is likely to happen in fits and starts on multiple timescales when heterogeneity is accounted for. It is perhaps in such a scenario that one
should look for answers to the questions raised under (1), (2) and (3) above.

\section*{Methods}
\subsection*{\bf A. Generating extremal ACSs of length $L$.}
In the following we describe three different algorithms used for generating a set of reactions that provide a pathway to produce a molecule of a given length $L$ from the food set. \\
{\bf Algorithm 1: Incremental, smallest-stepsize, deterministic construction}: Each molecule of size $n$ ($n=2,3,\ldots,L$) is produced in the reaction $\mathrm{A}(1) + \mathrm{A}(n-1) \rightarrow \mathrm{A}(n)$. All such reactions for $n=2,3,\ldots,L$ are included in the ACS.\\
{\bf Algorithm 2: Shortest path, top-down, deterministic construction}: Start with $\mathrm{A}(L)$. If $L$ is even, it is produced in the reaction $\mathrm{A}(L/2) + \mathrm{A}(L/2) \rightarrow \mathrm{A}(L)$. If $L$ is odd, say $L=2m+1$, then $\mathrm{A}(L)$ is produced in the reaction $\mathrm{A}(m) + \mathrm{A}(m+1) \rightarrow \mathrm{A}(L)$. The same algorithm is used to select a reaction for the production of the precursor(s) of $\mathrm{A}(L)$ (namely, for $\mathrm{A}(L/2)$ if $L$ is even, and for each of $\mathrm{A}(m)$ and $\mathrm{A}(m+1)$ if $L$ is odd ($=2m+1$)), and is iterated for their precursors, etc., until the only reactant appearing in the reactions is the food set molecule, A(1). All the production reactions mentioned above are included in the ACS.\\
{\bf Algorithm 3: Incremental, random construction}: In this method, starting from the food set $\mathcal{F}$ the set of reactants $R_k$ is sequentially enlarged step by step ($\mathcal{F}=R_0 \subset R_1 \subset R_2 \subset \ldots $) to include larger molecules and a reaction chosen randomly until a product of length $L$ is obtained. $R_k$ denotes the set of reactants at step $k$ in the algorithm, and $L_k$ the size of the largest molecule in $R_k$. At step $k$, pick a pair of molecules (say $X$ and $Y$) randomly from $R_k$, and determine the product $Z$ formed if they were to be ligated ($X$ being the same as $Y$ is allowed). If the size of this product, $L(Z)$, is $\leq L_k$ or $> L$, discard the pair and choose another pair. If $L_k < L(Z) < L$, add the ligation reaction $X + Y \rightarrow Z$ to the ACS, define $R_{k+1} = R_k \cup \{ Z \} $, and iterate the procedure. If $L(Z) = L$, add the ligation reaction $X + Y \rightarrow Z$ to the ACS, and stop. Initially ($k=0$) the reactant set is just the food set, $R_0 = \mathcal{F}$.

To complete the extremal ACS, we assign $\mathrm{A}(L)$ as the catalyst of all the reactions generated using any of the above algorithms. For the simulations reported in the paper, the reverse of each reaction is also included as a reaction catalyzed by the same catalyst. For concreteness we have described the Algorithms 1 and 2 for the single monomer case; their generalizations to $f=2$ have also been considered by us. Algorithm 3, as described above, can be used for any $f$.

\subsection*{\bf B. Generating a cascade of nested ACSs}
{\bf Algorithm 4: Incremental, random construction of a sequence of reaction sets with maximal overlap}:
We construct successive sets, or `generations', $g$ in number, $P_1, P_2, \ldots P_g$, of product molecules, starting from the food set $\mathcal {F} (\equiv P_0)$. Each generation $P_k$ $(k=1,2,\ldots g)$ has a pre-specifed number of molecules, $n_k$ ($n_1$, $n_2$, $\ldots n_g$ need to be specified before running the algorithm). At step $k$ of the cascade an ACS $S_k$ is constructed from the previous ACS $S_{k-1}$ by adding reactions between molecules belonging to a reactant set $R_k$ consisting of all the products of the previous generations and the food set, $R_k = P_0 \cup P_1 \cup P_2 \ldots P_{k-1}$. Let $L_k$ denote the size of the largest molecule in $R_k$. At the beginning of the $k^\mathrm{th}$ step, $P_k$ is empty and the set of reactions in $S_k$ is the same as in $S_{k-1}$, except that the catalysts of the reactions in $S_k$ are not yet assigned. ($S_0$ is the empty set.) To construct $P_k$ and $S_k$, pick a molecule $X$ at random from the previous generation of products $P_{k-1}$ and another molecule $Y$ at random from $R_k$, and determine the product $Z$ formed if they were to be ligated ($X$ being the same as $Y$ is allowed). If the size of this product, $L(Z)$, is $\leq L_k$, discard the pair and choose another pair. If $L(Z) > L_k$, add the molecule $Z$ to $P_k$ and the ligation reaction $X + Y \rightarrow Z$ to $S_k$. Repeat this procedure until $n_k$ molecules are added to $P_k$ and $n_k$ reactions are added to $S_k$. Assign catalysts to each reaction in $S_k$ (which includes the reactions inherited from $S_{k-1}$ and the new $n_k$ reactions) randomly from $P_k$. This completes the $k^\mathrm{th}$ step. To get the full cascade with $g$ generations this process is carried out for $k=1,2, \ldots g$. By construction the set of reactions (ignoring the catalyst) of each $S_k$ is fully contained in that of $S_{k+1}$ but the catalysts are different, being drawn from $P_k$ for $S_k$ and $P_{k+1}$ for $S_{k+1}$. The union of the $S_k$'s constitutes the set of reactions in the catalyzed chemistry. Note that in this catalyzed chemistry reactions have multiple catalysts, the multiplicity declining for reactions producing higher generation molecules.

The size of the food set constrains how large $n_k$ can be. For $f=1$, $n_1 = 1$ and $P_1= \{ A(2) \} $ as the only product one can make from a reaction in the food set is $\mathrm {A}(1) + \mathrm {A}(1) \rightarrow \mathrm {A}(2)$, and $n_2$ can only have values 1 and 2 as the only products one can produce in the second generation (from reactants in $R_2 = \{ \mathrm {A}(1),\mathrm {A}(2) \})$ are A(3) and A(4), etc. Similarly for $f=2$, $n_1$ can be only 1,2 or 3, as reactions among the two food set molecules $(1,0)$ and $(0,1)$ can only produce three molecules $(2,0)$, $(1,1)$ and $(0,2)$.

\noindent {\bf Algorithm 5: Incremental, random construction of a sequence of reaction sets with partial overlap}:
In this algorithm we decide on a sequence of lengths, $L_1, L_2, \ldots, L_g$, and generate an extremal ACS, denoted $S_i$, of length $L_i$ using Algorithm 3 for each $i =1,2,\ldots,g$. For each $i$ a different random number seed is used to initialize Algorithm 3. The union of the $S_i$'s constitutes the set of reactions in the catalyzed chemistry.

\section*{Acknowledgments}
We would like to thank Eric Smith for drawing our attention to the
paper of Ohtsuki and Nowak (ref. \cite{Ohtsuki2009}) and Shobhit
Mahajan, Amitabha Mukherjee and Awadhesh Prasad for discussions. VG
acknowledges Senior Research Fellowships from the University Grants
Commission and the Department of Biotechnology (DBT), Government of
India. SJ acknowledges research g\textbf{}rants from DBT.

\bibliography{giri_jain}

\begin{thebibliography}{10}
\providecommand{\url}[1]{\texttt{#1}}
\providecommand{\urlprefix}{URL }
\expandafter\ifx\csname urlstyle\endcsname\relax
  \providecommand{\doi}[1]{doi:\discretionary{}{}{}#1}\else
  \providecommand{\doi}{doi:\discretionary{}{}{}\begingroup
  \urlstyle{rm}\Url}\fi
\providecommand{\bibAnnoteFile}[1]{%
  \IfFileExists{#1}{\begin{quotation}\noindent\textsc{Key:} #1\\
  \textsc{Annotation:}\ \input{#1}\end{quotation}}{}}
\providecommand{\bibAnnote}[2]{%
  \begin{quotation}\noindent\textsc{Key:} #1\\
  \textsc{Annotation:}\ #2\end{quotation}}
\providecommand{\eprint}[2][]{\url{#2}}

\bibitem{Miller1953}
Miller SL (1953) {A production of amino acids under possible primitive earth
  conditions}.
\newblock Science 117: 528--529.
\bibAnnoteFile{Miller1953}

\bibitem{Miller1959}
Miller SL, Urey HC (1959) {Organic compound synthesis on the primitive earth}.
\newblock Science 130: 245--251.
\bibAnnoteFile{Miller1959}

\bibitem{Ricardo2004}
Ricardo A, Carrigan MA, Olcott AN, Benner SA (2004) {Borate minerals stabilize
  ribose.}
\newblock Science (New York, NY) 303: 196.
\bibAnnoteFile{Ricardo2004}

\bibitem{Powner2009}
Powner MW, Gerland B, Sutherland JD (2009) {Synthesis of activated pyrimidine
  ribonucleotides in prebiotically plausible conditions.}
\newblock Nature 459: 239--242.
\bibAnnoteFile{Powner2009}

\bibitem{Parker2011}
Parker ET, Cleaves HJ, Dworkin JP, Glavin DP, Callahan M, et~al. (2011)
  {Primordial synthesis of amines and amino acids in a 1958 Miller H2S-rich
  spark discharge experiment.}
\newblock Proceedings of the National Academy of Sciences of the United States
  of America 108: 5526--5531.
\bibAnnoteFile{Parker2011}

\bibitem{Rode1999}
Rode BM (1999) {Peptides and the origin of life.}
\newblock Peptides 20: 773--786.
\bibAnnoteFile{Rode1999}

\bibitem{Ferris2004}
Ferris JP, Joshi PC, Wang K, Miyakawa S, Huang W (2004) {Catalysis in prebiotic
  chemistry : Application to the synthesis of RNA oligomers}.
\newblock Advances in Space Research 33: 100--105.
\bibAnnoteFile{Ferris2004}

\bibitem{Orgel2004}
Orgel LE (2004) {Prebiotic chemistry and the origin of the RNA world}.
\newblock Critical Reviews in Biochemistry 39: 99--123.
\bibAnnoteFile{Orgel2004}

\bibitem{Budin2010}
Budin I, Szostak JW (2010) {Expanding roles for diverse physical phenomena
  during the origin of life.}
\newblock Annual Review of Biophysics 39: 245--263.
\bibAnnoteFile{Budin2010}

\bibitem{Barbas2008}
Barbas CF (2008) {Organocatalysis lost: modern chemistry, ancient chemistry,
  and an unseen biosynthetic apparatus.}
\newblock Angewandte Chemie (International ed in English) 47: 42--47.
\bibAnnoteFile{Barbas2008}

\bibitem{MacMillan2008}
MacMillan DWC (2008) {The advent and development of organocatalysis.}
\newblock Nature 455: 304--308.
\bibAnnoteFile{MacMillan2008}

\bibitem{Bertelsen2009}
Bertelsen S, Jorgensen KA (2009) {Organocatalysis--after the gold rush.}
\newblock Chemical Society reviews 38: 2178--2189.
\bibAnnoteFile{Bertelsen2009}

\bibitem{Severin1997}
Severin K, Lee DH, Kennan AJ, Ghadiri MR (1997) {A synthetic peptide ligase.}
\newblock Nature 389: 706--709.
\bibAnnoteFile{Severin1997}

\bibitem{Brack2007}
Brack A (2007) {From interstellar amino acids to prebiotic catalytic peptides:
  A review.}
\newblock Chemistry \& Biodiversity 4: 665--679.
\bibAnnoteFile{Brack2007}

\bibitem{Cech1986a}
Cech TR, Bass BL (1986) {Biological catalysis by RNA.}
\newblock Annual review of biochemistry 55: 599--629.
\bibAnnoteFile{Cech1986a}

\bibitem{Symons1992}
Symons RH (1992) {Small catalytic RNAs.}
\newblock Annual review of biochemistry 61: 641--671.
\bibAnnoteFile{Symons1992}

\bibitem{Chen2007}
Chen X, Li N, Ellington AD (2007) {Ribozyme catalysis of metabolism in the RNA
  world}.
\newblock Chemistry \& Biodiversity 4: 633--655.
\bibAnnoteFile{Chen2007}

\bibitem{Srinivasan2009}
Srinivasan V, Morowitz HJ (2009) {Analysis of the intermediary metabolism of a
  reductive chemoautotroph}.
\newblock The Biological Bulletin 217: 222.
\bibAnnoteFile{Srinivasan2009}

\bibitem{Eigen1971}
Eigen M (1971) {Selforganization of matter and the evolution of biological
  macromolecules}.
\newblock Die Naturwissenschaften 58: 465--523.
\bibAnnoteFile{Eigen1971}

\bibitem{Kauffman1971}
Kauffman SA (1971) {Cellular homeostasis, epigenesis and replication in
  randomly aggregated macromolecular systems}.
\newblock Journal of Cybernetics 1: 71--96.
\bibAnnoteFile{Kauffman1971}

\bibitem{Rossler1971}
R\"{o}ssler OE (1971) {A system theoretic model for biogenesis.}
\newblock Zeitschrift f\"{u}r Naturforschung Teil B: Chemie, Biochemie,
  Biophysik, Biologie 26: 741--746.
\bibAnnoteFile{Rossler1971}

\bibitem{Farmer1986}
Farmer JD, Kauffman SA, Packard NH (1986) {Autocatalytic replication of
  polymers}.
\newblock Physica D: Nonlinear Phenomena 22: 50--67.
\bibAnnoteFile{Farmer1986}

\bibitem{Bagley1991}
Bagley RJ, Farmer JD (1991) {Spontaneous emergence of a metabolism}.
\newblock In: Langton CG, Taylor CE, Farmer JD, Rasmussen S, editors,
  Artificial Life II, Addison-Wesley. pp. 93--140.
\bibAnnoteFile{Bagley1991}

\bibitem{Bagley1991a}
Bagley RJ, Farmer JD, Fontana W (1991) {Evolution of a metabolism}.
\newblock In: Langton CG, Taylor CE, Farmer JD, Rasmussen S, editors,
  Artificial Life II, Addison-Wesley. pp. 141--158.
\bibAnnoteFile{Bagley1991a}

\bibitem{Stadler1993}
Stadler P, Fontana W, Miller J (1993) {Random catalytic reaction networks}.
\newblock Physica D: Nonlinear Phenomena 63: 378--392.
\bibAnnoteFile{Stadler1993}

\bibitem{Kauffman1993}
Kauffman SA (1993) {The origins of order: Self-organization and selection in
  evolution}.
\newblock New York: Oxford University Press.
\bibAnnoteFile{Kauffman1993}

\bibitem{Fontana1994}
Fontana W, Buss LW (1994) {What would be conserved if "the tape were played
  twice"?}
\newblock Proceedings of the National Academy of Sciences of the United States
  of America 91: 757--761.
\bibAnnoteFile{Fontana1994}

\bibitem{Jain1998}
Jain S, Krishna S (1998) {Autocatalytic sets and the growth of complexity in an
  evolutionary model}.
\newblock Physical Review Letters 81: 5684--5687.
\bibAnnoteFile{Jain1998}

\bibitem{Jain2001}
Jain S, Krishna S (2001) {A model for the emergence of cooperation,
  interdependence, and structure in evolving networks.}
\newblock Proceedings of the National Academy of Sciences of the United States
  of America 98: 543--547.
\bibAnnoteFile{Jain2001}

\bibitem{Hanel2005}
Hanel R, Kauffman S, Thurner S (2005) {Phase transition in random catalytic
  networks}.
\newblock Physical Review E 72: 1--7.
\bibAnnoteFile{Hanel2005}

\bibitem{Piedrafita2010}
Piedrafita G, Montero F, Mor\'{a}n F, C\'{a}rdenas ML, Cornish-Bowden A (2010)
  {A simple self-maintaining metabolic system: Robustness, autocatalysis,
  bistability}.
\newblock PLoS Computational Biology 6: e1000872.
\bibAnnoteFile{Piedrafita2010}

\bibitem{Eigen1977}
Eigen M, Schuster P (1977) {The hypercycle: A principle of natural
  self-organizsation. Part A: Emergence of the hypercycle}.
\newblock Die Naturwissenschaften 64: 541--565.
\bibAnnoteFile{Eigen1977}

\bibitem{Szathmary2006}
Szathm\'{a}ry E (2006) {The origin of replicators and reproducers.}
\newblock Philosophical transactions of the Royal Society B 361: 1761--1776.
\bibAnnoteFile{Szathmary2006}

\bibitem{Ohtsuki2009}
Ohtsuki H, Nowak MA (2009) {Prelife catalysts and replicators.}
\newblock Proceedings of the Royal Society B 276: 3783--3790.
\bibAnnoteFile{Ohtsuki2009}

\bibitem{Segre2000}
Segr\'{e} D, Ben-Eli D, Lancet D (2000) {Compositional genomes: prebiotic
  information transfer in mutually catalytic noncovalent assemblies.}
\newblock Proceedings of the National Academy of Sciences of the United States
  of America 97: 4112--4117.
\bibAnnoteFile{Segre2000}

\bibitem{Furusawa2006}
Furusawa C, Kaneko K (2006) {Evolutionary origin of power-laws in a biochemical
  reaction network: Embedding the distribution of abundance into topology}.
\newblock Physical Review E 73: 1--7.
\bibAnnoteFile{Furusawa2006}

\bibitem{Carletti2008}
Carletti T, Serra R, Poli I, Villani M, Filisetti A (2008) {Sufficient
  conditions for emergent synchronization in protocell models.}
\newblock Journal of theoretical biology 254: 741--751.
\bibAnnoteFile{Carletti2008}

\bibitem{Kamimura2010}
Kamimura A, Kaneko K (2010) {Reproduction of a protocell by replication of a
  minority molecule in a catalytic reaction network}.
\newblock Physical Review Letters 105: 1--4.
\bibAnnoteFile{Kamimura2010}

\bibitem{Bagley1989}
Bagley RJ, Farmer JD, Kauffman SA, Packard NH, Perelson AS, et~al. (1989)
  {Modeling adaptive biological systems.}
\newblock Bio Systems 23: 113--137.
\bibAnnoteFile{Bagley1989}

\bibitem{Wachtershauser1990}
W\"{a}chtersh\"{a}user G (1990) {Evolution of the first metabolic cycles.}
\newblock Proceedings of the National Academy of Sciences of the United States
  of America 87: 200--204.
\bibAnnoteFile{Wachtershauser1990}

\bibitem{Morowitz2000}
Morowitz HJ, Kostelnik JD, Yang J, Cody GD (2000) {The origin of intermediary
  metabolism.}
\newblock Proceedings of the National Academy of Sciences of the United States
  of America 97: 7704--7708.
\bibAnnoteFile{Morowitz2000}

\bibitem{SUNDIALS}
Hindmarsh AC, Brown PN, Grant KE, Lee SL, Serban R, et~al. (2005) {SUNDIALS:
  Suite of nonlinear and differential/algebraic equation solvers}.
\newblock ACM Transactions on Mathematical Software 31: 363--396.
\bibAnnoteFile{SUNDIALS}

\bibitem{XPPAUT}
Ermentrout B (2002) {Simulating, Analyzing, and Animating Dynamical Systems: A
  Guide to XPPAUT for Researchers and Students}.
\newblock Philadelphia: Society for Industrial Mathematics, first edition, 290
  pp.
\newblock \urlprefix\url{http://www.math.pitt.edu/~bard/xpp/xpp.html}.
\bibAnnoteFile{XPPAUT}

\bibitem{Craciun2006}
Craciun G, Tang Y, Feinberg M (2006) {Understanding bistability in complex
  enzyme-driven reaction networks.}
\newblock Proceedings of the National Academy of Sciences of the United States
  of America 103: 8697--8702.
\bibAnnoteFile{Craciun2006}

\bibitem{Wilhelm2007}
Wilhelm T (2007) {Analysis of structures causing instabilities}.
\newblock Physical Review E 76: 011911.
\bibAnnoteFile{Wilhelm2007}

\bibitem{Ramakrishnan2008}
Ramakrishnan N, Bhalla US (2008) {Memory switches in chemical reaction space.}
\newblock PLoS computational biology 4: e1000122.
\bibAnnoteFile{Ramakrishnan2008}

\bibitem{Copley2007}
Copley SD, Smith E, Morowitz HJ (2007) {The origin of the RNA world:
  co-evolution of genes and metabolism.}
\newblock Bioorganic chemistry 35: 430--443.
\bibAnnoteFile{Copley2007}

\bibitem{Powner2011}
Powner MW, Sutherland JD (2011) {Prebiotic chemistry: a new modus operandi.}
\newblock Philosophical transactions of the Royal Society B 366: 2870--2877.
\bibAnnoteFile{Powner2011}

\bibitem{Morowitz1999}
Morowitz HJ (1999) {A theory of biochemical organization, metabolic pathways,
  and evolution}.
\newblock Complexity 4: 39--53.
\bibAnnoteFile{Morowitz1999}

\bibitem{Steel2000}
Steel M (2000) {The emergence of a self-catalysing structure in abstract
  origin-of-life models}.
\newblock Applied Mathematics Letters 13: 91--95.
\bibAnnoteFile{Steel2000}

\bibitem{Hordijk2011}
Hordijk W, Kauffman SA, Steel M (2011) {Required levels of catalysis for
  emergence of autocatalytic sets in models of chemical reaction systems.}
\newblock International journal of molecular sciences 12: 3085--3101.
\bibAnnoteFile{Hordijk2011}

\bibitem{Kun2008}
Kun A, Papp B, Szathm\'{a}ry E (2008) {Computational identification of
  obligatorily autocatalytic replicators embedded in metabolic networks.}
\newblock Genome biology 9: R51.
\bibAnnoteFile{Kun2008}

\bibitem{Orgel2000}
Orgel LE (2000) {Self-organizing biochemical cycles.}
\newblock Proceedings of the National Academy of Sciences of the United States
  of America 97: 12503--12507.
\bibAnnoteFile{Orgel2000}

\bibitem{Orgel2008}
Orgel LE (2008) {The implausibility of metabolic cycles on the prebiotic
  earth.}
\newblock PLoS biology 6: e18.
\bibAnnoteFile{Orgel2008}

\bibitem{Wales2004}
Wales DJ (2004) {Energy landscapes: Applications to clusters, biomolecules and
  glasses}.
\newblock Cambridge University Press, 692 pp.
\newblock \doi{10.2277/0521814154}.
\bibAnnoteFile{Wales2004}

\end{thebibliography}


\section*{Supporting Information}
\noindent\textbf{Appendix S1\\Details of the general model and explicit examples for $f=1$ and 2.}

\ \\
\noindent\textbf{Appendix S2\\$N$-independence of results at large $N$.}

\ \\
\noindent\textbf{Appendix S3\\Dimensionless rate equations and dependence of $\Lambda$ on dimensionless parameters.}

\ \\
\noindent\textbf{Table S1\\List of reactions and their catalysts in ACS(8,10) (referred in Fig. 4).}

\ \\
\noindent\textbf{Table S2\\List of reactions and their catalysts in ACS441 (referred in Fig. 11).}

\ \\
\noindent\textbf{Table S3\\List of reactions and their catalysts in ACS(36,28) (referred in Fig. 12).}

\ \\
\noindent\textbf{Table S4\\List of reactions and their catalysts in ACS(18,27) (referred in Fig. 13).}

\newpage
\setcounter{section}{0}
\setcounter{equation}{0}
\makeatletter
\renewcommand{\theequation}{S1.\@arabic\c@equation}
\makeatother

\section*{\huge{Supporting Information: Appendix S1\\Details of the general model and explicit examples for $f=1$ and 2}}
\section{Rate equations in a general case}
When there are $f$ food set species, ${\mathcal F} = \{{\mathrm A}_1, {\mathrm A}_2, \ldots, {\mathrm A}_f\}$, a general molecule is represented as ${\mathrm A} = (a_1, a_2, \ldots, a_f)$, where $a_i$ (a non-negative integer) is the number of monomers of type ${\mathrm A}_i$ contained in A. The `size' or `length' $n$ of the molecule is defined as the total number of monomers of all types in it: $n = \sum_i a_i$. The food set molecules themselves are represented by the $f$-tuples $(1,0,0,\ldots,0)$, $(0,1,0,0,\ldots,0)$, $\ldots$, $(0,0,\ldots,0,1)$.

A general reaction pair is given by ${\mathrm B} + {\mathrm C} \rightleftharpoons {\mathrm A}$, where ${\mathrm A}=(a_1, a_2, \ldots, a_f)$, ${\mathrm B}=(b_1, b_2, \ldots, b_f)$ and ${\mathrm C}=(c_1, c_2, \ldots, c_f)$, with $a_i = b_i+c_i$. If $x_{\mathrm A}$ denotes the concentration of the molecule A, the rate equations are given by $\dot x_{\mathrm{A}} = 0$ if $\mathrm{A} \in \mathcal{F}$; otherwise,
\begin{equation}
    \label{rateequation-generalmodel}
    \dot x_{\mathrm{A}} = \sum_{(\mathrm{B,C}) \in {\mathcal Q}_{\mathrm A}} v_{\mathrm{BC}} - \sum_{\mathrm B, B \ne A}  v_{\mathrm{AB}} - 2 v_{\mathrm{AA}} - \phi_{\mathrm{A}} x_{\mathrm{A}},
\end{equation}
where $v_{\mathrm{XY}}  = k^F_{\mathrm{XY}}x_{\mathrm X} x_{\mathrm Y} - k^R_{\mathrm{XY}}x_{\mathrm{Z}}$ is the net forward flux of the reaction pair $\mathrm{X} + \mathrm{Y} \rightleftharpoons \mathrm{Z}$ with forward rate constant $k^F_{\mathrm{XY}}$ and reverse rate constant $k^R_{\mathrm{XY}}$, $\phi_{\mathrm A}$ is the loss rate of A, and ${\mathcal Q}_{\mathrm A}$ represents the set of unordered pairs of molecules which can combine together to form A (${\mathcal Q}_{\mathrm A} = \{({\mathrm B},{\mathrm C}): b_i+c_i = a_i, \mathrm{\ for\  all\ } i=1,2,\ldots,f\}$).

\section{\label{f1-chemistry}Rate equations for a chemistry with $f=1$}
For $f=1$, instead of using a `1-tuple', we represent the molecules of size $n$ by the notation A($n$) for clarity.
Following Eq. (\ref{rateequation-generalmodel}), rate equations for the system are given by $\dot x_1 = 0$ and for $n=2,3,\ldots$,
\begin{align}
    \label{rateequationflux-f1}
    \dot x_n =& \sum_{i \leq j,i+j=n}v_{ij} - \sum_{i=1,i \neq n}^{\infty}v_{in} - 2v_{nn} - \phi_n x_n \\
    \label{rateequation-f1}
    =& \sum_{i \leq j,i+j=n}\left(k^F_{ij}x_ix_j - k^R_{ij}x_n\right) - \sum_{i=1,i \neq n}^{\infty}\left(k^F_{in}x_ix_n - k^R_{in}x_{i+n}\right) - 2\left(k^F_{nn}x_n^2 - k^R_{nn}x_{2n}\right) - \phi_n x_n.
\end{align}

\noindent Explicitly,
\begin{align*}
    \dot x_2 =& \left(k^F_{11}x_1x_1 - k^R_{11}x_{2}\right) - \left(k^F_{12}x_1x_2 - k^R_{12}x_{3}\right) - 2\left(k^F_{22}x_2^2 - k^R_{22}x_{4}\right) - \left(k^F_{32}x_3x_2 - k^R_{32}x_{5}\right) - \ldots - \phi_2 x_2 \\
    \dot x_3 =& \left(k^F_{12}x_1x_2 - k^R_{12}x_{3}\right) - \left(k^F_{13}x_1x_3 - k^R_{13}x_{4}\right) - \left(k^F_{23}x_2x_3 - k^R_{23}x_{5}\right) - 2\left(k^F_{33}x_3x_3 - k^R_{33}x_{6}\right) - \left(k^F_{43}x_4x_3 - k^R_{43}x_7\right) \\
     & - \ldots - \phi_3 x_3 \\
    \dot x_4 =& \left(k^F_{13}x_1x_3 - k^R_{13}x_{4}\right) + \left(k^F_{22}x_2x_2 - k^R_{22}x_{4}\right)
    - \left(k^F_{14}x_1x_4 - k^R_{14}x_{5}\right) - \left(k^F_{24}x_2x_4 - k^R_{24}x_{6}\right) - \left(k^F_{34}x_3x_4 - k^R_{34}x_{7}\right) \\
     &  - 2\left(k^F_{44}x_4^2 - k^R_{44}x_{8}\right) - \left(k^F_{54}x_5x_4 - k^R_{54}x_{9}\right) - \ldots - \phi_4 x_4 \\
    \vdots
\end{align*}

\noindent\textbf{Truncation of the chemistry for numerical simulations:}
The model has no upper limit on the largest molecule that can be produced in the chemistry but for the purposes of numerical simulation we assume that the largest molecule that can be produced is of size $N$. For the finite chemistry we exclude all the ligation reactions (and their reverse) that produce molecules of size larger than $N$. This results in $N-1$ coupled ordinary differential equations in $f=1$ case, given by:
\begin{eqnarray}
    \label{rateequation-N}
    \dot x_n &=& \sum_{i \leq j,i+j=n}\left(k^F_{ij}x_ix_j - k^R_{ij}x_n\right) - \sum_{i=1,i \neq n}^{(N-n)}\left(k^F_{in}x_ix_n - k^R_{in}x_{i+n}\right) - 2\left(k^F_{nn}x_n^2 - k^R_{nn}x_{2n}\right) - \phi_n x_n
\end{eqnarray}
for $n=2,3,\ldots,N$. Note that $\infty$ is replaced by $(N-n)$ in the second summation so that the largest molecule produced is of length $N$.

These ODEs were integrated using CVODE library of SUNDIALS package\footnote{https://computation.llnl.gov/casc/sundials/main.html} \cite{SUNDIALS} and (for smaller $N$) XPPAUT \cite{XPPAUT}. The dependence of steady state concentrations on $N$ is discussed separately in Supporting Appendix S2.

\subsection{An example with $f=1, N=6$}
The fully connected spontaneous chemistry has 9 reaction pairs of forward and reverse reactions given in the following table:

\noindent
\begin{small}
\begin{tabularx}{\textwidth}{|X|X|X|}
  \hline
  R1: A(1) + A(1) \reactionrevarrow{\ensuremath{k^F_{11}}}{\ensuremath{k^R_{11}}} A(2) &
  R4: A(2) + A(2) \reactionrevarrow{\ensuremath{k^F_{22}}}{\ensuremath{k^R_{22}}} A(4) &
  R7: A(1) + A(5) \reactionrevarrow{\ensuremath{k^F_{15}}}{\ensuremath{k^R_{15}}} A(6) \\
  R2: A(1) + A(2) \reactionrevarrow{\ensuremath{k^F_{12}}}{\ensuremath{k^R_{12}}} A(3) &
  R5: A(1) + A(4) \reactionrevarrow{\ensuremath{k^F_{14}}}{\ensuremath{k^R_{14}}} A(5) &
  R8: A(2) + A(4) \reactionrevarrow{\ensuremath{k^F_{24}}}{\ensuremath{k^R_{24}}} A(6) \\
  R3: A(1) + A(3) \reactionrevarrow{\ensuremath{k^F_{13}}}{\ensuremath{k^R_{13}}} A(4) &
  R6: A(2) + A(3) \reactionrevarrow{\ensuremath{k^F_{23}}}{\ensuremath{k^R_{23}}} A(5) &
  R9: A(3) + A(3) \reactionrevarrow{\ensuremath{k^F_{33}}}{\ensuremath{k^R_{33}}} A(6) \\
  \hline
\end{tabularx}
\end{small}

From Eq. (\ref{rateequation-N}), the 5 equations defining the dynamics are:
\begin{subequations}
 \begin{align}
    \nonumber
    \dot x_2 =& (k^F_{11} x_1^2 - k^R_{11} x_2)- (k^F_{12} x_1x_2 - k^R_{12}x_3) - 2(k^F_{22}x_2^2 - k^R_{22}x_4) - (k^F_{23}x_2x_3 - k^R_{23}x_5) - (k^F_{24}x_2x_4 - k^R_{24}x_6) \\
              & - \phi_2 x_2, \\
    \dot x_3 =& (k^F_{12} x_1x_2 - k^R_{12} x_3) - (k^F_{13} x_1x_3 - k^R_{13}x_4) - (k^F_{23} x_2x_3 - k^R_{23}x_5) - 2(k^F_{33}x_3^2 - k^R_{33}x_6) - \phi_3 x_3, \\
    \dot x_4 =& (k^F_{13} x_1x_3 - k^R_{13} x_4) + (k^F_{22} x_2^2 - k^R_{22}x_4) - (k^F_{14} x_1x_4 - k^R_{14}x_5) - (k^F_{24}x_2x_4 - k^R_{24}x_6) - \phi_4 x_4, \\
    \dot x_5 =& (k^F_{14} x_1x_4 - k^R_{14} x_5) + (k^F_{23} x_2x_3 - k^R_{23}x_5) - (k^F_{15} x_1x_5 - k^R_{15}x_6) - \phi_5 x_5, \\
    \dot x_6 =& (k^F_{15} x_1x_5 - k^R_{15} x_6) + (k^F_{24} x_2x_4 - k^R_{24}x_6) + (k^F_{33} x_3^2 - k^R_{33}x_6) - \phi_6 x_6.
 \end{align}
\end{subequations}

\subsection{\label{f1-catalyst}An example of inclusion of catalyst}
Say reaction pairs R1 and R4 of the previous example are catalyzed by ${\mathrm A}(4)$ with catalytic strengths $\kappa^{11}_4$ and $\kappa^{22}_4$ respectively. Then the equations for $x_2$ and $x_4$ get modified to
\begin{subequations}
 \begin{align}
    \nonumber
    \dot x_2 =& (1 + \kappa^{11}_4x_4)(k^F_{11} x_1^2 - k^R_{11} x_2)- (k^F_{12} x_1x_2 - k^R_{12}x_3) - 2(1 + \kappa^{22}_4x_4)(k^F_{22}x_2^2 - k^R_{22}x_4) - (k^F_{23}x_2x_3 - k^R_{23}x_5) \\  & - (k^F_{24}x_2x_4 - k^R_{24}x_6) - \phi_2 x_2, \\
    \nonumber
    \dot x_4 =& (k^F_{13} x_1x_3 - k^R_{13} x_4) + (1 + \kappa^{22}_4x_4)(k^F_{22} x_2^2 - k^R_{22}x_4) - (k^F_{14} x_1x_4 - k^R_{14}x_5) - (k^F_{24}x_2x_4 - k^R_{24}x_6) \\  & - \phi_4 x_4.
 \end{align}
\end{subequations}

The equations for $\dot x_3$, $\dot x_5$ and $\dot x_6$ remain as before. This chemistry, but with $N$ extended to 15 is discussed in the context of bistability in Fig. 6 of main text. There all the spontaneous rate constants and $x_1$ have been chosen to be unity, all the $\phi_n$'s are equal and $\kappa^{11}_4 = \kappa^{22}_4 = \kappa$.

\section{\label{f2-chemistry}Rate equations for a chemistry with $f=2$}
For a $f=2$ chemistry, ${\mathcal F} = \{{\mathrm A}_1, {\mathrm A}_2\}$, a general molecule is given by $\mathrm{A} = (a_1, a_2)$, and the monomers are given by 2-tuples (1,0) and (0,1). There are $(n+1)$ molecules of length $n$, given by 2-tuples: $(n,0),(n-1,1),\ldots,(1,n-1),(0,n)$. A finite chemistry, \ie, a chemistry in which the largest molecules that can be produced is of length $N$, has a total of $\frac{(N+1)(N+2)}{2}-1$ molecules (including monomers), and hence $\frac{(N+1)(N+2)}{2}-3$ rate equations.

\subsection{An example with $f=2$, $N=3$}
The fully connected spontaneous chemistry has 9 reaction pairs given in the following table:

\noindent
\begin{small}
\begin{tabularx}{\textwidth}{|X|X|X|}
  \hline
  R1: (0,1) + (0,1) \reactionrevarrow{}{} (0,2) &
  R4: (0,1) + (0,2) \reactionrevarrow{}{} (0,3) &
  R7: (0,1) + (2,0) \reactionrevarrow{}{} (2,1) \\
  R2: (0,1) + (1,0) \reactionrevarrow{}{} (1,1) &
  R5: (0,1) + (1,1) \reactionrevarrow{}{} (1,2) &
  R8: (1,0) + (1,1) \reactionrevarrow{}{} (2,1) \\
  R3: (1,0) + (1,0) \reactionrevarrow{}{} (2,0) &
  R6: (1,0) + (0,2) \reactionrevarrow{}{} (1,2) &
  R9: (1,0) + (2,0) \reactionrevarrow{}{} (3,0) \\
  \hline
\end{tabularx}
\end{small}

Using Eq. (\ref{rateequation-generalmodel}) suitably modified for a finite chemistry (as for $f=1$ case in Section \ref{f1-chemistry}), we can write the following rate equations:
\begin{subequations}
\label{f2-fluxequations}
\begin{align}
    \dot x_{(0,2)} =& v_{\mathrm R1} - v_{\mathrm R4} - v_{\mathrm R6} - \phi_{(0,2)} x_{(0,2)} \\
    \label{f2-fluxequations-11}\dot x_{(1,1)} =& v_{\mathrm R2} - v_{\mathrm R5} - v_{\mathrm R8} - \phi_{(1,1)} x_{(1,1)} \\
    \dot x_{(2,0)} =& v_{\mathrm R3} - v_{\mathrm R7} - v_{\mathrm R9} - \phi_{(2,0)} x_{(2,0)} \\
    \dot x_{(0,3)} =& v_{\mathrm R4} - \phi_{(0,3)} x_{(0,3)} \\
    \dot x_{(1,2)} =& v_{\mathrm R5} + v_{\mathrm R6} - \phi_{(1,2)} x_{(1,2)} \\
    \dot x_{(2,1)} =& v_{\mathrm R7} + v_{\mathrm R8} - \phi_{(2,1)} x_{(2,1)} \\
    \dot x_{(3,0)} =& v_{\mathrm R9} - \phi_{(3,0)} x_{(3,0)}.
\end{align}
\end{subequations}
$v_{\mathrm Ri}$ is the net forward flux of the reaction pair R$i$ (as given in the table above).
Eqs. (\ref{f2-fluxequations}) can be expanded as to get the equations in $x$'s. For example, Eq. (\ref{f2-fluxequations-11}) becomes
\begin{align}
    \nonumber
    \dot x_{(1,1)} =& \left(k^F_{\mathrm R2}x_{(0,1)}x_{(1,0)} - k^R_{\mathrm R2}x_{(1,1)}\right) - \left(k^F_{\mathrm R5}x_{(1,0)}x_{(1,1)} - k^R_{\mathrm R5}x_{(2,1)}\right) - \left(k^F_{\mathrm R8}x_{(0,1)}x_{(1,1)} - k^R_{\mathrm R8}x_{(1,2)}\right)\\
        & - \phi_{(1,1)} x_{(1,1)}
\end{align}
where, $k^F_{{\mathrm R}i}$ and $k^R_{{\mathrm R}i}$ are the forward and reverse rate constant of reaction ${\mathrm R}i$, respectively.

When the catalysts are included, the reactions that involve catalysts are changed in the same way as in $f=1$ case (see Section \ref{f1-catalyst} of this appendix).

\newpage
\setcounter{section}{0}
\setcounter{equation}{0}
\makeatletter
\renewcommand{\theequation}{S2.\@arabic\c@equation}
\makeatother
\section*{\huge{Supporting Information: Appendix S2\\$N$-independence of results at large $N$}}
\noindent While the chemistry considered in this model is an infinite one, numerical exploration requires us to work with a finite set of molecules that participate in the chemistry. This introduces an additional parameter $N$, the size of the largest molecule produced in the chemistry, in the model (for details see Appendix S1). Here we present evidence showing that certain important properties of the model become essentially $N$-independent at large enough $N$.

\section{Uncatalyzed chemistry}

The steady state concentrations for an uncatalyzed chemistry can be approximated by an exponential function, $x_n = ce^{-\gamma n} = c\Lambda^n$, where $c$ and $\Lambda=e^{-\gamma}$ are constants (see Eq. (3) in the main text). We calculate $\Lambda$ from the slope of the straight line fit on the plot of the log of the steady state concentrations, $x_n$, versus length, $n$. We found that for sufficiently large $N$, $\Lambda$ becomes independent of $N$; see Fig. \ref{Nvslamda}.

When $\Lambda<1$, the concentrations of the large molecules are small. It is evident that if the concentrations of the large molecules produced in the chemistry are so small that their contributions to the dynamics of the (relatively) smaller molecules is negligible, then, including molecules larger than those already considered in such a chemistry would cause no difference to the results.

\section{Catalyzed chemistry}

For chemistries with catalyzed reactions, including ACSs, we find that for sufficiently large $N$, the numerical results are independent of $N$. In Fig. \ref{N-depend-acs} we show how the steady state concentrations in the chemistry that includes ACS65 (defined by Eq. (5) in main text) depend upon $N$.

The results for the chemistry with $f=2$ were also found to be $N$-independent (for sufficiently large values of $N$). We find that for $f=2$ the $N$-independence is reached at much smaller values of $N$ than for $f=1$.

\newpage
\setcounter{section}{0}
\setcounter{equation}{0}
\makeatletter
\renewcommand{\theequation}{S3.\@arabic\c@equation}
\makeatother

\section*{\huge{Supporting Information: Appendix S3\\Dimensionless rate equations and dependence of $\Lambda$ on dimensionless parameters}}
\noindent Eq. (2) in the main text can be cast in dimensionless form by introducing a concentration scale $\omega$ and a time scale $\tau$. We discuss below the case of a homogeneous spontaneous chemistry.

Define dimensionless quantities $u_n \equiv x_n/\omega$, $t' \equiv t/\tau$, $k_f' \equiv k_f \omega\tau$, $k_r' \equiv k_r\tau$ and $\phi' \equiv \phi\tau$. Then
\begin{equation}
    \dot x_n = \frac{dx_n}{dt} = \frac{d(u_n\omega)}{d(t'\tau)} = \frac{\omega}{\tau}\frac{du_n}{dt'},
\end{equation}
and Eq. (2) in the main text becomes
\begin{eqnarray}\label{rateeqndimless}
    \nonumber
    \frac{\omega}{\tau}\frac{du_n}{dt'} &=& \sum_{i \leq j,i+j=n}\left(\frac{k_f'}{\omega\tau}(\omega u_i)(\omega u_j) - \frac{k_r'}{\tau}(\omega u_n)\right) - \sum_{i=1,i \neq n}^{\infty}\left(\frac{k_f'}{\omega\tau} (\omega u_i)(\omega u_n) - \frac{k_r'}{\tau}(\omega u_{i+n})\right) \\
    & & - 2\left(\frac{k_f'}{\omega\tau}(\omega u_n)^2 - \frac{k_r'}{\tau}(\omega u_{2n})\right) - \frac{\phi'}{\tau} (\omega u_n) \\
    \frac{du_n}{dt'} &=& \sum_{i \leq j,i+j=n}\left(k_f' u_i u_j - k_r'u_n\right) - \sum_{i=1,i \neq n}^{\infty}\left(k_f'u_i u_n - k_r'u_{i+n}\right) - 2\left(k_f'u_n^2 - k_r'u_{2n}\right) - \phi' u_n
\end{eqnarray}

Without loss of generality one can choose $\omega = x_1 = A$ and (whenever $k_r \neq 0$) $\tau = 1/k_r$. Then the dimensionless concentration variables satisfy the same equations as before, but with $A = k_r =1$. There are now only
two independent dimensionless parameters, $k_f'$ and $\phi'$. The dependence on all 4 parameters can be recovered at the end by replacing $u_n$ by $x_n/A$, $t'$ by $tk_r$, $k_f'$ by $k_f A/k_r$ and $\phi'$ by $\phi/k_r$. The behaviour of $\Lambda$ as a function of $k_f$ and $\phi$ (keeping $A = k_r = 1$) is shown in Fig. \ref{Kf-Phi-Phasespace}.

One may also choose $\tau = 1/\phi$ (whenever $\phi \neq 0$). In that case the two independent dimensionless parameters will be $k_f'$ and $k_r'$. The behaviour of $\Lambda$ as a function of $k_f$ and $k_r$ (keeping $A = \phi = 1$)
is shown in Fig. \ref{Kf-Kr-Phasespace}.

When $k_r=0$ and $\phi=0$ both the above choices for scaling fail. In this case the system has no steady state for any finite $N$ as $\dot x_N$ is always positive.

\newpage
\section*{\huge{Supporting Information: Table S1\\List of reactions and their catalysts in ACS(8,10) (referred in Fig. 4 of main text)}}

\begin{center}
\begin{tabular}{|c|c|}
  \hline
 {\bf Reaction} & \hspace{0.1cm} {\bf Catalyst} \hspace{0.1cm} \\ \hline
 \hspace{0.5cm} $(0,1) + (1,0) \rightleftharpoons (1,1)$ \hspace{0.5cm} & $(1,3)$ \\ \hline
 $(1,0) + (1,0) \rightleftharpoons (2,0)$ &  $(3,6)$ \\ \hline
 $(0,1) + (1,1) \rightleftharpoons (1,2)$ &  $(5,6)$ \\ \hline
 $(0,1) + (1,2) \rightleftharpoons (1,3)$ &  $(8,10)$ \\ \hline
 $(1,1) + (1,2) \rightleftharpoons (2,3)$ &  $(1,3)$ \\ \hline
 $(1,3) + (2,3) \rightleftharpoons (3,6)$ &  $(1,2)$ \\ \hline
 $(2,0) + (3,6) \rightleftharpoons (5,6)$ &  $(2,3)$ \\ \hline
 $(2,3) + (5,6) \rightleftharpoons (7,9)$ &  $(3,6)$ \\ \hline
 $(1,1) + (7,9) \rightleftharpoons (8,10)$ &  $(2,3)$ \\ \hline
\end{tabular}
\end{center}
Note that this catalyzed chemistry forms an ACS. Each catalyst in the above list gets produced by a reaction of this chemistry. Also, every reactant is either a member of the set ${\mathcal F}$ or gets produced in this chemistry.

\newpage

\section*{\huge{Supporting Information: Table S2\\List of reactions and their catalysts in ACS441 (referred in Fig. 11 of main text)}}
The table lists all the reactions with their respective catalysts in the example of a catalyzed chemistry, quoted in the main text, containing a cascade of nested ACSs for $f=1$ generated using Algorithm 4. The steady state concentrations for this chemistry are displayed in Fig. 11. This chemistry was generated with $g=15$ and $n_1=1, n_2 = n_3 = \ldots = n_{15} = 2$.

Note that a general molecule for the $f=1$ case, represented by A($n$) in the main text, has been represented here by $n$ for brevity.

The molecules in various generations are as follows:
$P_0 = \{1\}$, $P_1 = \{2\}$, $P_2 = \{3,4\}$, $P_3 = \{6,7\}$, $P_4 = \{10,12\}$, $P_5 = \{17,18\}$, $P_6 = \{20,24\}$, $P_7 = \{25,48\}$, $P_8 = \{52,66\}$, $P_9 = \{67,69\}$, $P_{10} = \{77,84\}$, $P_{11} = \{144,168\}$, $P_{12} = \{221,288\}$, $P_{13} = \{305,336\}$, $P_{14} = \{372,389\}$, $P_{15} = \{397,441\}$.

The catalyst for a reaction listed under generation $P_k$ is added at step $k$ of algorithm. It is apparent from the reaction table that the ACSs are maximally overlapping, \ie, any ACS of generation $k$ contains all the reactions of generation $k-1$.

\begin{scriptsize}
\begin{center}
\begin{longtable}{|c|c|c|c|c|c|c|c|c|c|c|c|c|c|c|c|}
  \hline
  {\bf Reaction} & \multicolumn{15}{|c|}{\bf Catalyst added in generation} \\ \cline{2-16}
  & \hspace{0.1cm}{\bf $P_1$}\hspace{0.1cm}  & \hspace{0.1cm}{\bf $P_2$}\hspace{0.1cm} & \hspace{0.1cm}{\bf $P_3$}\hspace{0.1cm} & \hspace{0.1cm}{\bf $P_4$}\hspace{0.1cm} & \hspace{0.1cm}{\bf $P_5$}\hspace{0.1cm} & \hspace{0.1cm}{\bf $P_6$}\hspace{0.1cm} & \hspace{0.1cm}{\bf $P_7$}\hspace{0.1cm} & \hspace{0.1cm}{\bf $P_8$}\hspace{0.1cm} & \hspace{0.1cm}{\bf $P_9$}\hspace{0.1cm} & \hspace{0.05cm}{\bf $P_{10}$}\hspace{0.05cm} & \hspace{0.05cm}{\bf $P_{11}$}\hspace{0.05cm} & \hspace{0.05cm}{\bf $P_{12}$}\hspace{0.05cm} & \hspace{0.05cm}{\bf $P_{13}$}\hspace{0.05cm} & \hspace{0.05cm}{\bf $P_{14}$}\hspace{0.05cm} & \hspace{0.05cm}{\bf $P_{15}$}\hspace{0.05cm} \\ \hline
  \endfirsthead
  \multicolumn{16}{r}{\it $\ldots$ continued from previous page} \\ \hline
  {\bf Reaction} & \multicolumn{15}{|c|}{\bf Catalyst in generation} \\ \cline{2-16}
  & {\bf $P_1$}  & {\bf $P_2$} & {\bf $P_3$} & {\bf $P_4$} & {\bf $P_5$} & {\bf $P_6$} & {\bf $P_7$} & {\bf $P_8$} & {\bf $P_9$} & {\bf $P_{10}$} & {\bf $P_{11}$} & {\bf $P_{12}$} & {\bf $P_{13}$} & {\bf $P_{14}$} & {\bf $P_{15}$} \\ \hline
  \endhead
      \multicolumn{16}{l}{{\it continued on next page $\ldots$}} \\
  \endfoot
  \endlastfoot

  $1 + 1 \rightleftharpoons 2$ & 2 & 3 & 6 & 10 & 17 & 24 & 48 & 52 & 69 & 84 & 144 & 288 & 336 & 372 & 397\\ \hline
  $1 + 2 \rightleftharpoons 3$ & & 3 & 7 & 10 & 18 & 20 & 25 & 52 & 67 & 77 & 168 & 221 & 305 & 372 & 397  \\ \hline
  $2 + 2 \rightleftharpoons 4$ & & 4 & 7 & 10 & 18 & 20 & 25 & 66 & 67 & 77 & 168 & 221 & 336 & 372 & 397 \\ \hline
  $3 + 3 \rightleftharpoons 6$ & & & 6 & 12 & 17 & 24 & 25 & 66 & 67 & 84 & 144 & 221 & 336 & 372 & 441 \\ \hline
  $3 + 4 \rightleftharpoons 7$ & & & 7 & 12 & 18 & 24 & 48 & 52 & 67 & 84 & 144 & 221 & 336 & 372 & 397 \\ \hline
  $4 + 6 \rightleftharpoons 10$ & & & & 10 & 18 & 24 & 48 & 52 & 69 & 77 & 168 & 288 & 336 & 389 & 397 \\ \hline
  $6 + 6 \rightleftharpoons 12$ & & & & 12 & 17 & 20 & 48 & 66 & 69 & 77 & 144 & 221 & 305 & 372 & 441 \\ \hline
  $7 + 10 \rightleftharpoons 17$ & & & & & 17 & 24 & 25 & 52 & 69 & 84 & 168 & 221 & 336 & 389 & 441 \\ \hline
  $6 + 12 \rightleftharpoons 18$ & & & & & 18 & 20 & 48 & 66 & 69 & 77 & 168 & 288 & 336 & 372 & 441 \\ \hline
  $2 + 18 \rightleftharpoons 20$ & & & & & & 20 & 48 & 52 & 69 & 84 & 144 & 221 & 336 & 389 & 441 \\ \hline
  $7 + 17 \rightleftharpoons 24$ & & & & & & 20 & 25 & 66 & 69 & 77 & 168 & 221 & 336 & 372 & 397 \\ \hline
  $1 + 24 \rightleftharpoons 25$ & & & & & & & 48 & 52 & 69 & 84 & 144 & 288 & 305 & 372 & 441 \\ \hline
  $24 + 24 \rightleftharpoons 48$ & & & & & & & 48 & 66 & 67 & 77 & 144 & 221 & 305 & 389 & 441 \\ \hline
  $4 + 48 \rightleftharpoons 52$ & & & & & & & & 66 & 69 & 84 & 168 & 221 & 336 & 372 & 397 \\ \hline
  $18 + 48 \rightleftharpoons 66$ & & & & & & & & 66 & 69 & 84 & 144 & 288 & 336 & 389 & 441 \\ \hline
  $1 + 66 \rightleftharpoons 67$ & & & & & & & & & 67 & 77 & 168 & 221 & 336 & 389 & 441 \\ \hline
  $3 + 66 \rightleftharpoons 69$ & & & & & & & & & 67 & 84 & 168 & 221 & 305 & 372 & 441 \\ \hline
  $10 + 67 \rightleftharpoons 77$ & & & & & & & & & & 77 & 168 & 288 & 305 & 372 & 441 \\ \hline
  $17 + 67 \rightleftharpoons 84$ & & & & & & & & & &  77 & 144 & 221 & 305 & 389 & 397 \\ \hline
  $67 + 77 \rightleftharpoons 144$ & & & & & & & & & & & 144 & 288 & 305 & 389 & 397 \\ \hline
  $84 + 84 \rightleftharpoons 168$ & & & & & & & & & & & 144 & 221 & 336 & 389 & 441 \\ \hline
  $77 + 144 \rightleftharpoons 221$ & & & & & & & & & & & & 221 & 305 & 372 & 397 \\ \hline
  $144 + 144 \rightleftharpoons 288$ & & & & & & & & & & & & 221 & 336 & 372 & 441 \\ \hline
  $17 + 288 \rightleftharpoons 305$ & & & & & & & & & & & & & 305 & 372 & 441 \\ \hline
  $48 + 288 \rightleftharpoons 336$ & & & & & & & & & & & & & 336 & 389 & 397 \\ \hline
  $67 + 305 \rightleftharpoons 372$ & & & & & & & & & & & & & & 389 & 397 \\ \hline
  $84 + 305 \rightleftharpoons 389$ & & & & & & & & & & & & & & 389 & 397 \\ \hline
  $25 + 372 \rightleftharpoons 397$ & & & & & & & & & & & & & & & 441 \\ \hline
  $69 + 372 \rightleftharpoons 441$ & & & & & & & & & & & & & & & 397 \\ \hline
\end{longtable}
\end{center}
\end{scriptsize}

\newpage
\section*{\huge{Supporting Information: Table S3\\List of reactions and their catalysts in ACS(36,28) (referred in Fig. 12 of main text)}}
The table lists all the reactions with their respective catalysts in the example of a catalyzed chemistry, quoted in the main text, containing a cascade of nested ACSs for $f=2$ generated using Algorithm 4. The steady state concentrations for this chemistry are displayed in Fig. 12. This chemistry was generated with $g=7$ and $n_k=3$.

The molecules in various generations are as follows:
$P_0 = \{(1,0),(0,1)\}$, $P_1 = \{(1,1),(0,2),(2,0)\}$, $P_2 = \{(1,3),(2,2),(3,0)\}$, $P_3 = \{(2,6),(2,3),(5,2)\}$, $P_4 = \{(4,6),(7,4),(7,8)\}$, $P_5 = \{(8,12),(6,12),(14,8)\}$, $P_6 = \{(14,24),(22,20),(11,12)\}$, $P_7 = \{(29,28),(36,28), (24,26)\}$.

The catalyst for a reaction listed under generation $P_k$ is added at step $k$ of algorithm. It is apparent from the reaction table that the ACSs are maximally overlapping, \ie, any ACS of generation $k$ contains all the reactions of generation $k-1$.

\begin{small}
\begin{center}
\begin{longtable}{|c|c|c|c|c|c|c|c|}
  \hline
  \hspace{0.2cm}{\bf Reaction}\hspace{0.2cm} & \multicolumn{7}{|c|}{\bf Catalyst added in generation} \\ \cline{2-8}
  & \hspace{0.2cm}{\bf $P_1$}\hspace{0.2cm}  & \hspace{0.2cm}{\bf $P_2$}\hspace{0.2cm} & \hspace{0.2cm}{\bf $P_3$}\hspace{0.2cm} & \hspace{0.2cm}{\bf $P_4$}\hspace{0.2cm} & \hspace{0.2cm}{\bf $P_5$}\hspace{0.2cm} & \hspace{0.2cm}{\bf $P_6$}\hspace{0.2cm} & \hspace{0.2cm}{\bf $P_7$}\hspace{0.2cm} \\ \hline
  \endfirsthead
  \multicolumn{8}{r}{\it $\ldots$continued from previous page} \\ \hline
  {\bf Reaction} & \multicolumn{7}{|c|}{\bf Catalyst in generation} \\ \cline{2-8}
  & {\bf $P_1$}  & {\bf $P_2$} & {\bf $P_3$} & {\bf $P_4$} & {\bf $P_5$} & {\bf $P_6$} & {\bf $P_7$} \\ \hline
  \endhead
      \multicolumn{8}{l}{{\it continued on next page $\ldots$}} \\
  \endfoot
  \endlastfoot

  $(0,1) + (0,1) \rightleftharpoons  (0,2)$ & $(1,1)$ &  $(1,3)$ &  $(2,3)$ &  $(7,4)$ &  $(8,12)$ &  $(11,12)$ &  $(29,28)$ \\ \hline
  $(0,1) + (1,0) \rightleftharpoons  (1,1)$ & $(0,2)$ &  $(1,3)$ &  $(2,6)$ &  $(4,6)$ &  $(14,8)$ &  $(22,20)$ &  $(36,28)$ \\ \hline
  $(1,0) + (1,0) \rightleftharpoons  (2,0)$ & $(1,1)$ &  $(3,0)$ &  $(5,2)$ &  $(7,4)$ &  $(14,8)$ &  $(14,24)$ &  $(29,28)$ \\ \hline
  $(1,0) + (2,0) \rightleftharpoons  (3,0)$ & & $(2,2)$ &  $(5,2)$ &  $(4,6)$ &  $(14,8)$ &  $(14,24)$ &  $(24,26)$ \\ \hline
  $(1,1) + (0,2) \rightleftharpoons  (1,3)$ & & $(1,3)$ &  $(2,6)$ &  $(7,8)$ &  $(8,12)$ &  $(14,24)$ &  $(24,26)$ \\ \hline
  $(0,2) + (2,0) \rightleftharpoons  (2,2)$ & & $(1,3)$ &  $(2,6)$ &  $(7,8)$ &  $(6,12)$ &  $(22,20)$ &  $(29,28)$ \\ \hline
  $(0,1) + (2,2) \rightleftharpoons  (2,3)$ & & & $(2,6)$ &  $(7,4)$ &  $(14,8)$ &  $(22,20)$ &  $(24,26)$ \\ \hline
  $(3,0) + (2,2) \rightleftharpoons  (5,2)$ & & & $(2,3)$ &  $(7,8)$ &  $(6,12)$ &  $(11,12)$ &  $(29,28)$ \\ \hline
  $(1,3) + (1,3) \rightleftharpoons  (2,6)$ & & & $(5,2)$ &  $(7,8)$ &  $(14,8)$ &  $(22,20)$ &  $(24,26)$ \\ \hline
  $(2,0) + (2,6) \rightleftharpoons  (4,6)$ & & & & $(4,6)$ &  $(14,8)$ &  $(22,20)$ &  $(36,28)$ \\ \hline
  $(2,2) + (5,2) \rightleftharpoons  (7,4)$ & & & & $(7,4)$ &  $(6,12)$ &  $(14,24)$ &  $(36,28)$ \\ \hline
  $(5,2) + (2,6) \rightleftharpoons  (7,8)$ & & & & $(7,4)$ &  $(14,8)$ &  $(14,24)$ &  $(36,28)$ \\ \hline
  $(2,6) + (4,6) \rightleftharpoons  (6,12)$ & & & & & $(14,8)$ &  $(11,12)$ &  $(29,28)$ \\ \hline
  $(4,6) + (4,6) \rightleftharpoons  (8,12)$ & & & & & $(14,8)$ &  $(22,20)$ &  $(36,28)$ \\ \hline
  $(7,4) + (7,4) \rightleftharpoons  (14,8)$ & & & & & $(8,12)$ &  $(11,12)$ &  $(36,28)$ \\ \hline
  $(3,0) + (8,12) \rightleftharpoons  (11,12)$ & & & & & & $(22,20)$ &  $(36,28)$ \\ \hline
  $(8,12) + (6,12) \rightleftharpoons  (14,24)$ & & & & & & $(22,20)$ &  $(24,26)$ \\ \hline
  $(14,8) + (8,12) \rightleftharpoons  (22,20)$ & & & & & & $(11,12)$ &  $(29,28)$ \\ \hline
  $(2,6) + (22,20) \rightleftharpoons  (24,26)$ & & & & & & & $(29,28)$ \\ \hline
  $(7,8) + (22,20) \rightleftharpoons  (29,28)$ & & & & & & & $(29,28)$ \\ \hline
  $(14,8) + (22,20) \rightleftharpoons  (36,28)$ & & & & & & & $(29,28)$ \\ \hline
\end{longtable}
\end{center}
\end{small}

\newpage
\setcounter{table}{0}
\makeatletter
\renewcommand{\thetable}{S4\@alph\c@table}
\makeatother
\section*{\huge{Supporting Information: Table S4\\List of reactions and their catalysts in ACS(18,27) (referred in Fig. 13 of main text)}}

The table lists all the reactions with their respective catalysts and catalytic strengths in the example of a catalyzed chemistry, quoted in the main text, containing a cascade of partially overlapping ACSs for $f=2$ generated using Algorithm 5. The steady state concentrations for this chemistry are displayed in Fig. 13.

The catalyzed chemistry contains 10 generations of ACSs of lengths 3, 6, 10, 15, 19, 25, 30, 35, 40, and 45.

\begin{center}
\begin{longtable}{|c|c|}
  \caption{{\bf List of reactions in the ACSs of different length and their catalysts.} ACSs of increasing length (using Algorithm 5) are added in the chemistry. The length of the ACS and the catalytic strength of the catalyst are mentioned in the table.} \\
  \hline
  \hspace{2.5cm}{\bf Reaction}\hspace{2.5cm} & \hspace{1cm}{\bf Catalyst}\hspace{1cm} \\ \hline
  \endfirsthead
  \multicolumn{1}{l}{\bfseries \tablename\ \thetable{}} & \multicolumn{1}{r}{{\it $\ldots$ continued from previous page}} \\   \hline
  \hspace{2cm}{\bf Reaction}\hspace{2cm} & \hspace{1cm}{\bf Catalyst}\hspace{1cm} \\ \hline
  \endhead
    \multicolumn{2}{l}{{\it continued on next page $\ldots$}} \\
  \endfoot
  \endlastfoot

  \multicolumn{2}{|c|}{\bf Generation 1 of length 3 ($\kappa=1000$)} \\ \hline
  $(1,0) + (1,0) \rightleftharpoons (2,0)$ & $(3,0)$ \\ \hline
  $(2,0) + (1,0) \rightleftharpoons (3,0)$ & $(3,0)$ \\ \hline

  \multicolumn{2}{|c|}{\bf Generation 2 of length 6 ($\kappa=2000$)} \\ \hline
  $(1,0) + (1,0) \rightleftharpoons (2,0)$ & $(4,2)$ \\ \hline
  $(0,1) + (2,0) \rightleftharpoons (2,1)$ & $(4,2)$ \\ \hline
  $(2,1) + (0,1) \rightleftharpoons (2,2)$ & $(4,2)$ \\ \hline
  $(2,1) + (2,1) \rightleftharpoons (4,2)$ & $(4,2)$ \\ \hline

  \multicolumn{2}{|c|}{\bf Generation 3 of length 10 ($\kappa=4000$)} \\ \hline
  $(1,0) + (0,1) \rightleftharpoons (1,1)$ & $(4,6)$ \\ \hline
  $(1,1) + (1,0) \rightleftharpoons (2,1)$ & $(4,6)$ \\ \hline
  $(1,1) + (1,1) \rightleftharpoons (2,2)$ & $(4,6)$ \\ \hline
  $(0,1) + (2,2) \rightleftharpoons (2,3)$ & $(4,6)$ \\ \hline
  $(2,3) + (2,3) \rightleftharpoons (4,6)$ & $(4,6)$ \\ \hline

  \multicolumn{2}{|c|}{\bf Generation 4 of length 15 ($\kappa=7000$)} \\ \hline
  $(0,1) + (0,1) \rightleftharpoons (0,2)$ & $(7,8)$ \\ \hline
  $(0,2) + (1,0) \rightleftharpoons (1,2)$ & $(7,8)$ \\ \hline
  $(1,2) + (1,0) \rightleftharpoons (2,2)$ & $(7,8)$ \\ \hline
  $(2,2) + (2,2) \rightleftharpoons (4,4)$ & $(7,8)$ \\ \hline
  $(1,2) + (4,4) \rightleftharpoons (5,6)$ & $(7,8)$ \\ \hline
  $(5,6) + (1,2) \rightleftharpoons (6,8)$ & $(7,8)$ \\ \hline
  $(5,6) + (2,2) \rightleftharpoons (7,8)$ & $(7,8)$ \\ \hline

  \multicolumn{2}{|c|}{\bf Generation 5 of length 19 ($\kappa=10000$)} \\ \hline
  $(0,1) + (1,0) \rightleftharpoons (1,1)$ & $(7,12)$ \\ \hline
  $(1,1) + (0,1) \rightleftharpoons (1,2)$ & $(7,12)$ \\ \hline
  $(1,2) + (1,2) \rightleftharpoons (2,4)$ & $(7,12)$ \\ \hline
  $(1,1) + (2,4) \rightleftharpoons (3,5)$ & $(7,12)$ \\ \hline
  $(2,4) + (3,5) \rightleftharpoons (5,9)$ & $(7,12)$ \\ \hline
  $(1,1) + (5,9) \rightleftharpoons (6,10)$ & $(7,12)$ \\ \hline
  $(1,2) + (6,10) \rightleftharpoons (7,12)$ & $(7,12)$ \\ \hline

  \multicolumn{2}{|c|}{\bf Generation 6 of length 25 ($\kappa=15000$)} \\ \hline
  $(0,1) + (1,0) \rightleftharpoons (1,1)$ & $(12,13)$ \\ \hline
  $(1,1) + (1,0) \rightleftharpoons (2,1)$ & $(12,13)$ \\ \hline
  $(2,1) + (0,1) \rightleftharpoons (2,2)$ & $(12,13)$ \\ \hline
  $(2,2) + (0,1) \rightleftharpoons (2,3)$ & $(12,13)$ \\ \hline
  $(1,0) + (2,3) \rightleftharpoons (3,3)$ & $(12,13)$ \\ \hline
  $(2,1) + (3,3) \rightleftharpoons (5,4)$ & $(12,13)$ \\ \hline
  $(2,3) + (5,4) \rightleftharpoons (7,7)$ & $(12,13)$ \\ \hline
  $(7,7) + (2,2) \rightleftharpoons (9,9)$ & $(12,13)$ \\ \hline
  $(7,7) + (3,3) \rightleftharpoons (10,10)$ & $(12,13)$ \\ \hline
  $(9,9) + (2,3) \rightleftharpoons (11,12)$ & $(12,13)$ \\ \hline
  $(11,12) + (1,1) \rightleftharpoons (12,13)$ & $(12,13)$ \\ \hline

  \multicolumn{2}{|c|}{\bf Generation 7 of length 30 ($\kappa=20000$)} \\ \hline
  $(0,1) + (1,0) \rightleftharpoons (1,1)$ & $(13,17)$ \\ \hline
  $(1,1) + (0,1) \rightleftharpoons (1,2)$ & $(13,17)$ \\ \hline
  $(1,2) + (1,1) \rightleftharpoons (2,3)$ & $(13,17)$ \\ \hline
  $(1,0) + (2,3) \rightleftharpoons (3,3)$ & $(13,17)$ \\ \hline
  $(3,3) + (1,1) \rightleftharpoons (4,4)$ & $(13,17)$ \\ \hline
  $(3,3) + (1,2) \rightleftharpoons (4,5)$ & $(13,17)$ \\ \hline
  $(2,3) + (4,5) \rightleftharpoons (6,8)$ & $(13,17)$ \\ \hline
  $(2,3) + (6,8) \rightleftharpoons (8,11)$ & $(13,17)$ \\ \hline
  $(6,8) + (4,4) \rightleftharpoons (10,12)$ & $(13,17)$ \\ \hline
  $(10,12) + (2,3) \rightleftharpoons (12,15)$ & $(13,17)$ \\ \hline
  $(12,15) + (1,2) \rightleftharpoons (13,17)$ & $(13,17)$ \\ \hline

  \multicolumn{2}{|c|}{\bf Generation 8 of length 35 ($\kappa=27000$)} \\ \hline
  $(1,0) + (0,1) \rightleftharpoons (1,1)$ & $(15,20)$ \\ \hline
  $(0,1) + (1,1) \rightleftharpoons (1,2)$ & $(15,20)$ \\ \hline
  $(1,1) + (1,2) \rightleftharpoons (2,3)$ & $(15,20)$ \\ \hline
  $(1,2) + (2,3) \rightleftharpoons (3,5)$ & $(15,20)$ \\ \hline
  $(3,5) + (1,1) \rightleftharpoons (4,6)$ & $(15,20)$ \\ \hline
  $(1,1) + (4,6) \rightleftharpoons (5,7)$ & $(15,20)$ \\ \hline
  $(3,5) + (4,6) \rightleftharpoons (7,11)$ & $(15,20)$ \\ \hline
  $(7,11) + (4,6) \rightleftharpoons (11,17)$ & $(15,20)$ \\ \hline
  $(1,0) + (11,17) \rightleftharpoons (12,17)$ & $(15,20)$ \\ \hline
  $(1,1) + (12,17) \rightleftharpoons (13,18)$ & $(15,20)$ \\ \hline
  $(13,18) + (1,2) \rightleftharpoons (14,20)$ & $(15,20)$ \\ \hline
  $(1,0) + (14,20) \rightleftharpoons (15,20)$ & $(15,20)$ \\ \hline

  \multicolumn{2}{|c|}{\bf Generation 9 of length 40 ($\kappa=35000$)} \\ \hline
  $(0,1) + (0,1) \rightleftharpoons (0,2)$ & $(14,26)$ \\ \hline
  $(1,0) + (0,2) \rightleftharpoons (1,2)$ & $(14,26)$ \\ \hline
  $(1,2) + (1,2) \rightleftharpoons (2,4)$ & $(14,26)$ \\ \hline
  $(1,2) + (2,4) \rightleftharpoons (3,6)$ & $(14,26)$ \\ \hline
  $(3,6) + (1,0) \rightleftharpoons (4,6)$ & $(14,26)$ \\ \hline
  $(3,6) + (2,4) \rightleftharpoons (5,10)$ & $(14,26)$ \\ \hline
  $(5,10) + (1,2) \rightleftharpoons (6,12)$ & $(14,26)$ \\ \hline
  $(6,12) + (2,4) \rightleftharpoons (8,16)$ & $(14,26)$ \\ \hline
  $(2,4) + (8,16) \rightleftharpoons (10,20)$ & $(14,26)$ \\ \hline
  $(0,1) + (10,20) \rightleftharpoons (10,21)$ & $(14,26)$ \\ \hline
  $(10,21) + (0,2) \rightleftharpoons (10,23)$ & $(14,26)$ \\ \hline
  $(0,2) + (10,23) \rightleftharpoons (10,25)$ & $(14,26)$ \\ \hline
  $(10,23) + (1,2) \rightleftharpoons (11,25)$ & $(14,26)$ \\ \hline
  $(8,16) + (5,10) \rightleftharpoons (13,26)$ & $(14,26)$ \\ \hline
  $(4,6) + (10,20) \rightleftharpoons (14,26)$ & $(14,26)$ \\ \hline

  \multicolumn{2}{|c|}{\bf Generation 10 of length 45 ($\kappa=50000$)} \\ \hline
  $(0,1) + (1,0) \rightleftharpoons (1,1)$ & $(18,27)$ \\ \hline
  $(0,1) + (1,1) \rightleftharpoons (1,2)$ & $(18,27)$ \\ \hline
  $(1,2) + (1,0) \rightleftharpoons (2,2)$ & $(18,27)$ \\ \hline
  $(1,2) + (1,1) \rightleftharpoons (2,3)$ & $(18,27)$ \\ \hline
  $(1,2) + (2,2) \rightleftharpoons (3,4)$ & $(18,27)$ \\ \hline
  $(2,3) + (2,3) \rightleftharpoons (4,6)$ & $(18,27)$ \\ \hline
  $(4,6) + (1,1) \rightleftharpoons (5,7)$ & $(18,27)$ \\ \hline
  $(4,6) + (3,4) \rightleftharpoons (7,10)$ & $(18,27)$ \\ \hline
  $(7,10) + (7,10) \rightleftharpoons (14,20)$ & $(18,27)$ \\ \hline
  $(0,1) + (14,20) \rightleftharpoons (14,21)$ & $(18,27)$ \\ \hline
  $(14,21) + (2,2) \rightleftharpoons (16,23)$ & $(18,27)$ \\ \hline
  $(2,3) + (16,23) \rightleftharpoons (18,26)$ & $(18,27)$ \\ \hline
  $(14,21) + (4,6) \rightleftharpoons (18,27)$ & $(18,27)$ \\ \hline
\end{longtable}
\end{center}

\begin{small}
\begin{center}
\begin{longtable}{|c|c|c|c|c|c|c|c|c|c|c|}
  \caption {{\bf List of reactions in the catalyzed chemistry and their catalysts.} The table lists all the reactions that are part of the catalyzed chemistry with all its catalysts. The catalysts that belong to different generations ($G_1$ to $G_{10}$) have been separated in different columns. It is easy to see from this table the amount of overlap between any two nested ACSs. For example, between the ACSs $G_5$ and $G_6$ which contain, respectively, 7 and 11 reactions, only one is common.} \\
  \hline
  \hspace{0.5cm}{\bf Reaction}\hspace{0.5cm} & \multicolumn{10}{|c|}{\bf Catalyst} \\ \cline{2-11}
  & \hspace{0.05cm}$G_1$\hspace{0.05cm} & \hspace{0.05cm}$G_2$\hspace{0.05cm} & \hspace{0.05cm}$G_3$\hspace{0.05cm} & \hspace{0.05cm}$G_4$\hspace{0.05cm} & \hspace{0.05cm}$G_5$\hspace{0.05cm} & \hspace{0.05cm}$G_6$\hspace{0.05cm} & \hspace{0.05cm}$G_7$\hspace{0.05cm} & \hspace{0.05cm}$G_8$\hspace{0.05cm} & \hspace{0.05cm}$G_9$\hspace{0.05cm} & $G_{10}$ \\ \hline
  \endfirsthead
    \multicolumn{1}{l}{\bfseries \tablename\ \thetable{}} & \multicolumn{10}{r}{{\it $\ldots$ continued from previous page}} \\   \hline
    \hspace{1cm}{\bf Reaction}\hspace{1cm} & \multicolumn{10}{|c|}{\bf Catalyst} \\ \cline{2-11}
    & $G_1$ & $G_2$ & $G_3$ & $G_4$ & $G_5$ & $G_6$ & $G_7$ & $G_8$ & $G_9$ & $G_{10}$ \\ \hline
  \endhead
    \multicolumn{11}{l}{{\it continued on next page $\ldots$}} \\
  \endfoot
  \endlastfoot
  $(0,1) + (0,1) \rightleftharpoons  (0,2)$ & & & & (7,8) & & & & & (14,26) & \\ \hline
  $(1,0) + (0,1) \rightleftharpoons  (1,1)$ & & & (4,6) & & (7,12) & (12,13) & (13,17) & (15,20) & & (18,27) \\ \hline
  $(1,0) + (1,0) \rightleftharpoons  (2,0)$ & (3,0) & (4,2) & & & & & & & & \\ \hline
  $(0,2) + (1,0) \rightleftharpoons  (1,2)$ & & & & (7,8) & & & & & (14,26) & \\ \hline
  $(1,1) + (0,1) \rightleftharpoons  (1,2)$ & & & & & (7,12) & & (13,17) & (15,20) & & (18,27) \\ \hline
  $(0,1) + (2,0) \rightleftharpoons  (2,1)$ & & (4,2) & & & & & & & & \\ \hline
  $(1,1) + (1,0) \rightleftharpoons  (2,1)$ & & & (4,6) & & & (12,13) & & & & \\ \hline
  $(2,0) + (1,0) \rightleftharpoons  (3,0)$ & (3,0) & & & & & & & & & \\ \hline
  $(2,1) + (0,1) \rightleftharpoons  (2,2)$ & & (4,2) & & & & (12,13) & & & & \\ \hline
  $(1,1) + (1,1) \rightleftharpoons  (2,2)$ & & & (4,6) & & & & & & & \\ \hline
  $(1,2) + (1,0) \rightleftharpoons  (2,2)$ & & & & (7,8) & & & & & & (18,27) \\ \hline
  $(0,1) + (2,2) \rightleftharpoons  (2,3)$ & & & (4,6) & & & (12,13) & & & & \\ \hline
  $(1,2) + (1,1) \rightleftharpoons  (2,3)$ & & & & & & & (13,17) & (15,20) & & (18,27) \\ \hline
  $(1,2) + (1,2) \rightleftharpoons  (2,4)$ & & & & & (7,12) & & & & (14,26) & \\ \hline
  $(1,0) + (2,3) \rightleftharpoons  (3,3)$ & & & & & & (12,13) & (13,17) & & & \\ \hline
  $(2,1) + (2,1) \rightleftharpoons  (4,2)$ & & (4,2) & & & & & & & & \\ \hline
  $(1,2) + (2,2) \rightleftharpoons  (3,4)$ & & & & & & & & & & (18,27) \\ \hline
  $(1,1) + (2,4) \rightleftharpoons  (3,5)$ & & & & & (7,12) & & & & & \\ \hline
  $(1,2) + (2,3) \rightleftharpoons  (3,5)$ & & & & & & & & (15,20) & & \\ \hline
  $(2,2) + (2,2) \rightleftharpoons  (4,4)$ & & & & (7,8) & & & & & & \\ \hline
  $(3,3) + (1,1) \rightleftharpoons  (4,4)$ & & & & & & & (13,17) & & & \\ \hline
  $(1,2) + (2,4) \rightleftharpoons  (3,6)$ & & & & & & & & & (14,26) & \\ \hline
  $(3,3) + (1,2) \rightleftharpoons  (4,5)$ & & & & & & & (13,17) & & & \\ \hline
  $(2,1) + (3,3) \rightleftharpoons  (5,4)$ & & & & & & (12,13) & & & & \\ \hline
  $(2,3) + (2,3) \rightleftharpoons  (4,6)$ & & & (4,6) & & & & & & & (18,27) \\ \hline
  $(3,5) + (1,1) \rightleftharpoons  (4,6)$ & & & & & & & & (15,20) & & \\ \hline
  $(3,6) + (1,0) \rightleftharpoons  (4,6)$ & & & & & & & & & (14,26) & \\ \hline
  $(1,2) + (4,4) \rightleftharpoons  (5,6)$ & & & & (7,8) & & & & & & \\ \hline
  $(1,1) + (4,6) \rightleftharpoons  (5,7)$ & & & & & & & & (15,20) & & (18,27) \\ \hline
  $(2,4) + (3,5) \rightleftharpoons  (5,9)$ & & & & & (7,12) & & & & & \\ \hline
  $(5,6) + (1,2) \rightleftharpoons  (6,8)$ & & & & (7,8) & & & & & & \\ \hline
  $(2,3) + (4,5) \rightleftharpoons  (6,8)$ & & & & & & & (13,17) & & & \\ \hline
  $(2,3) + (5,4) \rightleftharpoons  (7,7)$ & & & & & & (12,13) & & & & \\ \hline
  $(3,6) + (2,4) \rightleftharpoons  (5,10)$ & & & & & & & & & (14,26) & \\ \hline
  $(5,6) + (2,2) \rightleftharpoons  (7,8)$ & & & & (7,8) & & & & & & \\ \hline
  $(1,1) + (5,9) \rightleftharpoons  (6,10)$ & & & & & (7,12) & & & & & \\ \hline
  $(4,6) + (3,4) \rightleftharpoons  (7,10)$ & & & & & & & & & & (18,27) \\ \hline
  $(5,10) + (1,2) \rightleftharpoons  (6,12)$ & & & & & & & & & (14,26) & \\ \hline
  $(3,5) + (4,6) \rightleftharpoons  (7,11)$ & & & & & & & & (15,20) & & \\ \hline
  $(7,7) + (2,2) \rightleftharpoons  (9,9)$ & & & & & & (12,13) & & & & \\ \hline
  $(1,2) + (6,10) \rightleftharpoons  (7,12)$ & & & & & (7,12) & & & & & \\ \hline
  $(2,3) + (6,8) \rightleftharpoons  (8,11)$ & & & & & & & (13,17) & & & \\ \hline
  $(7,7) + (3,3) \rightleftharpoons  (10,10)$ & & & & & & (12,13) & & & & \\ \hline
  $(6,8) + (4,4) \rightleftharpoons  (10,12)$ & & & & & & & (13,17) & & & \\ \hline
  $(9,9) + (2,3) \rightleftharpoons  (11,12)$ & & & & & & (12,13) & & & & \\ \hline
  $(6,12) + (2,4) \rightleftharpoons  (8,16)$ & & & & & & & & & (14,26) & \\ \hline
  $(11,12) + (1,1) \rightleftharpoons  (12,13)$ & & & & & & (12,13) & & & & \\ \hline
  $(10,12) + (2,3) \rightleftharpoons  (12,15)$ & & & & & & & (13,17) & & & \\ \hline
  $(7,11) + (4,6) \rightleftharpoons  (11,17)$ & & & & & & & & (15,20) & & \\ \hline
  $(1,0) + (11,17) \rightleftharpoons  (12,17)$ & & & & & & & & (15,20) & & \\ \hline
  $(2,4) + (8,16) \rightleftharpoons  (10,20)$ & & & & & & & & & (14,26) & \\ \hline
  $(12,15) + (1,2) \rightleftharpoons  (13,17)$ & & & & & & & (13,17) & & & \\ \hline
  $(0,1) + (10,20) \rightleftharpoons  (10,21)$ & & & & & & & & & (14,26) & \\ \hline
  $(1,1) + (12,17) \rightleftharpoons  (13,18)$ & & & & & & & & (15,20) & & \\ \hline
  $(10,21) + (0,2) \rightleftharpoons  (10,23)$ & & & & & & & & & (14,26) & \\ \hline
  $(13,18) + (1,2) \rightleftharpoons  (14,20)$ & & & & & & & & (15,20) & & \\ \hline
  $(7,10) + (7,10) \rightleftharpoons  (14,20)$ & & & & & & & & & & (18,27) \\ \hline
  $(0,2) + (10,23) \rightleftharpoons  (10,25)$ & & & & & & & & & (14,26) & \\ \hline
  $(0,1) + (14,20) \rightleftharpoons  (14,21)$ & & & & & & & & & & (18,27) \\ \hline
  $(1,0) + (14,20) \rightleftharpoons  (15,20)$ & & & & & & & & (15,20) & & \\ \hline
  $(10,23) + (1,2) \rightleftharpoons  (11,25)$ & & & & & & & & & (14,26) & \\ \hline
  $(8,16) + (5,10) \rightleftharpoons  (13,26)$ & & & & & & & & & (14,26) & \\ \hline
  $(14,21) + (2,2) \rightleftharpoons  (16,23)$ & & & & & & & & & & (18,27) \\ \hline
  $(4,6) + (10,20) \rightleftharpoons  (14,26)$ & & & & & & & & & (14,26) & \\ \hline
  $(2,3) + (16,23) \rightleftharpoons  (18,26)$ & & & & & & & & & & (18,27) \\ \hline
  $(14,21) + (4,6) \rightleftharpoons  (18,27)$ & & & & & & & & & & (18,27) \\ \hline
\end{longtable}
\end{center}
\end{small}

\newpage
\setcounter{figure}{0}
\makeatletter
\renewcommand{\thefigure}{\@arabic\c@figure}
\makeatother

\section*{Figure Legends}
\begin{figure}[ht]
    \begin{center}
        \includegraphics[height=3in,angle=-90]{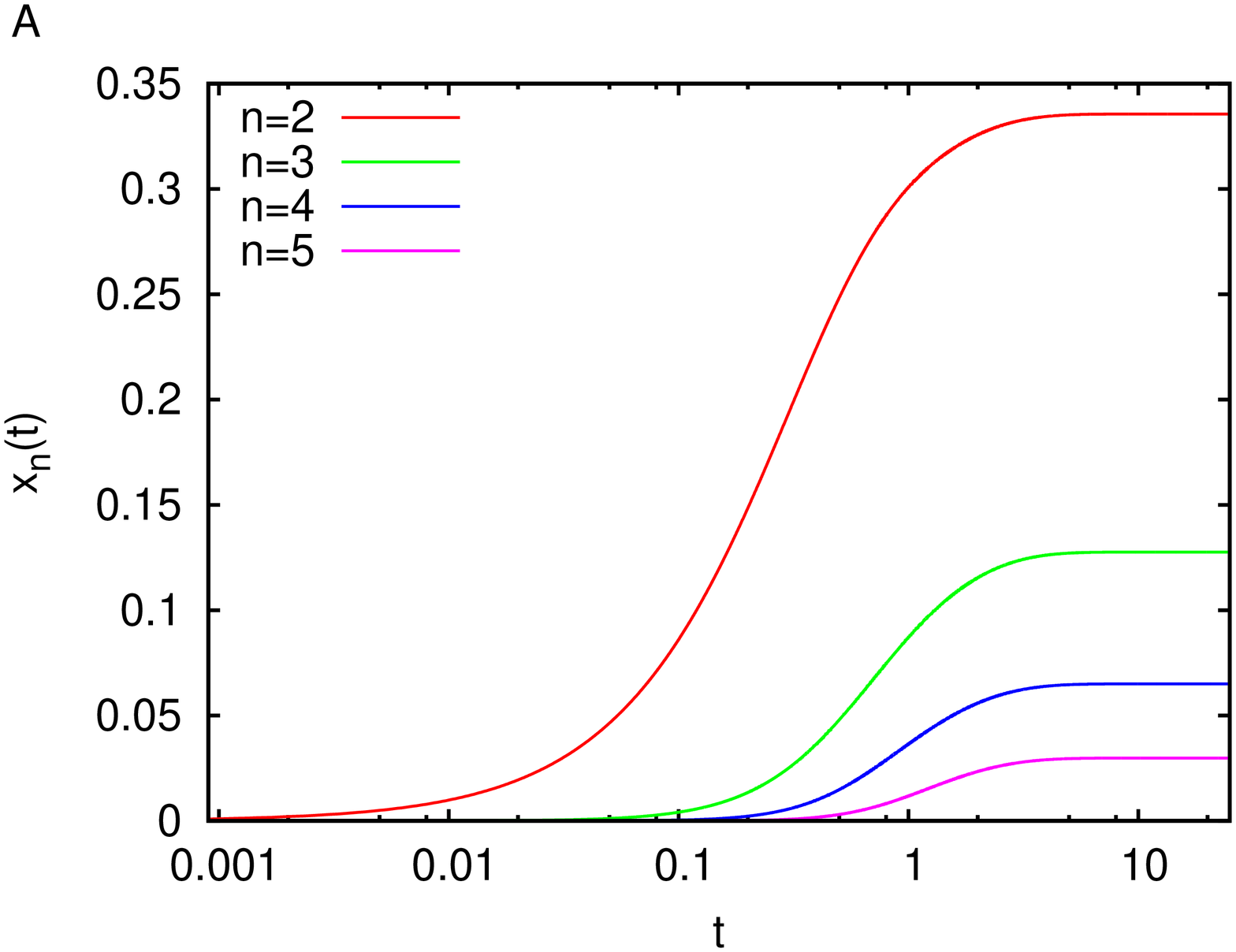}
        \includegraphics[height=3in,angle=-90]{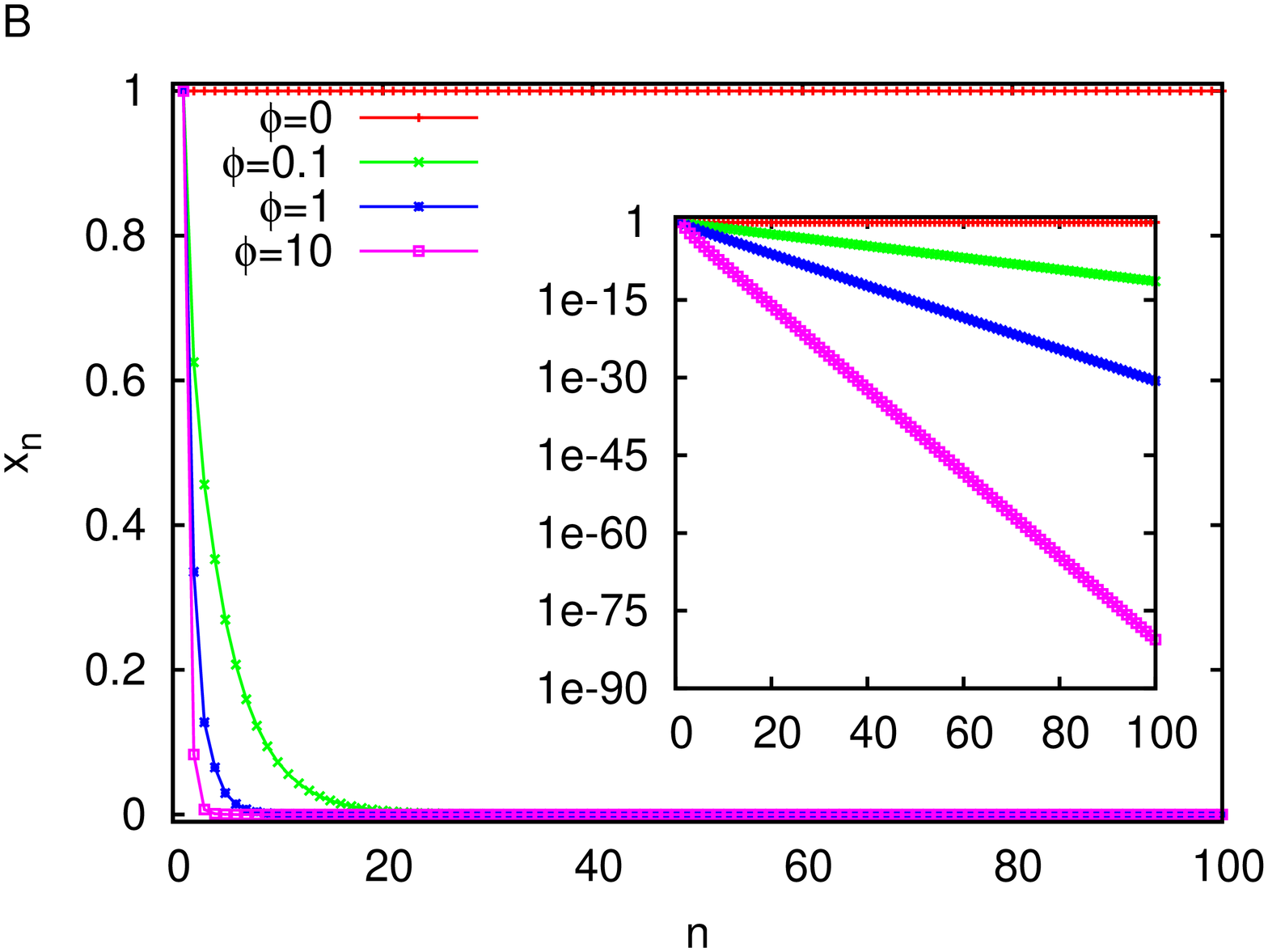}
    \end{center}
    \caption{{\bf Concentrations in uncatalyzed chemistries with a single food source.} (A) Evolution of concentrations with time for a chemistry with $k_f = k_r = A = \phi = 1$. For simulation purposes, the size of the largest molecule was taken to be $N=100$. (B) Steady state concentration as a function of molecule size. Parameters take the same values as in (A) except that four values of $\phi$ are shown, $\phi = 0, 0.1, 1, 10$. Inset shows the same on a semi-log plot; the straight lines are evidence of exponential damping of $x_n$ for large $n$ (Eq. (\ref{ss1})), with $\Lambda = 1, 0.77, 0.49, 0.16$ for the four cases, respectively. $\Lambda$ is computed from the slope of a straight line fit after ignoring the smaller molecules (up to $n=4$ in this case).}
    \label{noacs}
\end{figure}

\begin{figure}[ht]
    \begin{center}
        \includegraphics[height=3in,angle=-90,trim=2cm 0cm 2cm 0cm,clip=true]{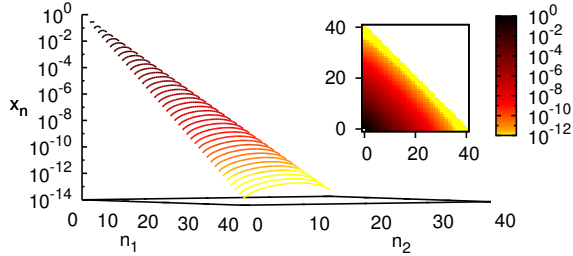}
    \end{center}
    \caption{{\bf Steady state concentration profile in an uncatalyzed chemistry with $f=2$.} The 3D plot shows the concentration $x_n$ of the molecule $n=(n_1,n_2)$ as a function of $n_1$ and $n_2$ in the steady state, for an uncatalyzed chemistry with $k_f = k_r = x_{(1,0)} = x_{(0,1)} = \phi = 1$, $N=40$. The inset shows a `top view' of the ($n_1$,$n_2$) plane with $x_n$ indicated in a colour map on a logarithmic scale.}
    \label{noacs-2d}
\end{figure}

\begin{figure}[ht]
    \begin{center}
        \includegraphics[height=2in,angle=-90]{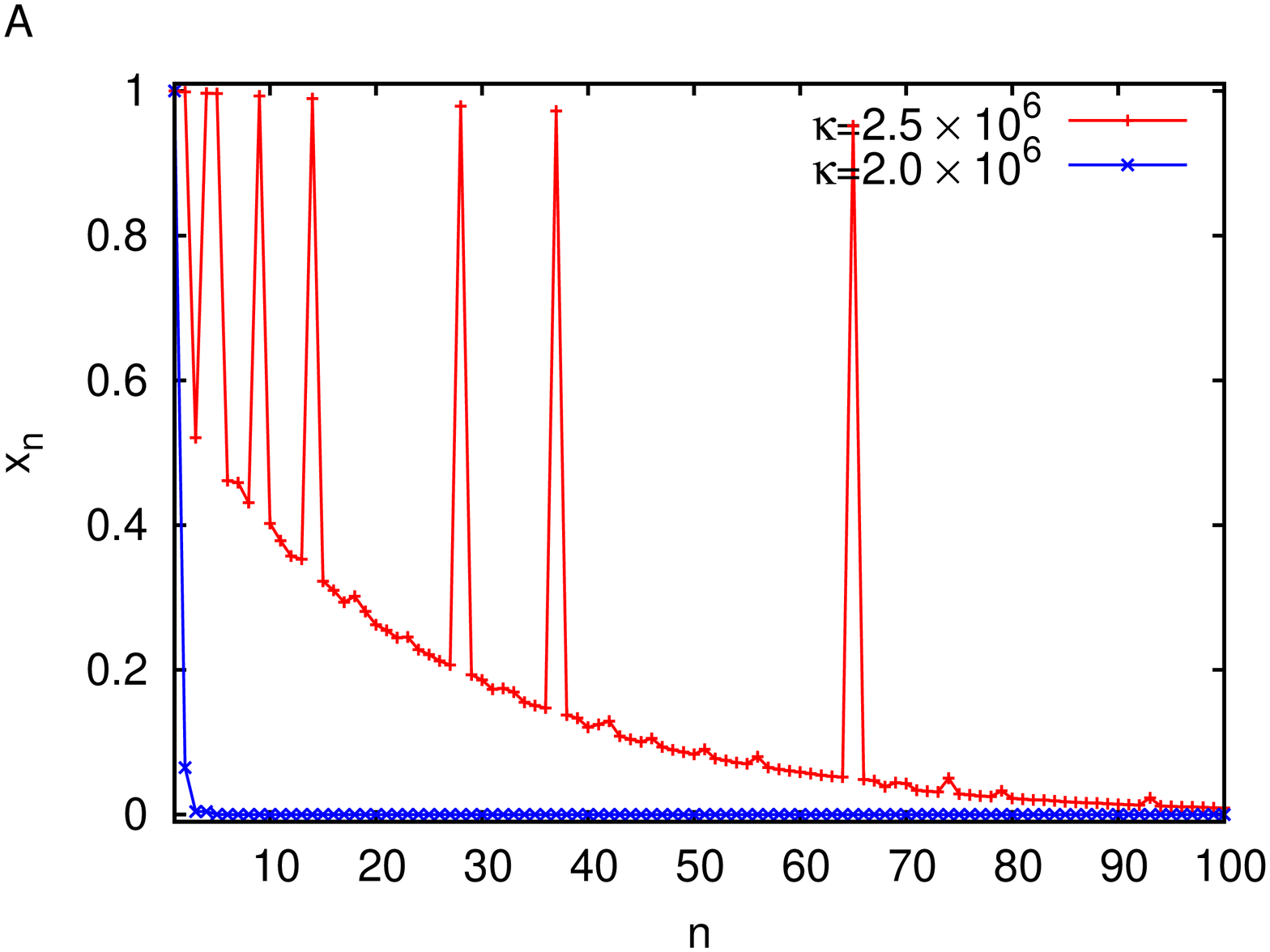}
        \includegraphics[height=2in,angle=-90]{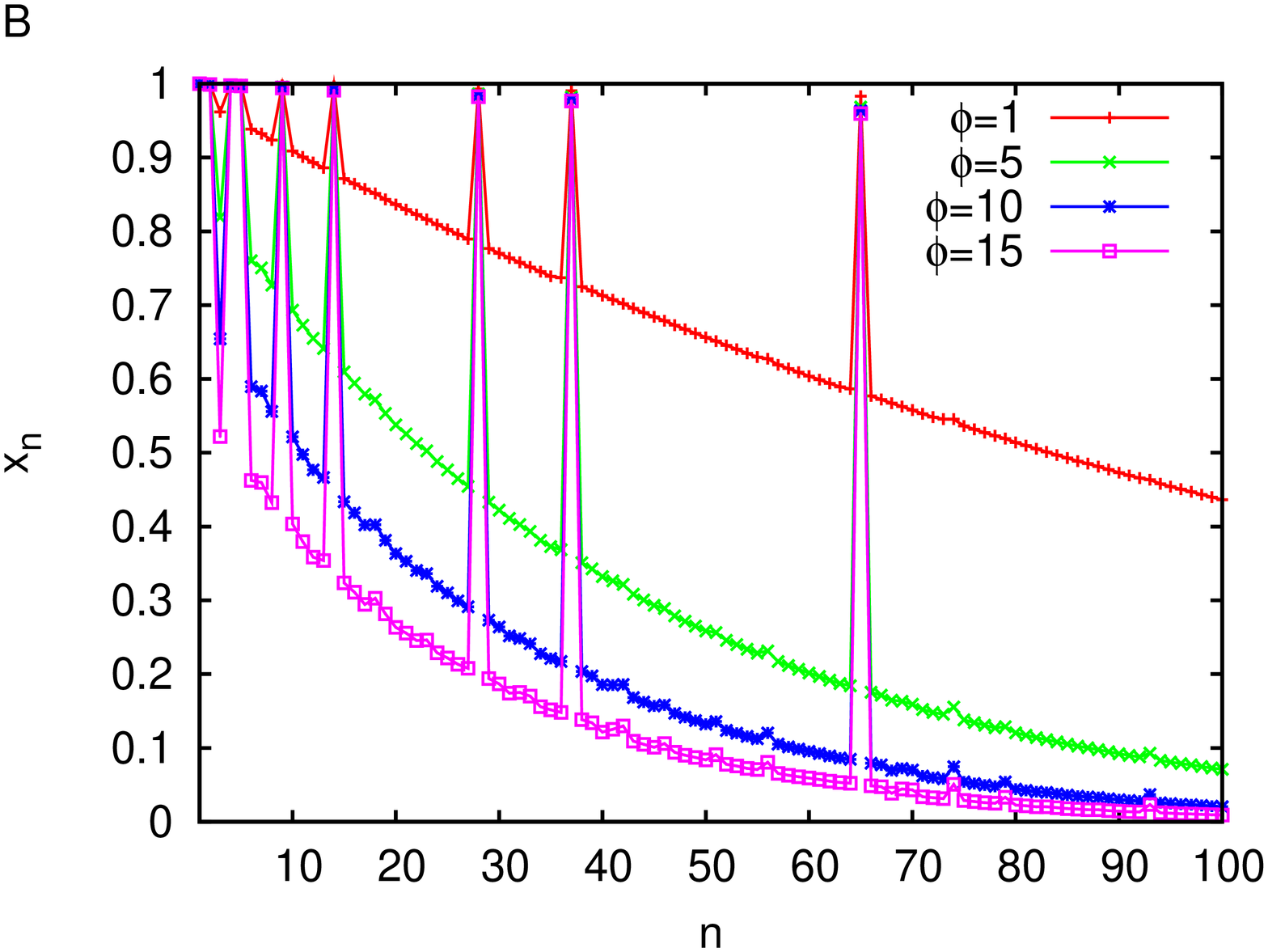}
        \includegraphics[height=2in,angle=-90]{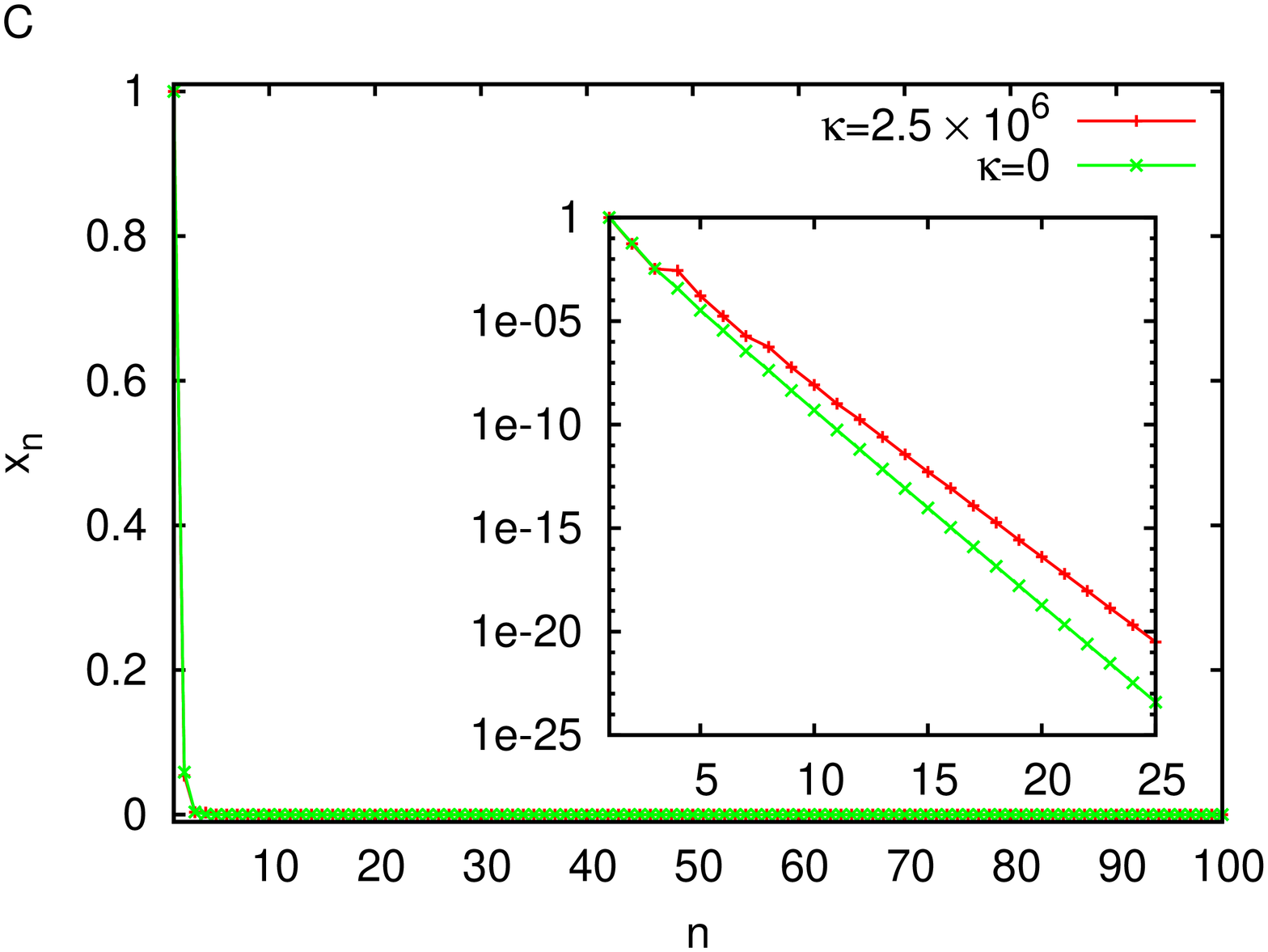}
    \end{center}
    \caption{{\bf Steady state concentration profile for ACS65 (Eqs. (\ref{acs65-definition})).} In all the cases $k_f = k_r = A = 1, N=100.$ (A) The concentration profile for two values of $\kappa$ for $\phi=15$. (B) The concentration profile for four values of $\phi$ for $\kappa = 3.0 \times 10^6$. (C) The concentration profile for $\kappa=2.5 \times 10^6$, $\phi=15$ but with reaction (\ref{acs65-rct1}) removed from the ACS (red curve) compared with the profile for the spontaneous chemistry, $\kappa=0$, $\phi=15$ (green curve). The inset shows the same with $x_n$ on a logarithmic scale. On the linear scale the two curves are indistinguishable.}
    \label{acs-eg}
\end{figure}

\begin{figure}[ht]
    \begin{center}
        \includegraphics[height=3in,angle=-90,trim=2cm 0cm 2cm 0cm,clip=true]{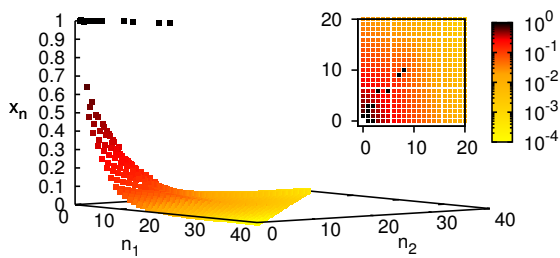}
    \end{center}
    \caption{{\bf Steady state concentration profile for ACS(8,10) in a chemistry with $f=2$.}  The molecules and reactions of the ACS are given in Supporting Table S1, the largest molecule being (8,10). $k_f = k_r = x_{(1,0)} = x_{(0,1)} = 1, \phi = 10, \kappa=10^6, N=40.$}
    \label{acs-eg-2d}
\end{figure}

\begin{figure}[ht]
    \begin{center}
        \includegraphics[height=3in,angle=-90]{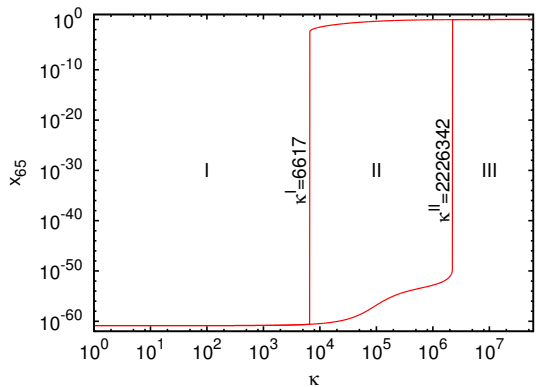}
    \end{center}
    \caption{{\bf Bistability in the dynamics of ACS65.} `Hysteresis curve' of the steady state concentration of A(65) versus $\kappa$ for $k_f = k_r = A = 1, \phi=15, N=100$. The curve is obtained by using two different initial conditions (i) the standard initial condition $x_n=0$ for all $n\geq2$, and (ii) a `high' initial condition $x_n=1$ for all $n\geq2$. In region I ($\kappa < \kappa^{I}=6617$) both initial conditions lead to a single fixed point in which $x_{65}$ is very low, $10^{-60}$. In region III ($\kappa > \kappa^{II} = 2226342$) both initial conditions again lead to a single fixed point but in this fixed point $x_{65}$ is high, close to unity. In region II ($\kappa^{I} \leq \kappa \leq \kappa^{II}$) the initial condition (i) leads to the lower fixed point and (ii) leads to the upper one. The transitions are very sharp, e.g., at $\kappa=2226341$ the system is numerically clearly seen in region II and at 2226343 in region III.}
    \label{acs-eg-bis}
\end{figure}

\begin{figure}[ht]
    \begin{center}
        \includegraphics[height=3in,angle=-90]{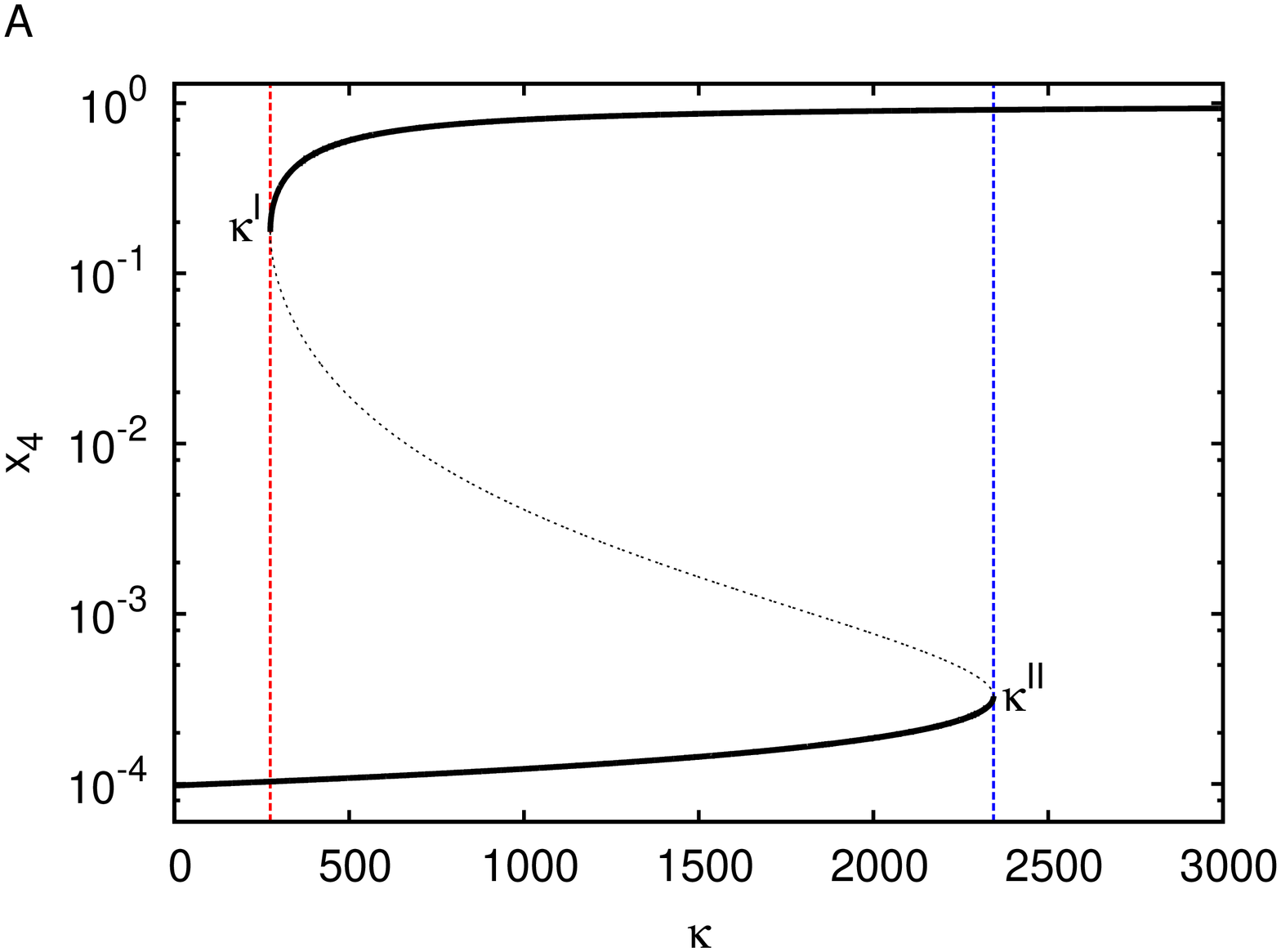}
        \includegraphics[height=3in,angle=-90]{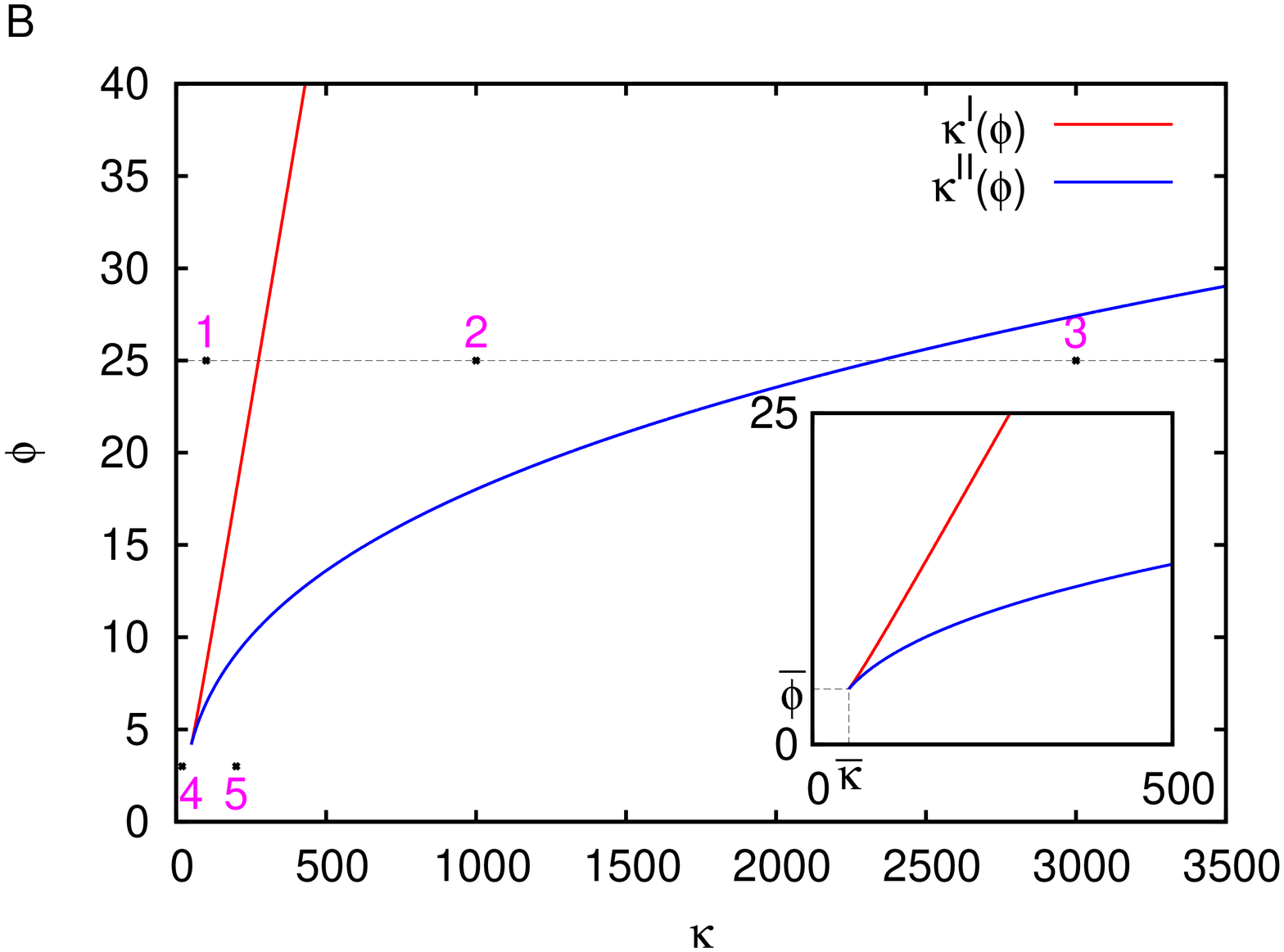}\\ \vspace{0.5cm}
        \includegraphics[height=2.75in,angle=-90]{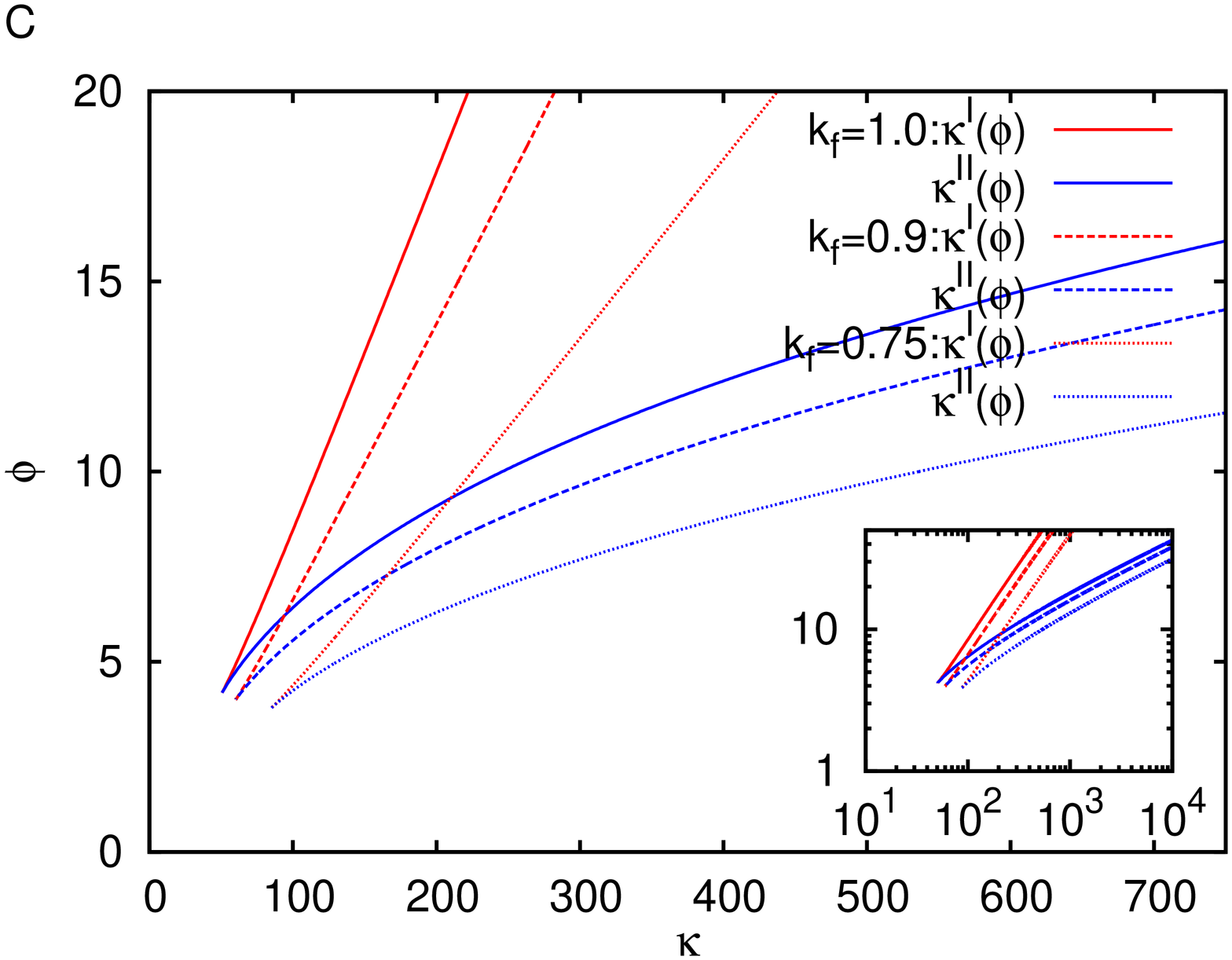}
        \includegraphics[width=1.75in]{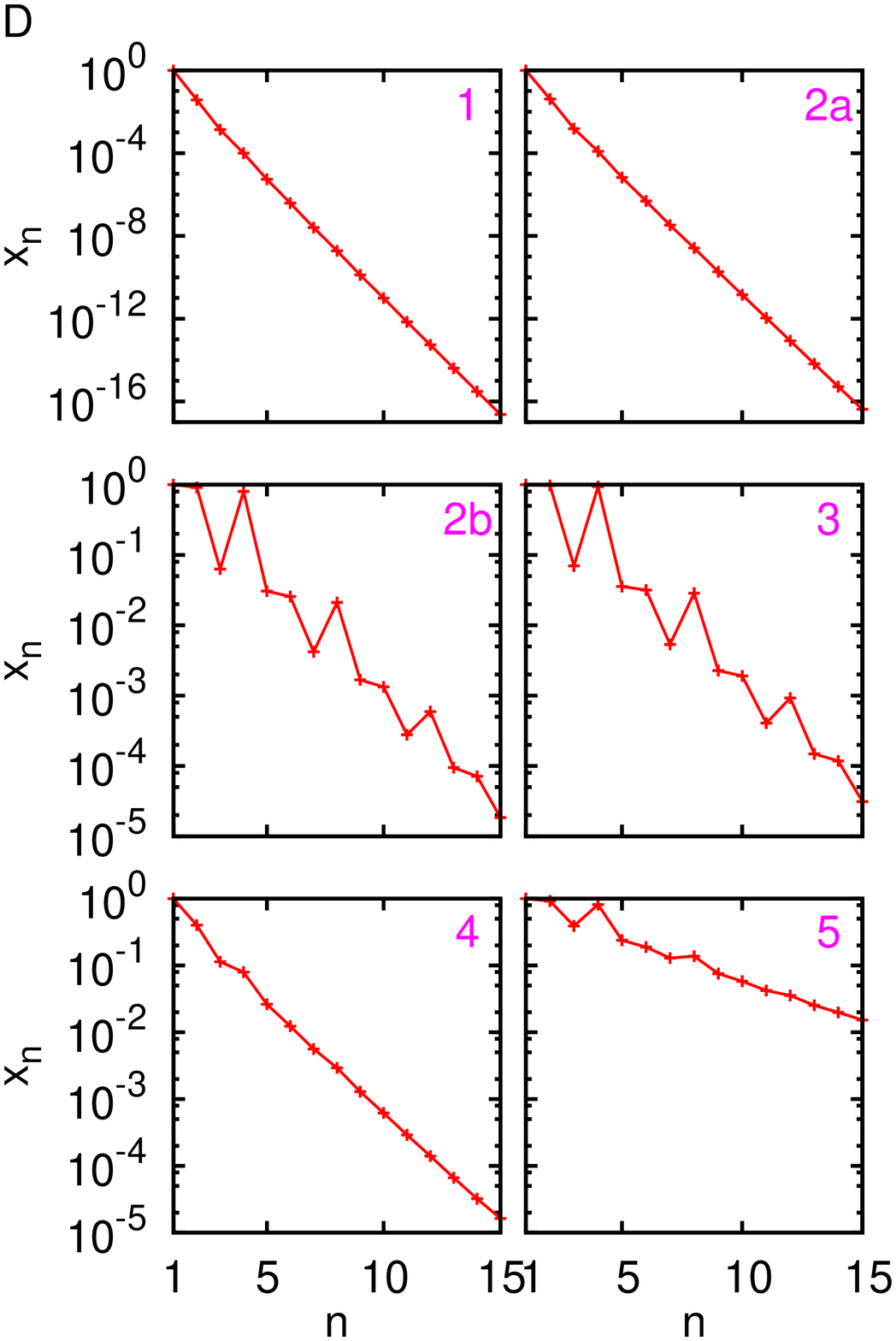}
    \end{center}
    \caption{{\bf Phase diagram and concentration profiles for ACS4.} (A) The steady state concentration $x_4$ versus $\kappa$ for $k_f = k_r = A = 1, \phi = 25, N=15$. The bistable region exists for the range $\kappa^{I} \leq \kappa \leq \kappa^{II}$ in which different initial conditions lead to two distinct steady state values of $x_4$. The solid black curves correspond to the two stable fixed points, and the dotted black curve to the unstable fixed point. (B) The dependence of $\kappa^{I}$ (red curve) and $\kappa^{II}$ (blue curve) on $\phi$ for $k_f = k_r = A = 1, N=15$. The bistable region lies between the two curves; in the rest of the phase space the system has a single fixed point. The inset shows the location of the critical point $(\bar{\kappa},\bar{\phi})$; there is no bistability for $\phi < \bar{\phi}$. (C) Dependence of the phase boundaries on $k_f$ for $k_r = A = 1, N = 15$, with the inset showing the behaviour on a log-log plot. (D) The steady state concentration profile of molecules shown at five representative points in the phase space (numbered 1 through 5 and marked in (B)). Note that at the phase point 2 that lies between the $\kappa^{I}$ and $\kappa^{II}$ curves there are two steady state profiles corresponding to the two stable fixed points of the system. The figure marked 2a shows the profile starting from the standard initial condition, and 2b from the initial condition where $x_n = 1$ for all $n$.}
    \label{acs4}
\end{figure}

\begin{figure}[ht]
    \begin{center}
        \includegraphics[height=3in,angle=-90]{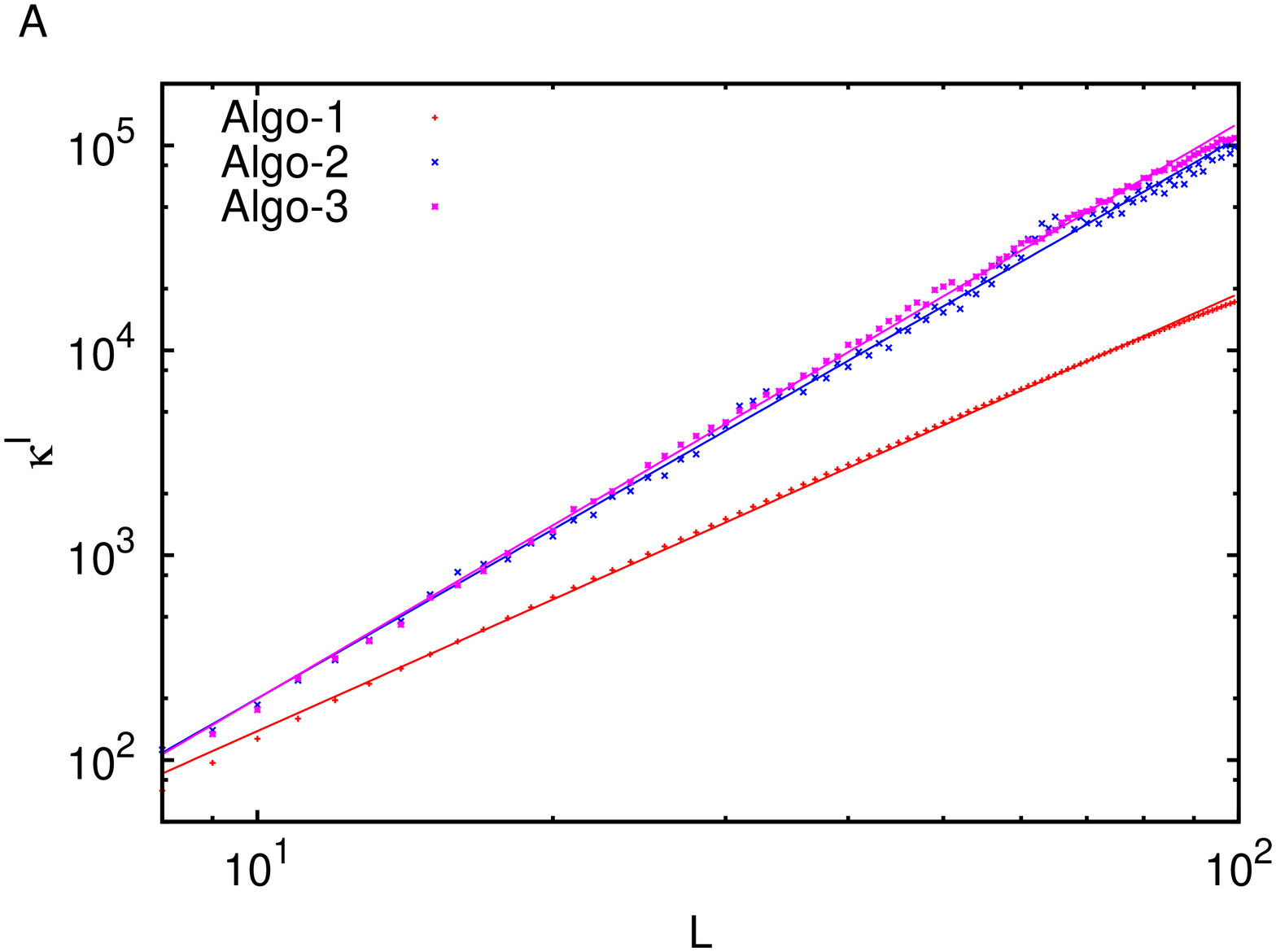}
        \includegraphics[height=3in,angle=-90]{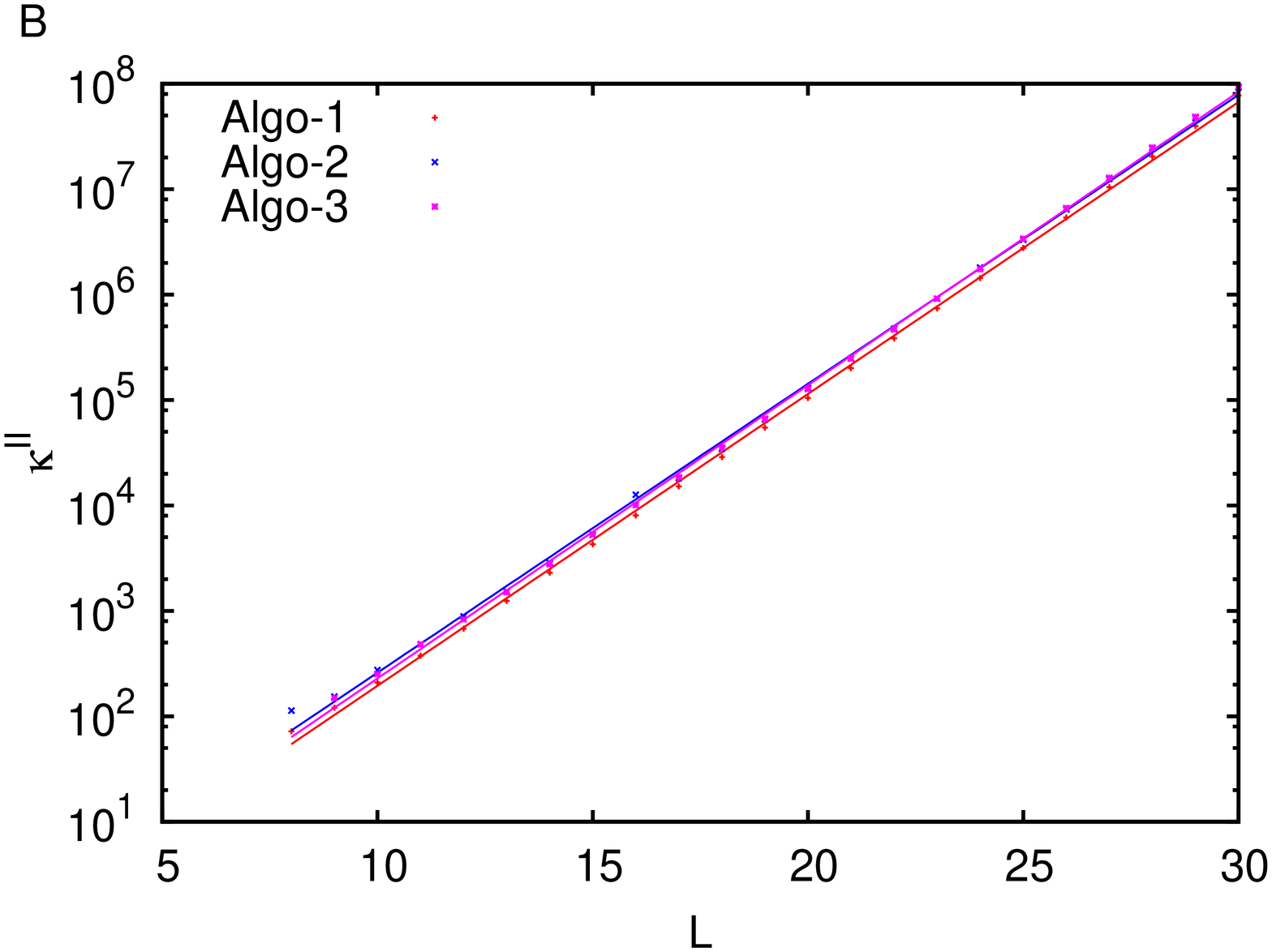}
    \end{center}
    \caption{{\bf The dependence of the bistable region on catalyst length $L$.} (A) The dependence of $\kappa^{I}$ on $L$. (B) The dependence of $\kappa^{II}$ on $L$. Simulations were done for extremal ACSs of length $L$ generated by three algorithms (see Methods), represented in the figure by different colours. For each $L$ the ACS in question has the property that the largest molecule produced in the ACS has $L$ monomers and catalyzes all the reactions in the ACS. All simulations were done for $k_f = k_r = A = \phi = 1$. $N=100$ in all cases except the $\kappa^{I}$ curve for Algorithm 1, where $N=200$, because in this case `finite-$N$' effects were quite significant at $N=100$. The figures suggest an approximate power law growth of $\kappa^{I}$ and exponential growth of $\kappa^{II}$ with $L$.}
    \label{kappa1-2-vs-L}
\end{figure}

\begin{figure}[ht]
    \begin{center}
        \includegraphics[height=5in,angle=-90,trim=2cm 0cm 2cm 0cm,clip=true]{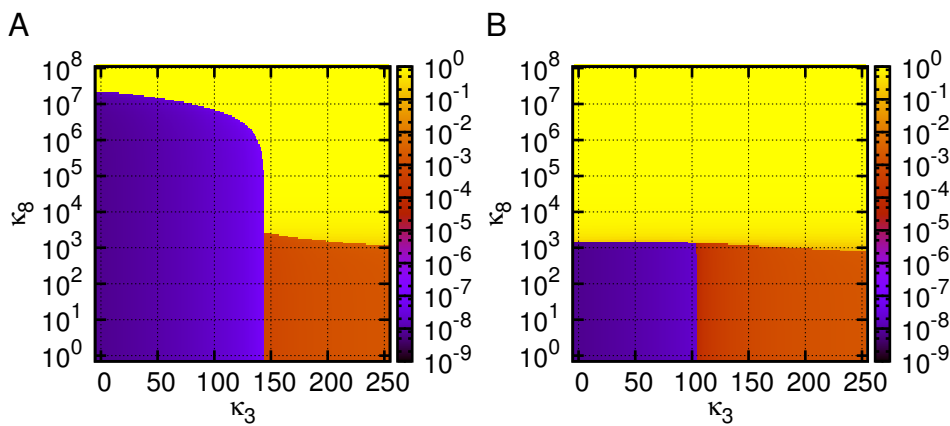}
    \end{center}
    \caption{{\bf Reinforcement of a larger ACS by a smaller one: The case of ACS3+8.} The figure shows the steady state concentration $x_8$ (in colour coding as indicated) for two different initial conditions as a function of $\kappa_3$ and $\kappa_8$, the catalytic strengths of $\mathrm{A}(3)$ and $\mathrm{A}(8)$ respectively. All simulations were done for $k_f = k_r = A = 1, \phi = 20, N=100$. The two figures (A) and (B) differ in the initial condition of the dynamics. (A) The standard initial condition, (B) initial condition $x_n = 1$ for all $n=2,3,\ldots,N$.}
    \label{nested-3-8}
\end{figure}

\begin{figure}[ht]
    \begin{center}
        \includegraphics[height=3in,angle=-90]{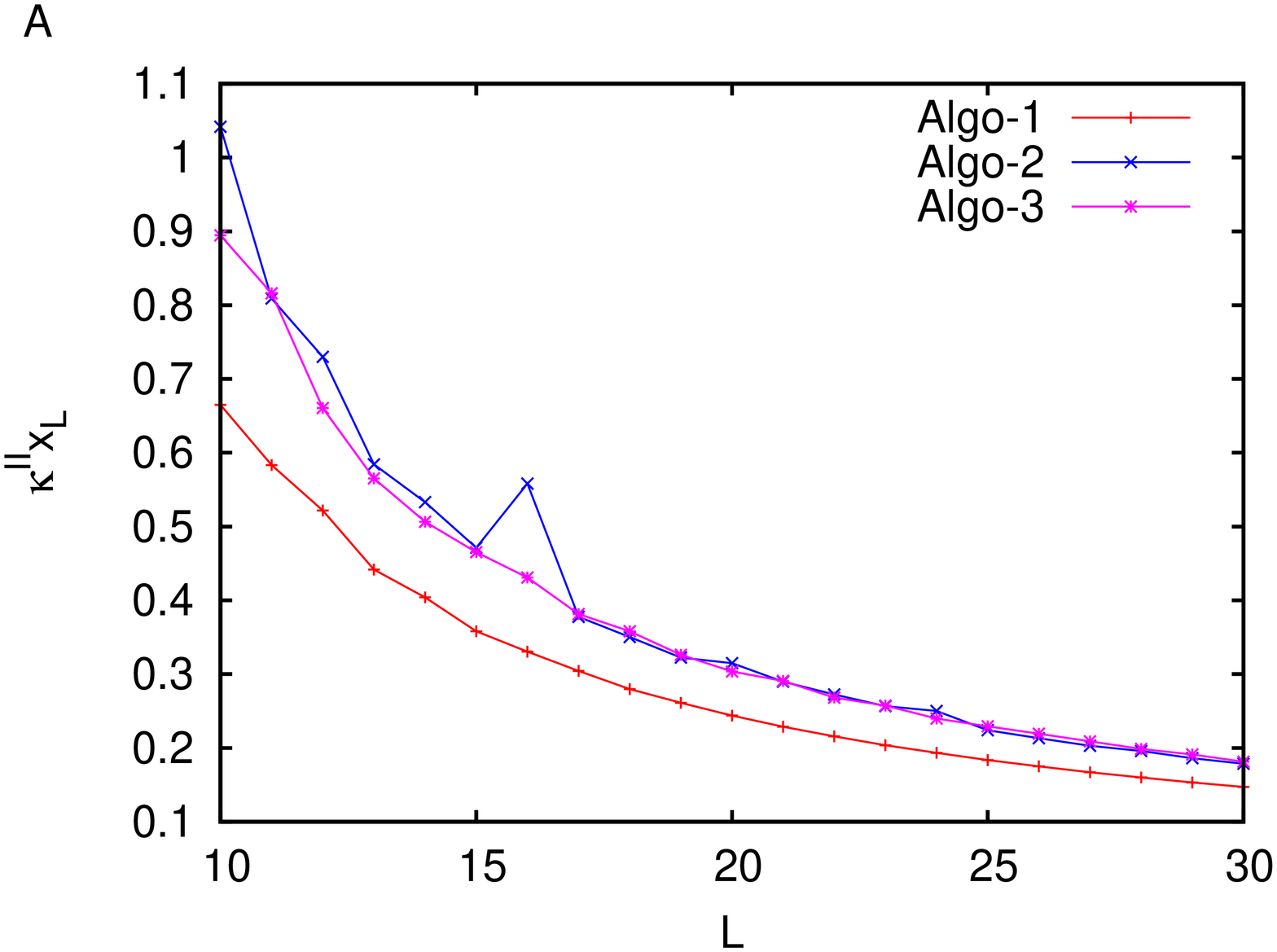}
        \includegraphics[height=3in,angle=-90]{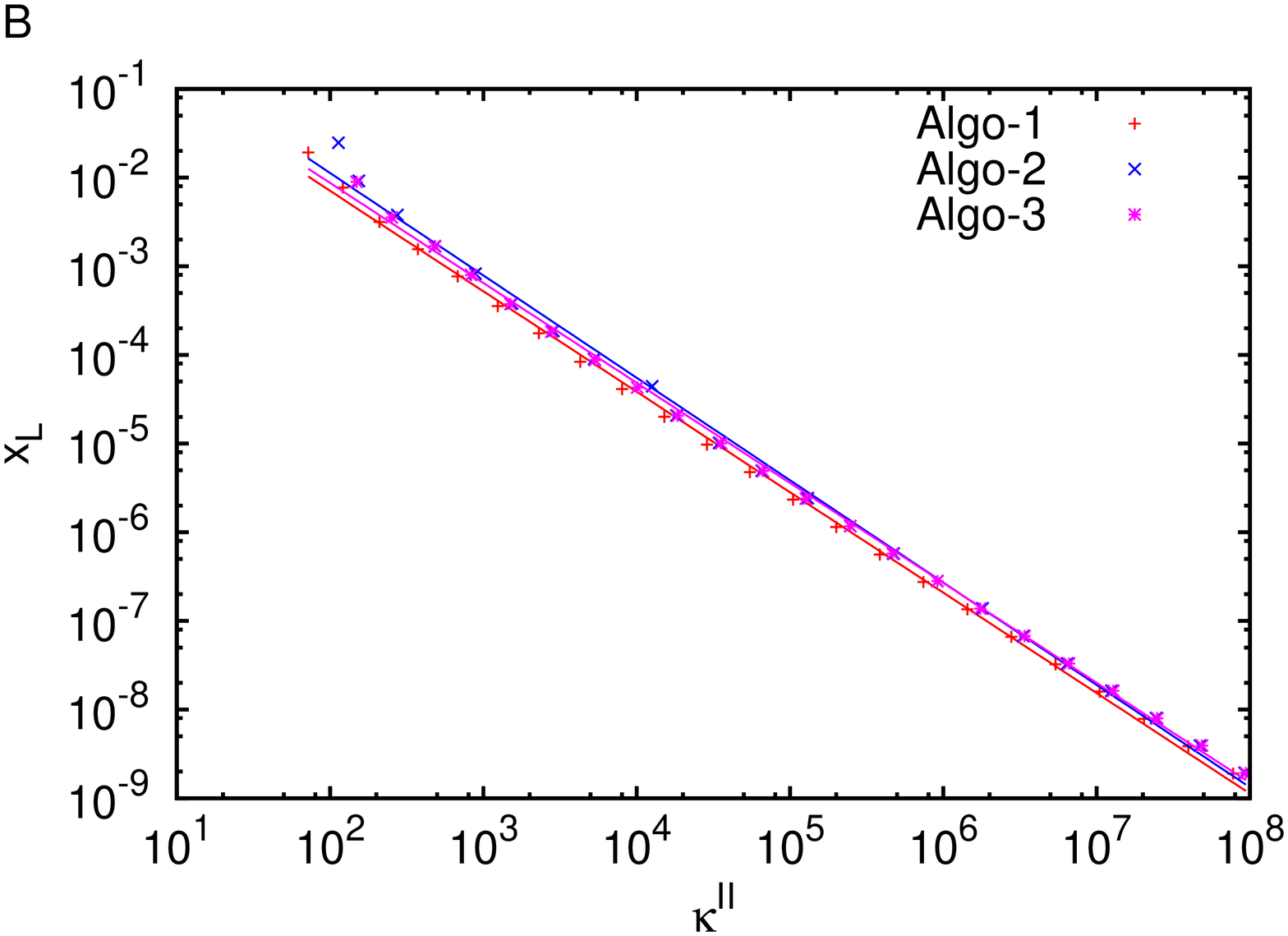}
    \end{center}
    \caption{{\bf The product $\kappa^{II}x_L$ is of order unity.} This figure is produced from the same data as was used for Fig. \ref{kappa1-2-vs-L}. Simulations were done for chemistries containing extremal ACSs of length $L$ generated by the three algorithms discussed earlier, represented in the figure by different colours. For this figure each chemistry was simulated at a value of $\kappa$ equal to the initiation threshold $\kappa^{II}$ corresponding to that chemistry, and the steady state concentration $x_L$ of the catalyst was determined in the low fixed point (starting from the standard initial condition). The parameters values are the same as in Fig. \ref{kappa1-2-vs-L}. (A) The product of $\kappa^{II}$ and $x_L$ as a function of $L$. (B) $x_L$ versus $\kappa^{II}$ on a log-log plot. The slopes of the fitted straight lines vary in the range -1.13 to -1.16 for the three algorithms (slope = -1 would have meant that $\kappa^{II}x_2$ is strictly constant. The figure shows that while each individual factor $\kappa^{II}$ and $x_L$ ranges over several orders of magnitude, their product, though not constant, is of order unity.}
    \label{kappa2-xL-relationship}
\end{figure}

\begin{figure}[ht]
    \begin{center}
        \includegraphics[height=5.5in,angle=-90,trim=2cm 0cm 5cm 0cm,clip=true]{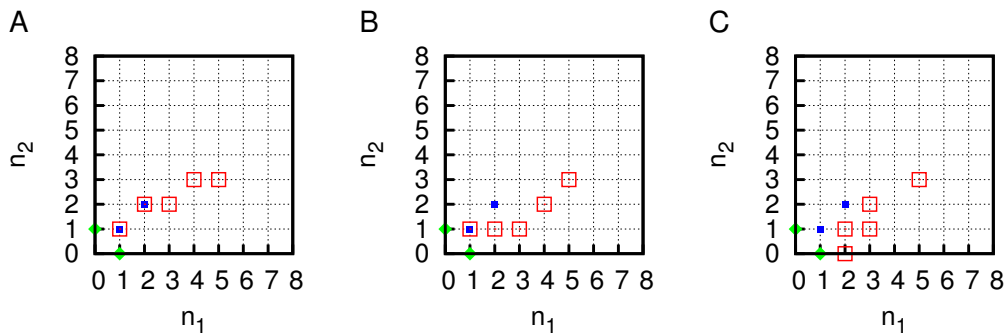}
    \end{center}
    \caption{{\bf Examples of nested ACS pairs with different degrees of overlap for $f=2$.} In the three cases the reaction sets have (A) maximal overlap, (B) partial overlap, (C) no overlap. The blue and red squares marking the grid points indicate the identity of molecules produced in the two ACSs; blue filled squares correspond to the products of the smaller ACS, red unfilled squares to those of the larger ACS. The $x$ and $y$ axes denote the number of monomers of type (1,0) and (0,1), respectively, in the molecules. The green rhombuses represent the two monomers.}
    \label{acs-2d-nested-4-8}
\end{figure}

\begin{figure}[ht]
    \begin{center}
        \includegraphics[height=2in,angle=-90]{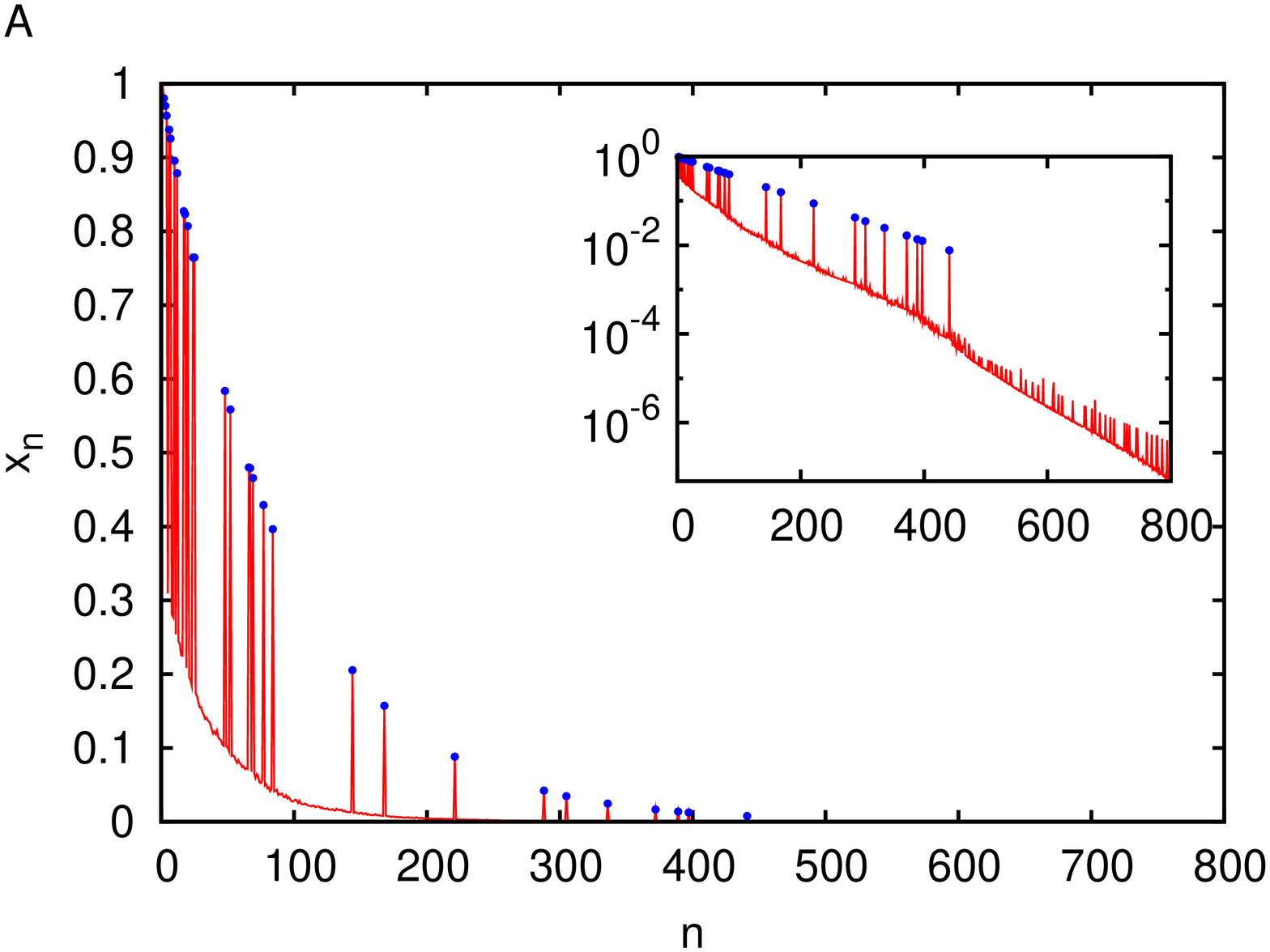}
        \includegraphics[height=2in,angle=-90]{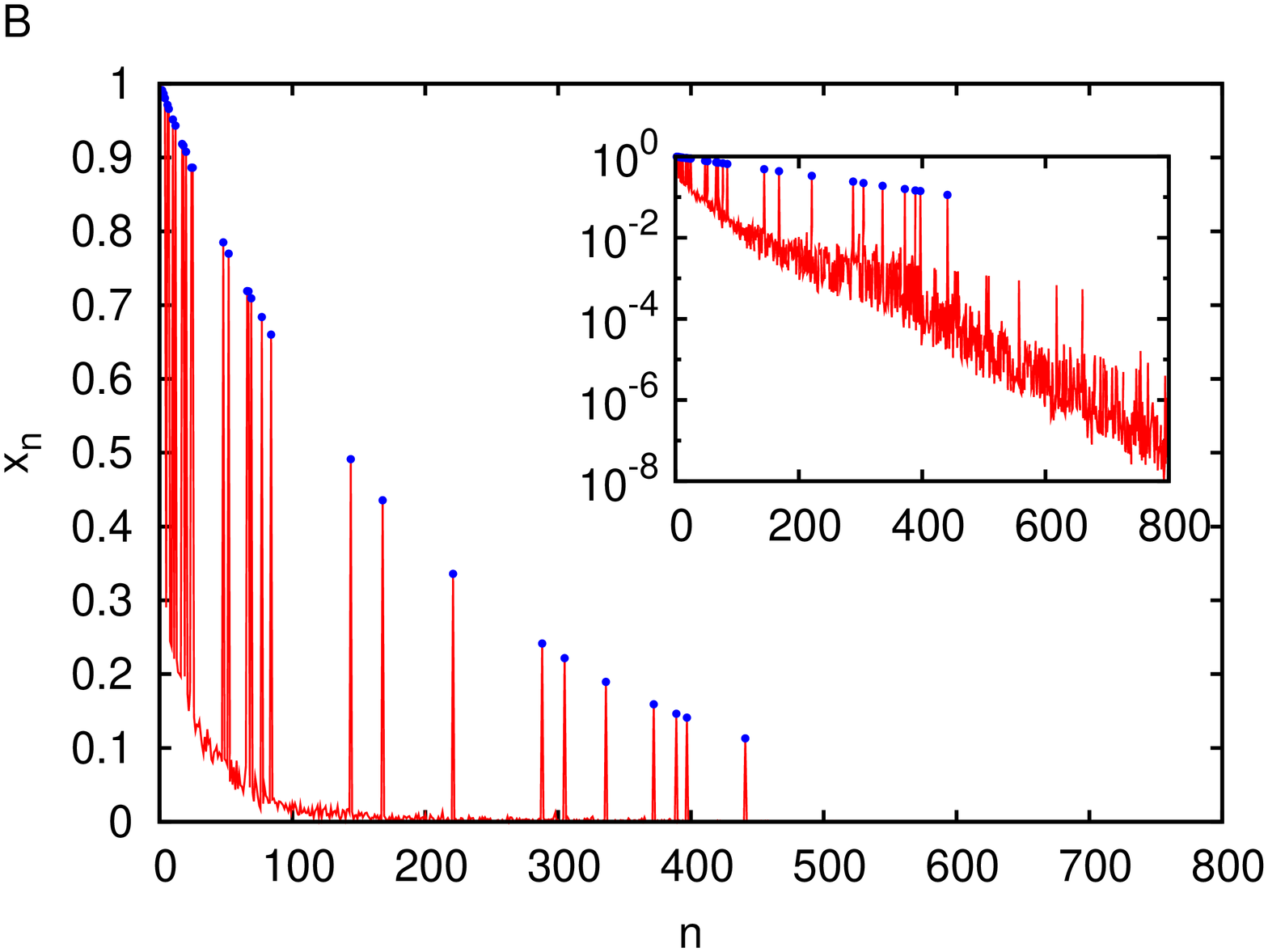}
        \includegraphics[height=2in,angle=-90]{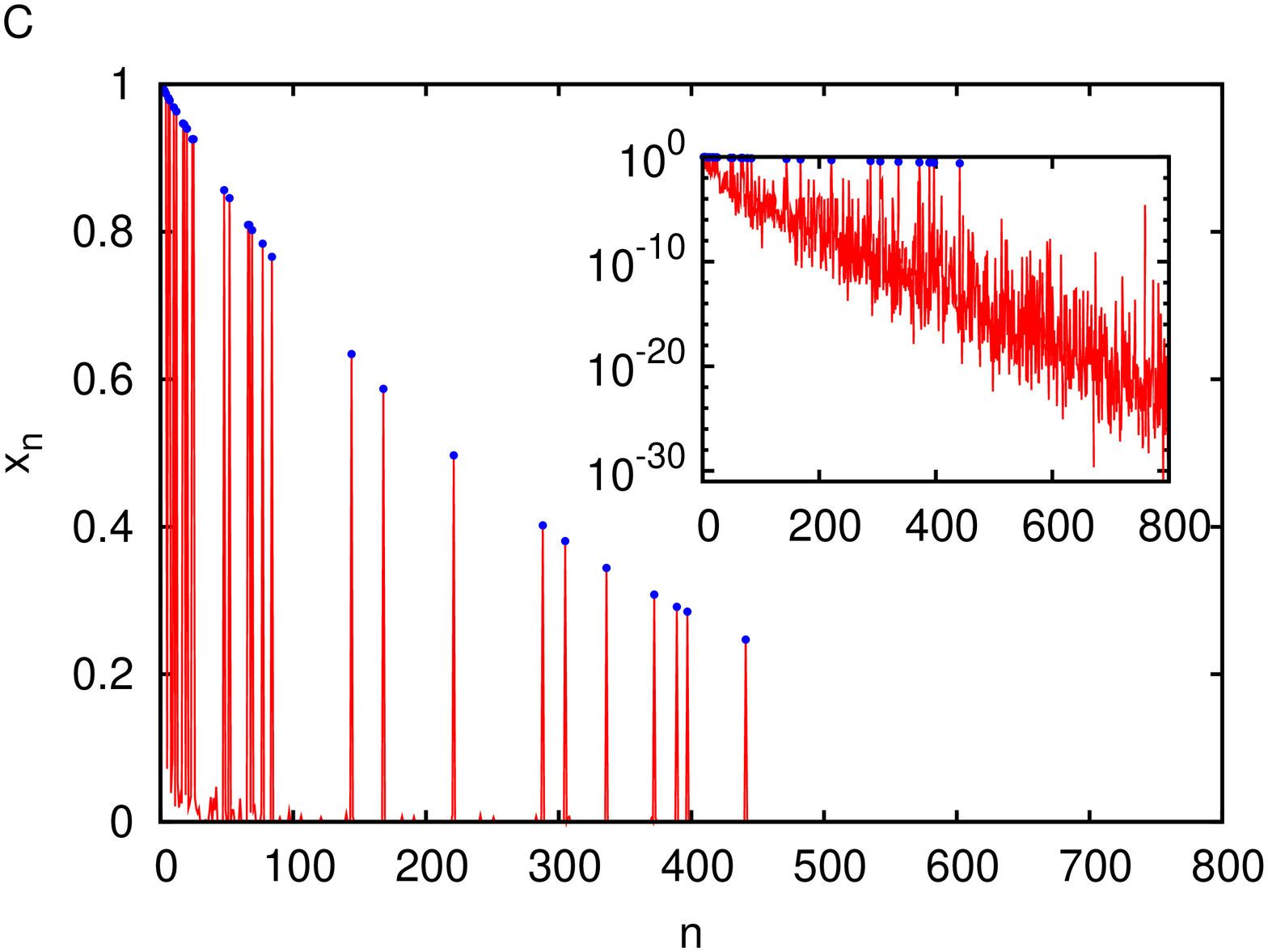}
    \end{center}
    \caption{{\bf The dominance of a cascade of nested ACSs with a molecule of size 441 (ACS441).} The molecules and reactions of this ACS are listed in Supporting Table S2. The red curves show the steady state concentration $x_n$ of all the molecules as a function of their size $n$, starting from the standard initial condition; blue dots show the concentrations of the ACS molecules. Insets show the same on a semi-log plot. It is evident that the large ACS molecules acquire a significant concentration. The catalytic strengths of the ACS molecules depend upon their size $n$ according to $\kappa(n) = 500 \times n^{1.5}$, and for all cases $k_f = k_r = 1$, $\phi = 50$, $N = 800$. The three figures differ in the level of sparseness of the spontaneous chemistry in which the ACS is embedded. The spontaneous chemistry in (A) is fully connected, in (B) has degree 20, and in (C) has degree 2.}
    \label{1d-algo4}
\end{figure}

\begin{figure}[ht]
    \begin{center}
        \includegraphics[height=3in,angle=-90]{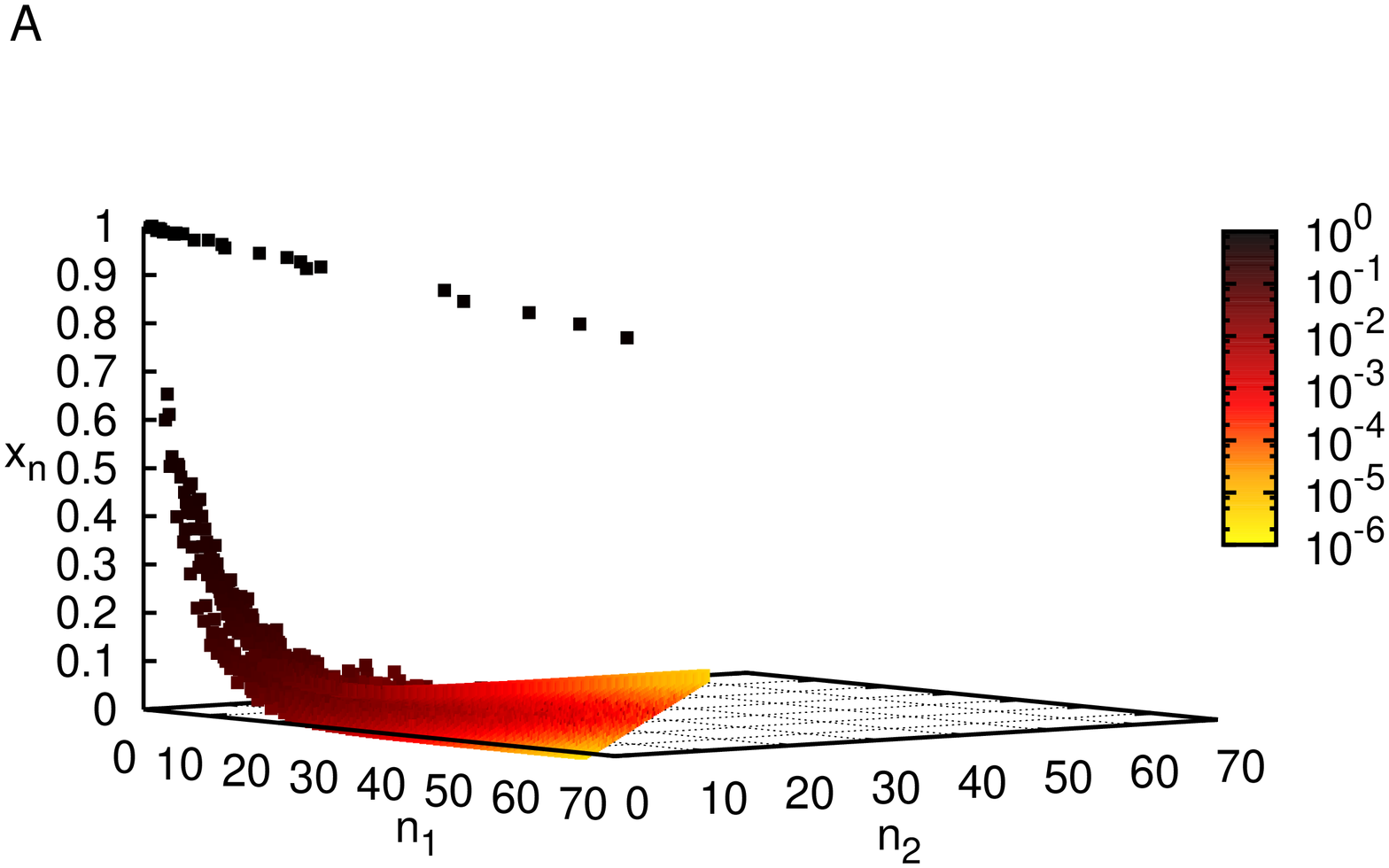}
        \includegraphics[height=3in,angle=-90]{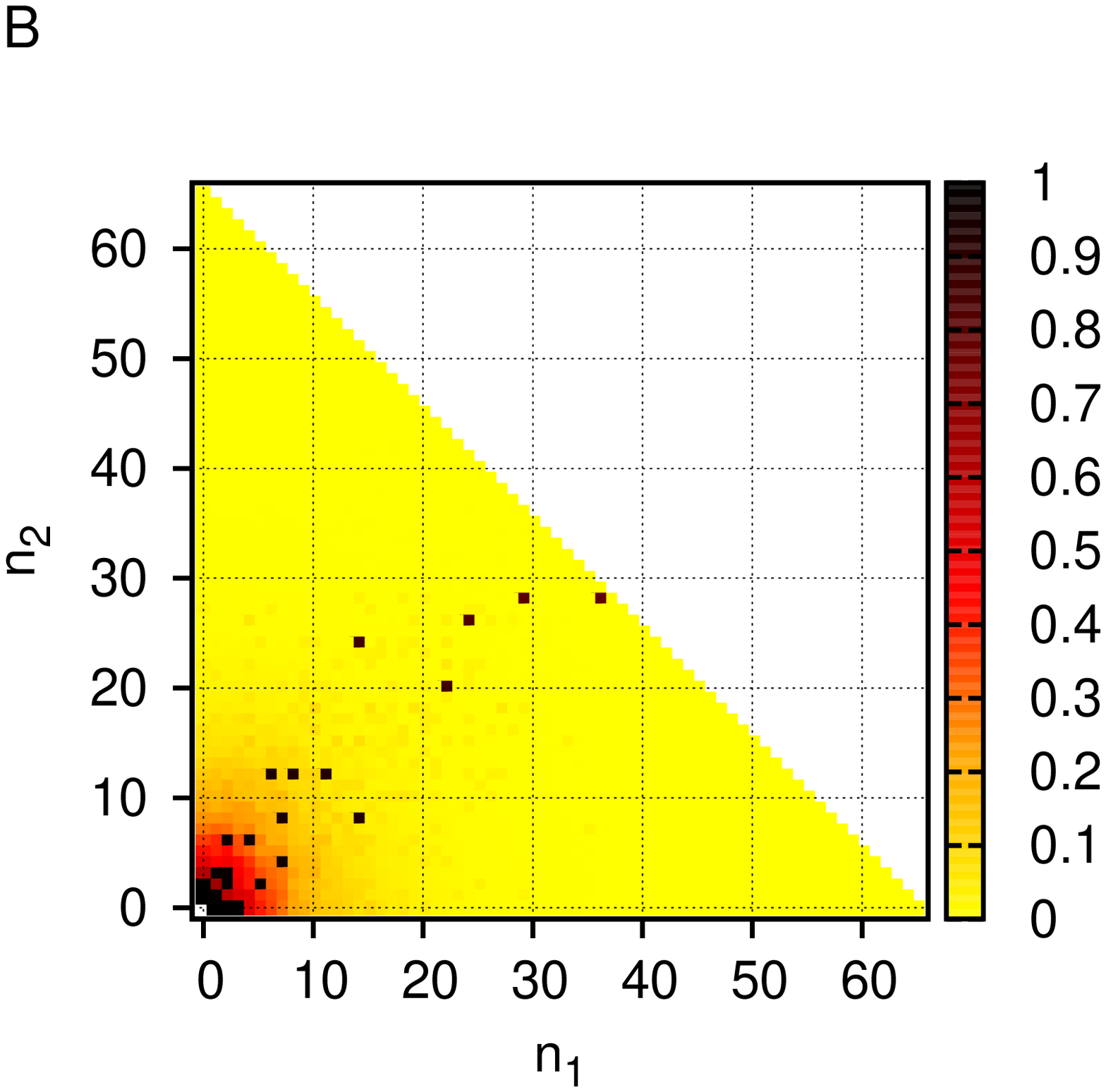}
    \end{center}
    \caption{{\bf Dominance of a cascade of nested ACSs with length 64 (ACS(36,28)) in a $f=2$ chemistry.} (A) 3D plot showing the steady state concentration $x_n$ of the molecule $n=(n_1,n_2)$ as a function of $n_1$ and $n_2$, starting from the standard initial condition. The colour coding is on a logarithmic scale of the concentration. (B) A `top view' of the same so that the ACS molecules and background are more clearly distinguished. The colour coding here is on a linear scale of concentration. The food set and ACS molecules have the highest concentrations and stand out as black dots. The catalytic strengths of the ACS molecules depend upon their size $L \equiv n_1 + n_2$ according to $\kappa(L) = 500 \times L^{1.5}$, and $k_f = k_r = 1$, $\phi = 10$, $N = 65$. The spontaneous chemistry has degree 20.}
    \label{2d-algo4}
\end{figure}

\begin{figure}[ht]
    \begin{center}
        \includegraphics[height=3.25in,angle=-90]{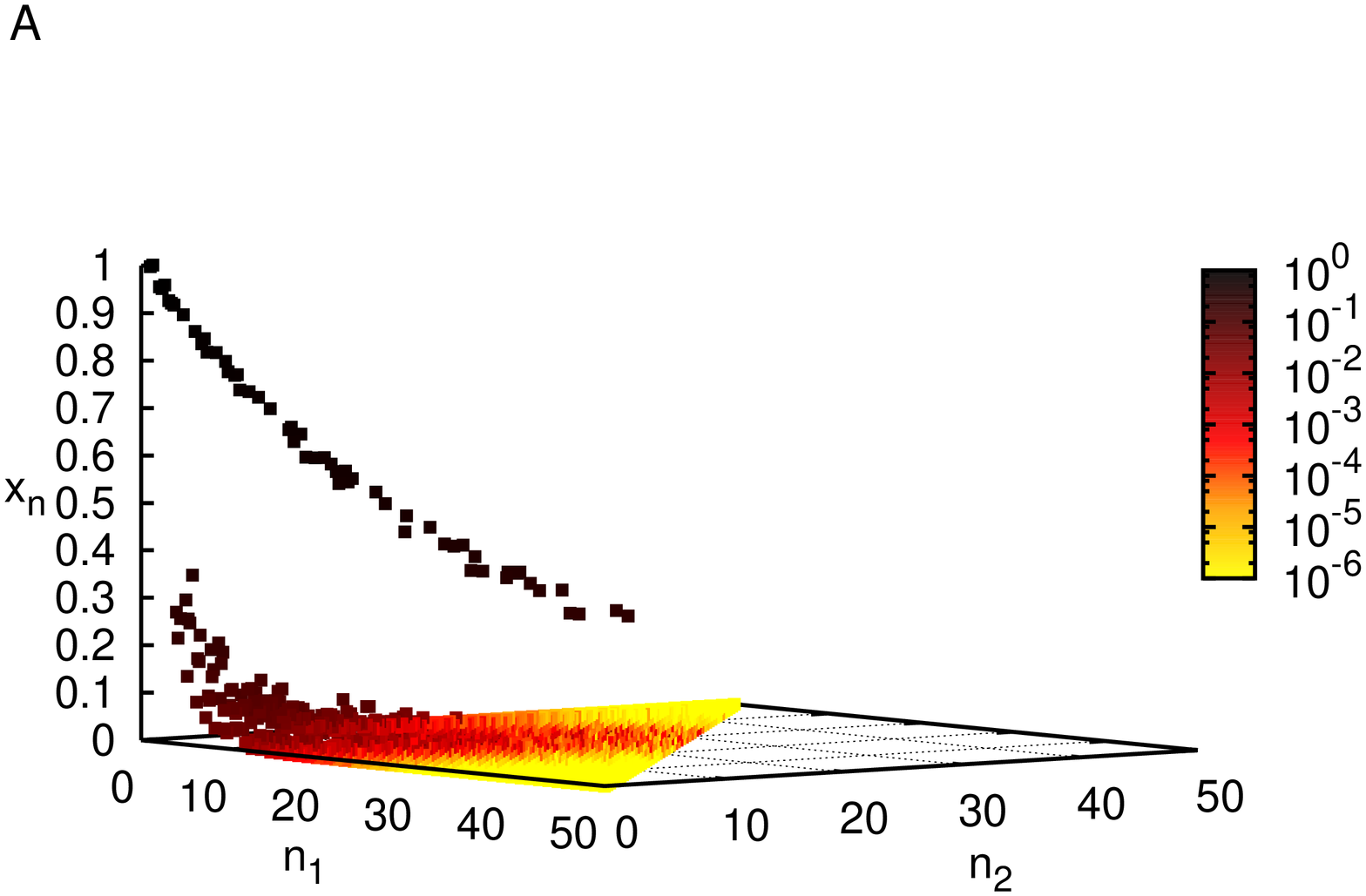}
        \includegraphics[height=3.15in,angle=-90]{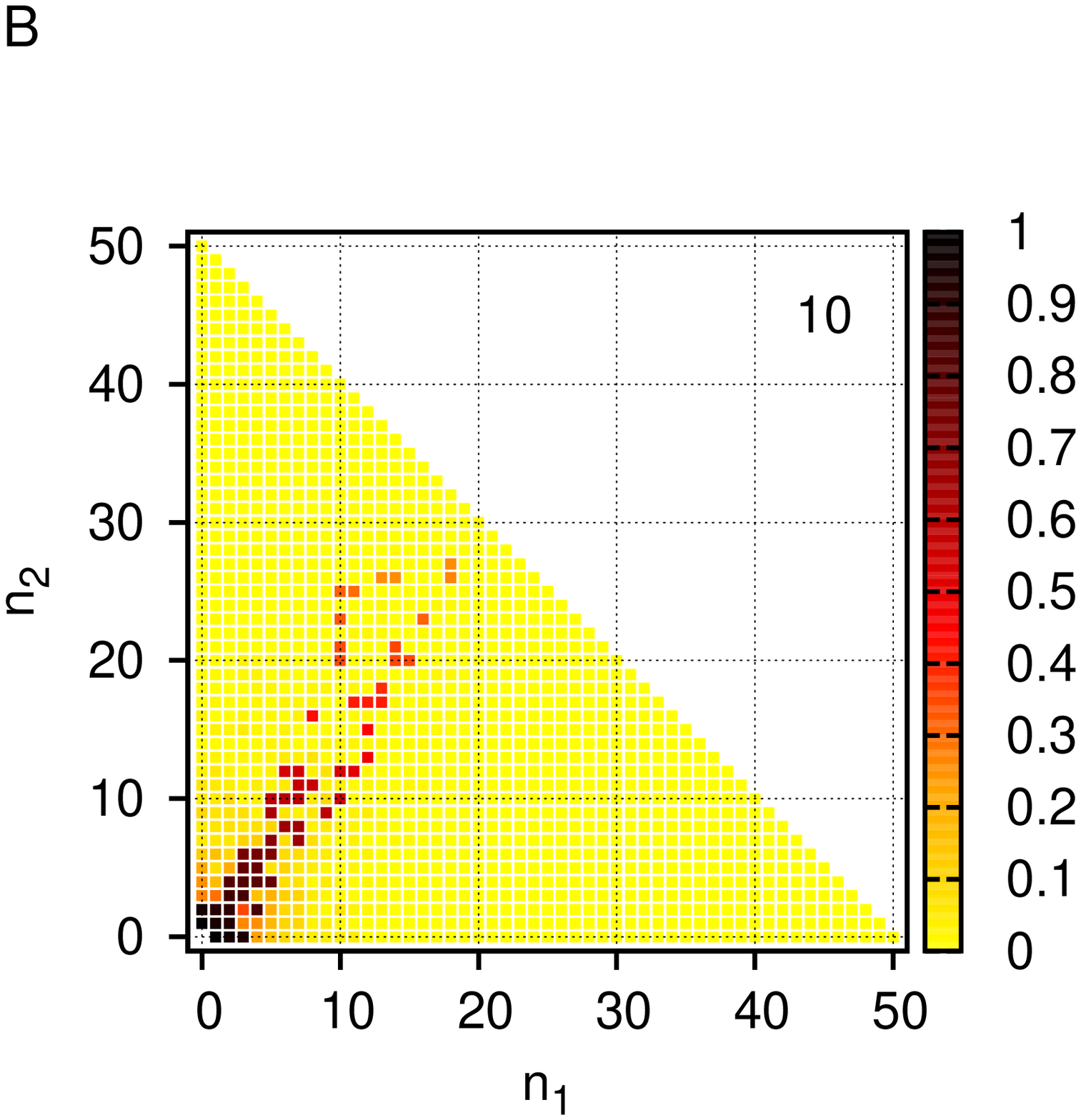}
        \includegraphics[height=3.25in,angle=-90]{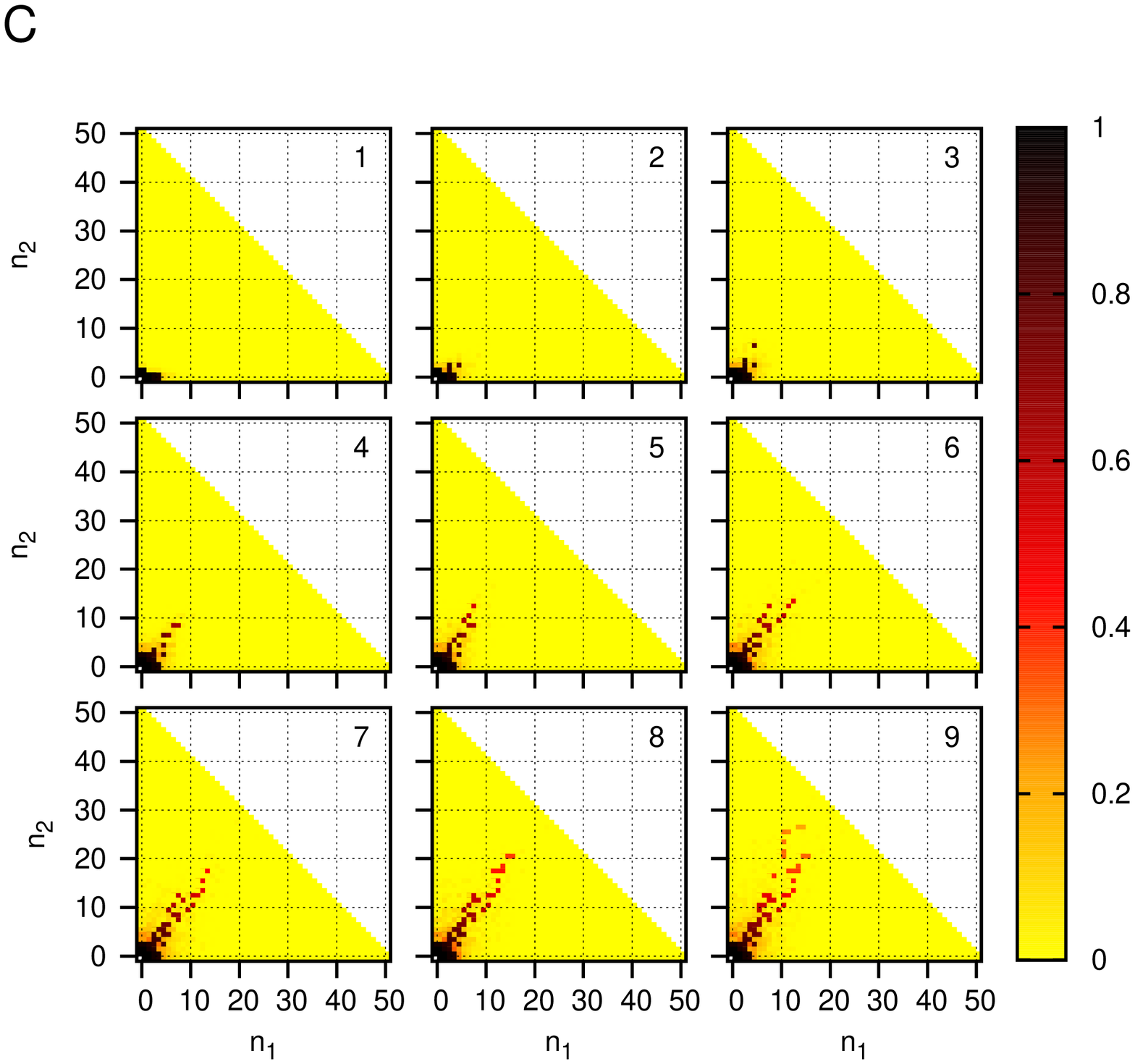}
    \end{center}
    \caption{{\bf Dominance of a cascade of partially overlapping nested ACSs (ACS(18,27)).} (A) Steady state concentration profile starting from the standard initial condition. (B) Top view of the same. (C) Sequence of steady state concentration profiles as each successive ACS is added to the chemistry. The legend for (A) is the same as for Fig. \ref{2d-algo4}A and for (B) and (C) the same as Fig. \ref{2d-algo4}B. $k_f = k_r = 1$, $\phi = 15$, $N = 50$, and the spontaneous chemistry has degree 5.}
    \label{2d-partial-nest}
\end{figure}

\setcounter{figure}{0}
\makeatletter
\renewcommand{\thefigure}{S2.\@arabic\c@figure}
\makeatother

\begin{figure}[h]\centering
\includegraphics[width=3in,angle=-90,trim=0.1cm 0.5cm 0cm 0cm,clip=true]{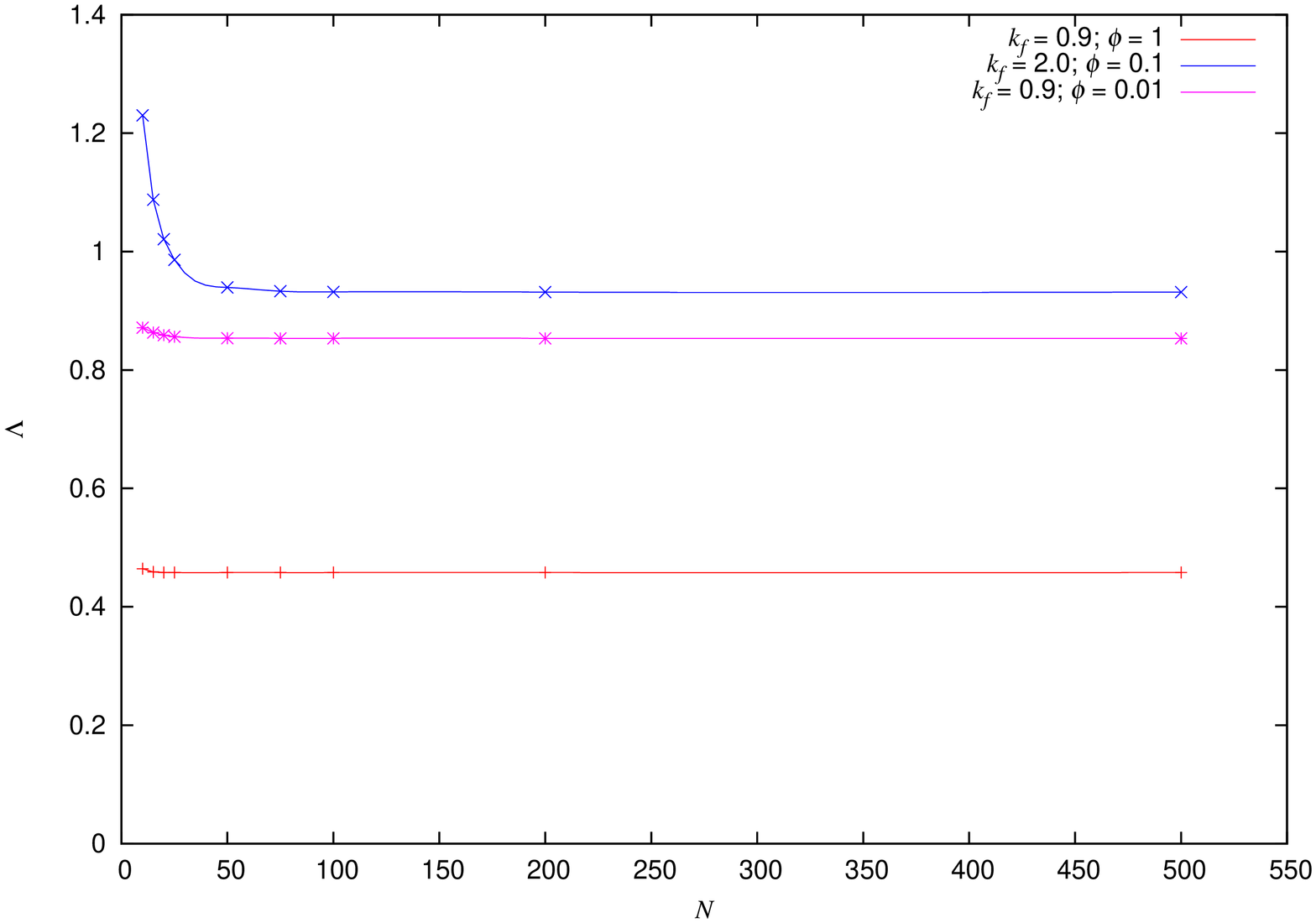}
\caption{\label{Nvslamda}{\bf Dependence of $\Lambda$ on $N$ for uncatalyzed chemistries.} $A=k_r=1$. We determine $\Lambda$ at different values of $N$ by fitting the steady state profiles to an exponential, $x_n = ce^{-\gamma n}$ (excluding the concentrations $x_1$ to $x_4$). We see a dependence of $\Lambda$ on $N$ for values of $N<50$, but $\Lambda$ becomes essentially independent of $N$ for $N>100$.
}
\end{figure}

\begin{figure}[h]\centering
\includegraphics[width=3in,angle=-90,trim=0.1cm 0.5cm 0cm 0cm,clip=true]{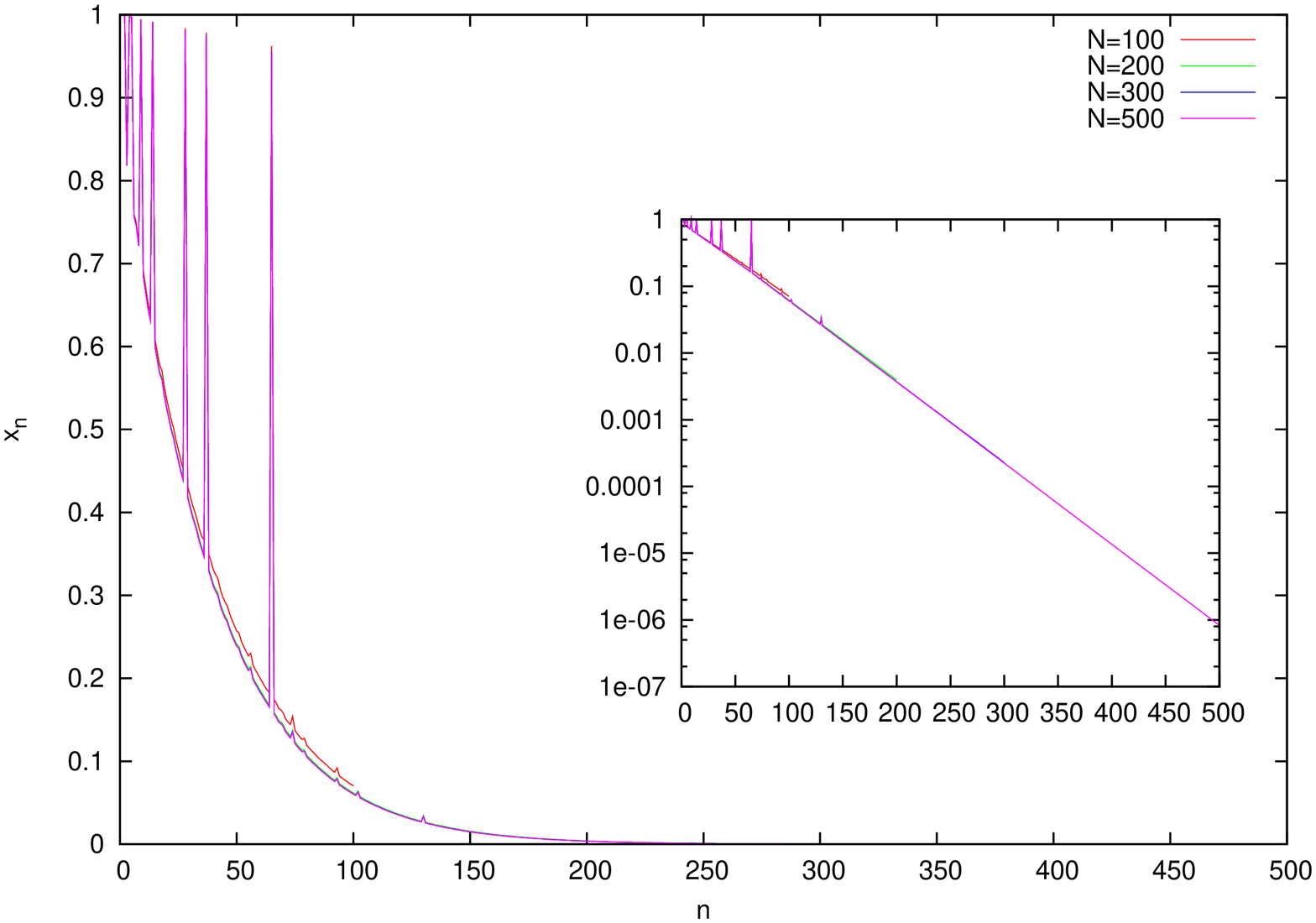}
\caption{\label{N-depend-acs}{\bf Dependence of steady state concentrations on $N$ for a chemistry that includes an ACS.} The figure shows the steady state concentrations for the chemistry that includes ACS65 (Eq. 5 in main text) for $N=100, 200, 300, 500$. In all cases $A = k_f = k_r = 1, \phi=5$. The steady state concentrations show an $N$-dependence upto $N=100$, but for $N\ge200$ the profile becomes $N$-independent.}
\end{figure}

\setcounter{figure}{0}
\makeatletter
\renewcommand{\thefigure}{S3.\@arabic\c@figure}
\makeatother

\begin{figure}
  \begin{center}
    \includegraphics[width=3in,angle=-90,trim=1.75cm 1.5cm 2cm 1cm,clip=true]{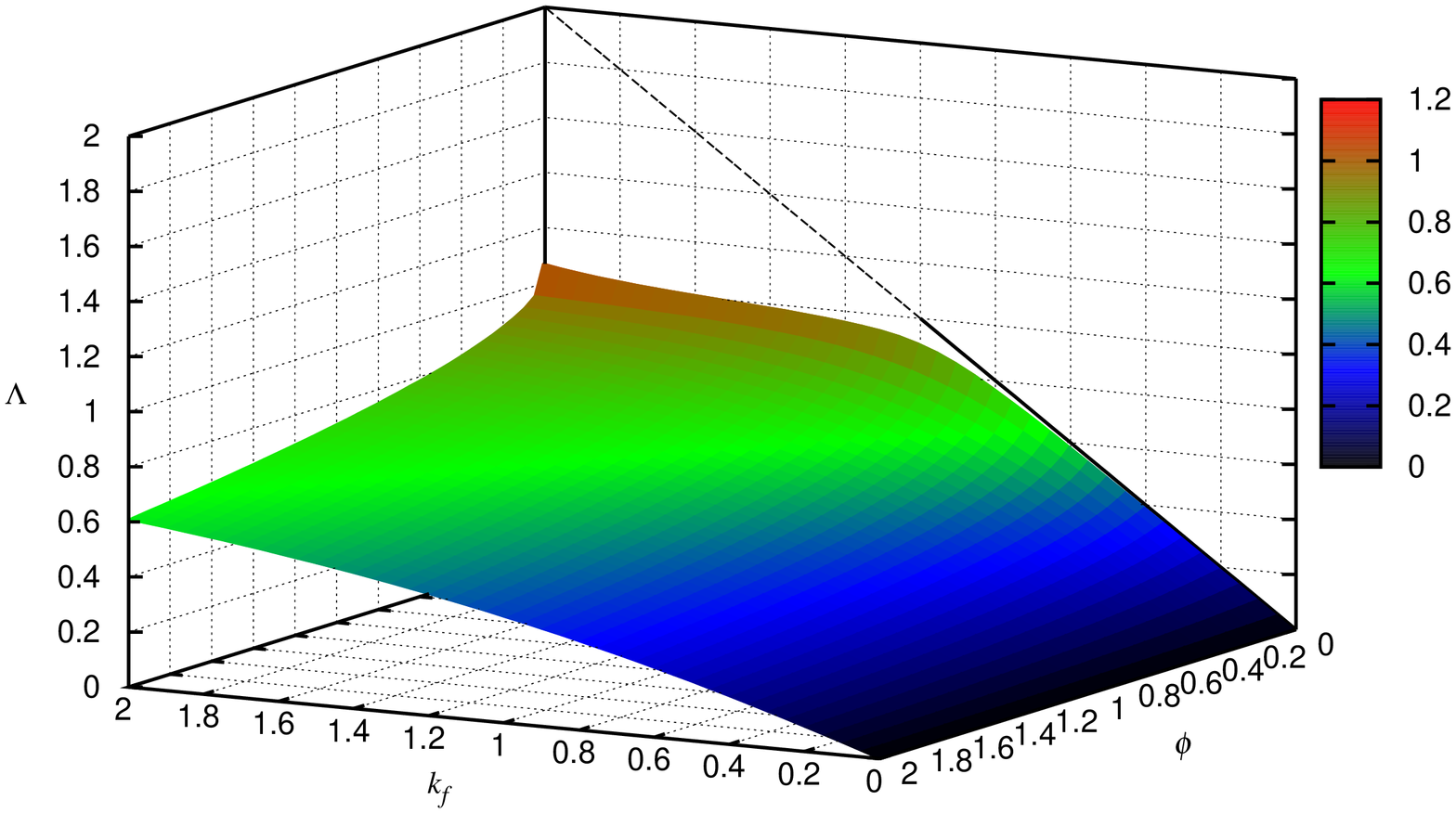}
  \end{center}
    \caption{{\bf Behaviour of $\Lambda$ as a function of $k_f$ and $\phi$.} The figure shows the dependence of $\Lambda$ on $k_f$ and $\phi$ for an uncatalyzed chemistry, keeping $A = k_r = 1$, $N=100$. The curved surface was made with parameter values in the range $0 \leq k_f \leq 2$, $0.01 \leq \phi \leq 2$. $\Lambda$ is found to be a monotonically increasing function of $k_f$ and a monotonically decreasing function of $\phi$. For $\phi=0$, there is an analytical solution $\Lambda = k_f$ (see Eq. (4) in the main text). This was verified numerically in the range $0 \leq k_f \leq 1$ (see solid line at $\phi=0$). In the region $k_f > 1$ the numerical integration does not converge at $\phi=0$ as the steady state solution (Eq. (4) of the main text) $x_n = A\Lambda^{n-1} = Ak_f^{n-1}$ is numerically very large for large $n$. The dotted extension of the line ($1 < k_f \le 2$) is simply the analytical result. Note that for most of the phase-space $\Lambda < 1$, except for very small values of $\phi$.}
  \label{Kf-Phi-Phasespace}
\end{figure}

\begin{figure}
  \begin{center}
    \includegraphics[width=3in,angle=-90,trim=1.75cm 1.5cm 2cm 1cm,clip=true]{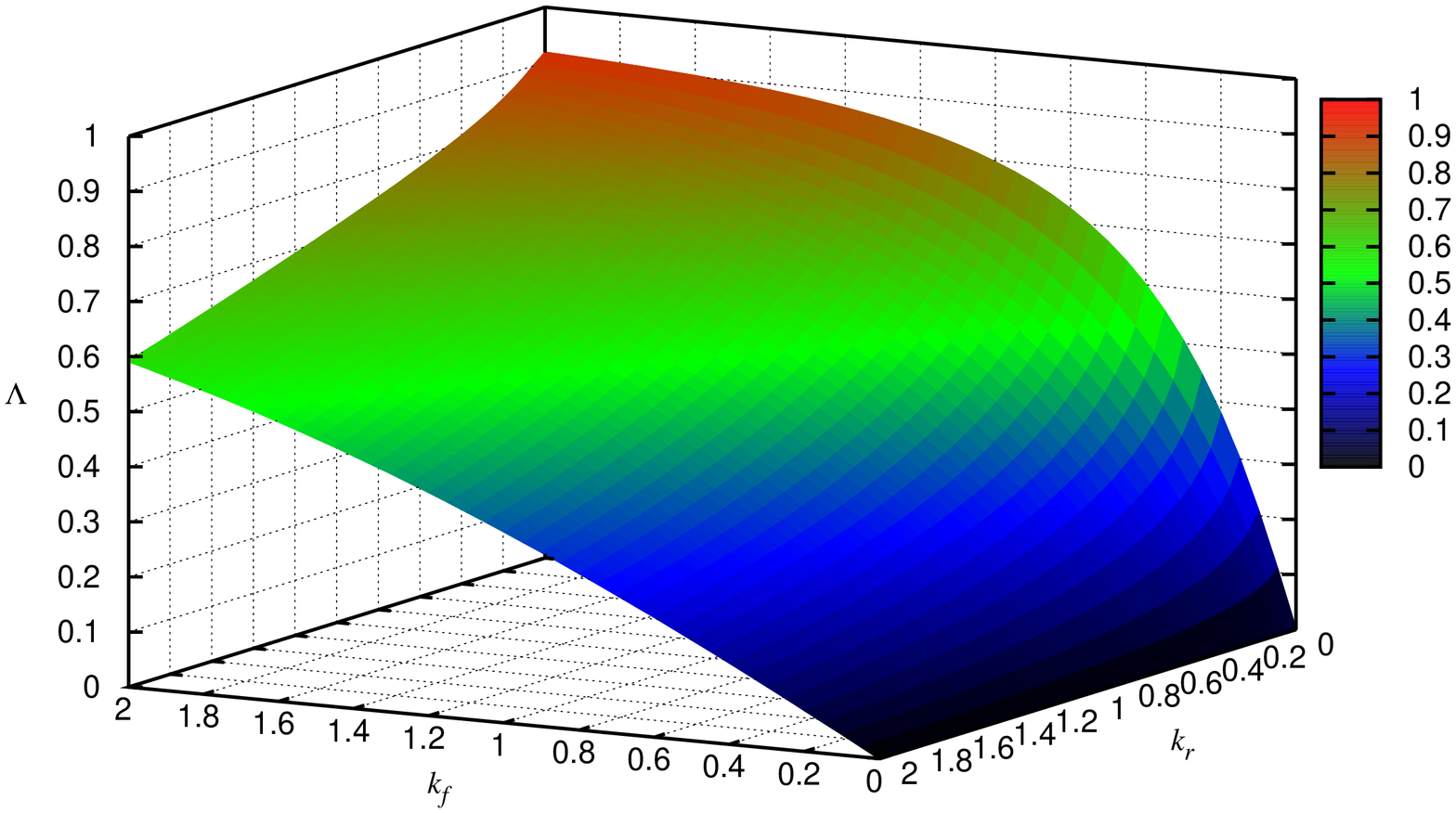}
  \end{center}
    \caption{{\bf Behaviour of $\Lambda$ as a function of $k_f$ and $k_r$.} The figure shows the dependence of $\Lambda$ on $k_f$ and $k_r$ for an uncatalyzed chemistry, keeping $A = \phi = 1$, $N=100$. $\Lambda$ is found to be a monotonically increasing function of $k_f$ and a monotonically decreasing function of $k_r$.}
  \label{Kf-Kr-Phasespace}
\end{figure}

\end{document}